%% file: ReviewPreprint.tex
\documentclass{ws-ijmpa-modified}

\begin{document}

\markboth{Robert M. Harris and Konstantinos Kousouris}
{Searches for Dijet Resonances at Hadron Colliders}

%
\catchline{}{}{}{}{}
%

\title{SEARCHES FOR DIJET RESONANCES AT HADRON COLLIDERS}

\author{ROBERT M. HARRIS}

\address{Fermilab\\
Batavia, Illinois   60510\\
United States of America\\
rharris@fnal.gov}

\author{KONSTANTINOS KOUSOURIS\footnote{
Also affiliated with Fermilab during this review.}}

\address{CERN\\
Geneva\\
Switzerland\\
Konstantinos.Kousouris@cern.ch}

\maketitle


\begin{abstract}
We review the experimental searches for new particles in the dijet mass spectrum conducted 
at the CERN $S\bar{p}pS$, the Fermilab Tevatron Collider, and the CERN Large Hadron Collider.
The theory of the QCD background and new particle signals is reviewed, with emphasis on the 
choices made by the experiments to model the background and signal. The experimental techniques, 
data, and results of dijet resonance searches at hadron colliders over the last quarter 
century are described and compared. Model independent and model specific limits on new 
particles decaying to dijets are reviewed, and a detailed comparison is made of the 
recently published limits from the ATLAS and CMS experiments.

\end{abstract}


\input{Introduction}

\input{Theory}

\input{Experiment}

\input{Conclusions}

\input{Acknowledgements}

\appendix

\input{AppCMSandATLAS}


\end{document}

%% file: Introduction.tex
\section{Introduction}	

\begin{figure}[hbt]
\centerline{
\psfig{file=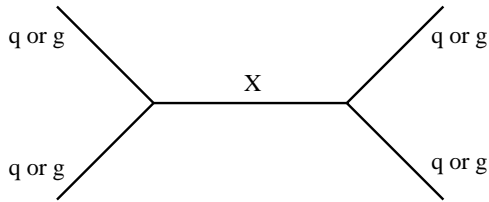,width=2.5in}
}
\vspace*{8pt} 
    \caption{ Diagram of dijet resonance in the $s-\text{channel}$. The initial state
and final states contain two partons (quarks, anti-quarks or
gluons) and the intermediate state contains a resonance $X$.
\label{feyn}
}
\end{figure}

Experiments at hadron colliders have used the dijet mass spectrum to 
search for new particles beyond the standard model. At the 
CERN $S\bar{p}pS$, the Fermilab Tevatron Collider, and the CERN Large 
Hadron Collider, with each successive advance in collision energy and 
integrated luminosity, progressively more energetic collisions of the partons 
in the incoming hadrons are produced and observed.  Each machine in its time 
has therefore probed the highest masses of dijet resonances: new 
particles that decay into two partons, giving two jets in the final state. The 
simple process the experiments have searched for is the 
$s-\text{channel}$ production and decay of dijet resonances shown in Fig.~\ref{feyn}.  

Here we review these experimental searches, their techniques, data, results,
and limits on new particles. In section~\ref{secTheory} we review the theory
of the QCD background and the models of new particle signals.  In 
section~\ref{secExperiment} we review the experiments from each collider 
in chronological order.  In section~\ref{secResonanceShapes} we review how each experiment modeled the
resonance shapes as a function of the dijet mass. In 
section~\ref{secData} we review the data of each experiment and how each
experiment modeled the QCD background.
In section~\ref{secLimit} we review the limits on dijet resonance masses 
published by each experiment, discussing the experimental uncertainties,
statistical procedures, and the cross section assumed for each model.
In section~\ref{secConclusions} we conclude with a few observations.
Also, in~\ref{sectionCompare}, we include details of the 
cross-section calculations for axigluons and excited quarks by ATLAS and CMS,    
which are necessary to understand the mass limits on these models from the two  
experiments.

%% file: Theory.tex
\section{Theory}\label{secTheory}

In this section we present the fundamental ingredients of the theory, which are necessary for the better understanding of the experimental searches presented in this review. In Section~\ref{sec:QCD} we describe some basic features of Quantum Chromodynamics (QCD), and in Section~\ref{sec:models} we present the theoretical models that predict partonic resonances and are quoted in the experimental searches. It should be noted, that the purpose of this section is not to give all the details of the models presented, but rather an overview of their features.

\subsection{Elements of QCD}\label{sec:QCD}

\subsubsection{The QCD Lagrangian}

Quantum Chromodynamics is the gauge field theory of the strong interaction between particles that carry the \textit{color} degree of freedom. The underlying symmetry group is the $\text{SU}(3)_C$, which makes QCD a non-Abelian theory. The profound implication of this property of QCD is that the gauge mediators (gluons) are colored and thus self interacting. The QCD Lagrangian is written as:  
\begin{equation}
  \mathcal{L}_{QCD} = \sum_i{\bar{q}_{i,a}\left(i\gamma^\mu\partial_\mu\delta_{ab}-g_s\gamma^\mu t^C_{ab} G_\mu^C-m_i\delta_{ab}\right)q_{i,b}}-\frac{1}{4}F_{\mu\nu}^AF^{\mu\nu,A},
\end{equation}
where $q_{i,a}$ represents the quark spinor of flavor $i$ and color $a=1\rightarrow 3$, $G_{\mu\nu}^A$ is the gluon field associated with the generator $t^A_{ab}$ ($A=1\rightarrow 8$), $g_s$ is the gauge coupling, and $F_{\mu\nu}^A$ is the gluon field tensor:
\begin{equation}
  F_{\mu\nu}^A = \partial_\mu G_\mu^A - \partial_\nu G_\nu^A - g_s f_{ABC}G^B_\mu G^C_\nu.
\end{equation}
The structure constants $f_{ABC}$ satisfy the relation:
\begin{equation}
  \left[t^A,t^B\right] = if_{ABC}t^C.
\end{equation}

The non-Abelian nature of QCD leads to two remarkable features: the \textit{confinement} and the \textit{asymptotic freedom}. As a result of the confinement, only color-singlet states can be directly observed, which means that quarks and gluons cannot be found free. The asymptotic freedom is the property where the running strong coupling constant decreases with increasing momentum transfer between the strongly interacting particles. This in turn means, that the hard-scattering of quarks and gluons can be described in a perturbative way.

\subsubsection{Formation of jets}

Because of the confinement, partons cannot be detected free. Instead, the experimental signatures of quarks and gluons are the 
\textit{jets}. A jet is a "spray" of highly collimated particles, primarily hadrons, but also photons and leptons. A jet is a not uniquely defined object, but the output of a well-defined mathematical rule (\textit{clustering algorithm}), which clusters the jet constituents, according to their kinematic properties. This procedure is based upon the features of QCD, which describe the transformation of a parton to a set of observable particles. The jet-formation steps are the following:
\begin{itemize}
  \item {\bf\it Parton branching}: each parton, whether a gluon or a quark, has a finite probability to split into two partons, which are emitted in small angles with respect to the direction of the initial parton. One feature of the parton branching is that the probability depends on the color factor related to the type of the involved partons. For $\text{gluon}\rightarrow\text{gluon},\text{gluon}$, $\text{gluon}\rightarrow\text{quark},\text{antiquark}$, and $\text{quark}\rightarrow\text{gluon},\text{quark}$ splittings, the color factors are $C_{gg} = 3$, $C_{qq} = \frac{4}{3}$, and $C_{qg} = \frac{1}{2}$ respectively. As a result, gluons systematically shower more than quarks. Another implication of parton branching in small angles is that throughout the process, partons are produced close to the direction of the initial partons, which results in a high degree of collimation of the final hadrons. It should be noted, that the parton branching is a perturbative procedure, which can be re-summed to all orders of the perturbation series, under certain assumptions.

\item {\bf\it Hadronization}: when the parton shower has evolved long enough, the energy of the partons is reduced, such that low-momentum transfer occur. In these conditions, the parton interactions become non-perturbative, and the phase of hadronization begins. During the hadronization, partons are combined into color-singlet states, thus forming the hadrons. While the hadronization procedure cannot be described perturbatively, the \text{local parton-hadron duality} ensures that the flow of quantum numbers at the hadron level, follows approximately the corresponding flow at parton level.

\item {\bf\it Underlying event \& out-of-cone showering}: the term {\it underlying event} in hadron collisions is used to describe the activity not related to the hard scattering, for example due to multiple parton interactions happening simultaneously. Since the definition of jets involve the clustering of hadrons which are sufficiently correlated, it can happen that particles originating from the soft interactions are clustered together with those coming from the hard-scattered parton shower. In the opposite direction, partons from the initial shower can be emitted in relatively large angles, and the associated hadrons may not be clustered in the resulting jets. This effect is commonly known as {\it out-of-cone showering}.      
\end{itemize} 
 
Despite the fact that the formation of jets is a complicated effect, certain conclusions can be drawn: to first approximation, the kinematical properties of a jet are the same as those of the original parton. However, the various effects involved, introduce an intrinsic resolution of the hadronic jet properties with respect to the parton properties.

\subsubsection{Kinematics of two-parton scattering}

Before the details of the strong interaction dynamics are discussed, it is useful to present the kinematical properties of a two-to-two parton scattering. In the topology of the $1+2\rightarrow 3+4$ scattering, some general kinematic relations hold, regardless of the details of the interaction. The Mandelstam variables of the process are defined as $\hat{s} = (p_1+p_2)^2$, $\hat{t} = (p_1-p_3)^2$, and $\hat{u} = (p_2-p_3)^2$, where $p_i$ are the four-momenta of the partons. For massless partons, the Mandelstam variables satisfy the relation $\hat{s}+\hat{t}+\hat{u}=0$ and two of those can be expressed as a function of the third one and the scattering angle $\theta^*$ in the center-of-mass frame:
\begin{equation}
  \hat{t} = -\frac{1}{2}\hat{s}\left(1-\cos\theta^*\right),\,\,\,\hat{u} = -\frac{1}{2}\hat{s}\left(1+\cos\theta^*\right).
\end{equation}
The rapidities\footnote{$y=\frac{1}{2}\ln\left(\frac{E+p_z}{E-p_z}\right)$} of the outgoing partons, in the center-of-mass frame, are opposite $(\pm y^*)$, due to transverse momentum conservation, and related to the scattering angle:
\begin{equation}
  \cos\theta^* = \tanh y^*.
\label{costheta}
\end{equation}
The Mandelstam variable $\hat{s}$ can be expressed in terms of the outgoing partons transverse momentum $p_T$ and $y^*$:
\begin{equation}
  \hat{s} = 4p_T^2\cosh^2y^*.
\end{equation}
In the laboratory frame, the rapidities $y_{3,4}$ of the outgoing partons are related to the rapidity of the center-of-mass frame $\bar{y}$ and to $y^*$ as:
\begin{equation}
  \bar{y} = \frac{y_3+y_4}{2},\,\,\,y^*=\frac{y_3-y_4}{2}.
\end{equation}
From the relations above, one can express the scattering angle at the center-of-mass frame as a function of the rapidities of the scattered partons at the laboratory frame:
\begin{equation}
  \cos\theta^* = \tanh\left(\frac{y_3-y_4}{2}\right)
\end{equation}
Each initial parton is carrying a fraction $x$ of the hadron momentum and the invariant mass of the two-parton system is expressed as:
\begin{equation}
  M^2=\hat{s}=x_1x_2s,
\end{equation}
where $x_{1,2}$ are the momentum fractions of the interacting partons and $\sqrt{s}$ is the colliding energy of the hadrons:
\begin{equation}
  x_1 = \frac{2p_T}{\sqrt{s}}\cosh y^*e^{\bar{y}},\,\,\,x_2 = \frac{2p_T}{\sqrt{s}}\cosh y^*e^{-\bar{y}}.
\end{equation}
Following from the relation above, the rapidity of the center-of-mass frame $\bar{y}$ can be expressed as a function of the momentum fractions:
\begin{equation}
  \bar{y} = \frac{1}{2}\ln{\frac{x_1}{x_2}}
\end{equation}

\subsubsection{Partonic cross sections}\label{QCDxsections}

The dynamics of the hard scatter of colliding hadrons is approximately described as a two-to-two process between massless partons. Because of the different structure and color factors of the interaction between the parton types, the matrix elements are different for each subprocess. The leading order (LO) amplitudes can be calculated analytically using the Feynman diagrams at tree level, and are summarized in Table~\ref{tab:LOXSEC}. The squared amplitudes are averaged (summed) over the initial (final) color and spin indices, and are expressed in terms of the Mandelstam variables. 

\begin{table*}[htbH]
  \tbl{Leading order matrix elements for 2-to-2 massless parton subprocesses
  from Ref.~\protect\refcite{LOxsec}. The color and spin states are averaged over the initial states and summed over the final ones.}
  {
  \begin{tabular}{|c|c|}
    \hline
    & \\ 
    Subprocess & $\mathcal{S}(ij\rightarrow kl) = \frac{\hat{s}^2}{\pi\alpha_s^2}\frac{d\hat{\sigma}}{d\hat{t}}(ij\rightarrow kl)$ \\
    & \\
    \hline
    & \\
    $q_1q_2\rightarrow q_1q_2$ & $\frac{4}{9}\frac{\hat{s}^2+\hat{u}^2}{\hat{t}^2}$ \\
    & \\
    $q_1\bar{q}_2\rightarrow q_1\bar{q}_2$ & $\frac{4}{9}\frac{\hat{s}^2+\hat{u}^2}{\hat{t}^2}$ \\
    & \\
    $qq\rightarrow qq$ & $\frac{4}{9}\left(\frac{\hat{s}^2+\hat{u}^2}{\hat{t}^2}+\frac{\hat{s}^2+\hat{t}^2}{\hat{u}^2}\right)-\frac{8}{27}\frac{\hat{s}^2}{\hat{u}\hat{t}}$ \\
    & \\
    $q_1\bar{q}_1\rightarrow q_2\bar{q}_2$ & $\frac{4}{9}\frac{\hat{t}^2+\hat{u}^2}{\hat{s}^2}$ \\
    & \\
    $q\bar{q}\rightarrow q\bar{q}$ & $\frac{4}{9}\left(\frac{\hat{s}^2+\hat{u}^2}{\hat{t}^2}+\frac{\hat{s}^2+\hat{t}^2}{\hat{u}^2}\right)-\frac{8}{27}\frac{\hat{u}^2}{\hat{s}\hat{t}}$ \\
    & \\
    $q\bar{q}\rightarrow gg$ & $\frac{32}{27}\frac{\hat{t}^2+\hat{u}^2}{\hat{t}\hat{u}}-\frac{8}{3}\frac{\hat{t}^2+\hat{u}^2}{\hat{s}^2}$ \\
    & \\
    $gg\rightarrow q\bar{q}$ & $\frac{1}{6}\frac{\hat{t}^2+\hat{u}^2}{\hat{t}\hat{u}}-\frac{3}{8}\frac{\hat{t}^2+\hat{u}^2}{\hat{s}^2}$ \\
    & \\
    $gq\rightarrow gq$ & $-\frac{4}{9}\frac{\hat{s}^2+\hat{u}^2}{\hat{s}\hat{u}}+\frac{\hat{u}^2+\hat{s}^2}{\hat{t}^2}$ \\
    & \\
    $gg\rightarrow gg$ & $\frac{9}{2}\left(3-\frac{\hat{t}\hat{u}}{\hat{s}^2}-\frac{\hat{s}\hat{u}}{\hat{t}^2}-\frac{\hat{s}\hat{t}}{\hat{u}^2}\right)$ \\ 
    & \\ 
    \hline
  \end{tabular}\label{tab:LOXSEC}}
\end{table*} 

\begin{figure}[phtb]
\centering
\psfig{file=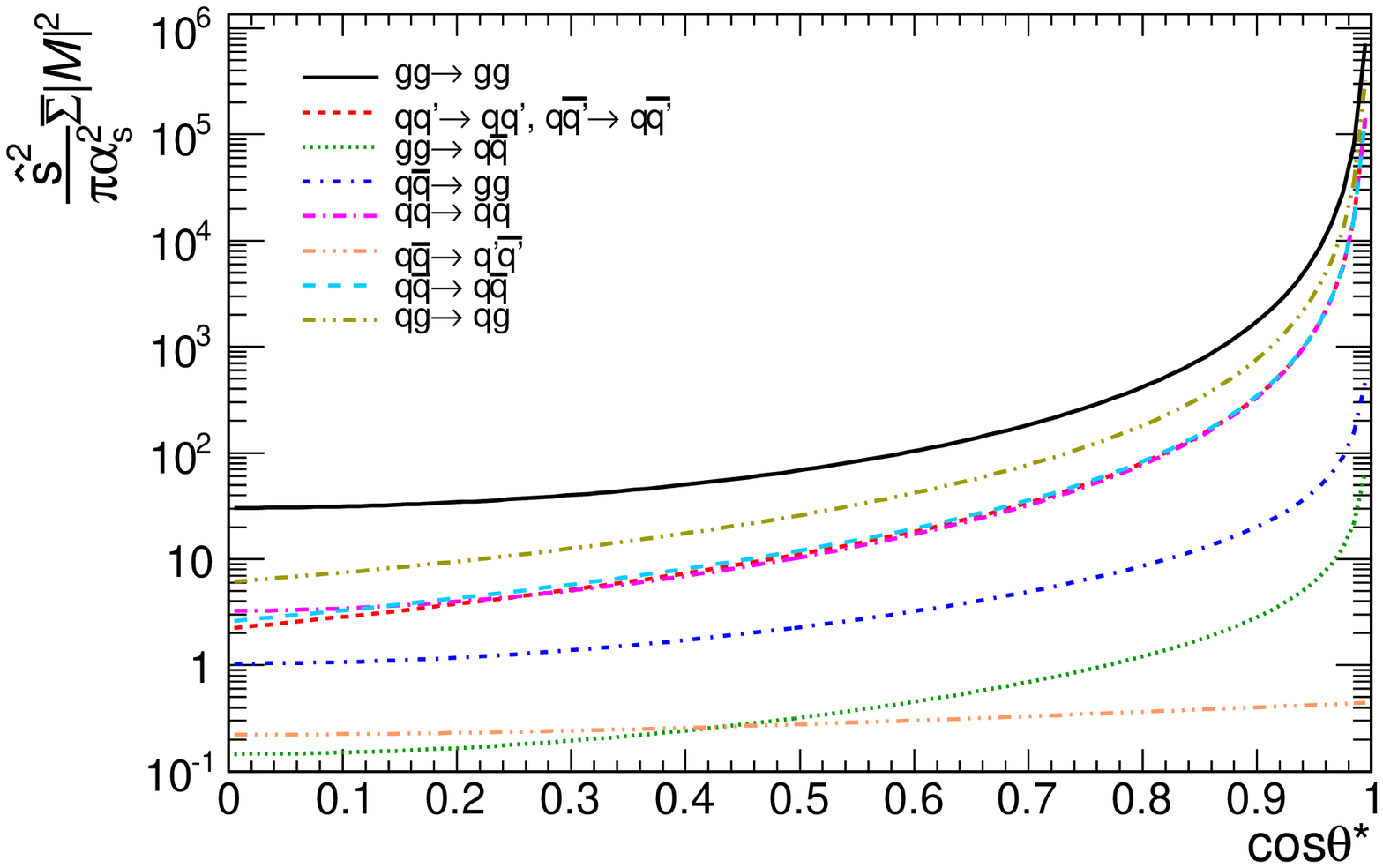, width=0.8\columnwidth}
\vspace*{8pt}
\caption{Leading order matrix elements for two-to-two massless parton scattering, as a function of $\cos\theta^*$.\label{fig:LOXSEC}}
\end{figure}

Figure~\ref{fig:LOXSEC} shows the matrix elements of the various subprocesses, at LO, as a function of $\cos\theta^*$. With the exception of one subprocess ($q_1\bar{q}_1\rightarrow q_2\bar{q}_2$), there is a characteristic $t-\text{channel}$ pole which enhances the two-parton scattering at small angles. Another important feature is the fact that, due to the larger color factor of gluons, the matrix element of the subprocesses with gluons in the initial state lead to larger values.   

\subsubsection{Hadronic cross sections}

In a hard scattering process, initiated by colliding hadrons, the experimentally measured cross section can be generally expressed in terms of the parton distribution functions (PDFs) $f(x)$ and the parton-parton scattering cross section $\hat{\sigma}$, summed over all the incoming and outgoing parton types (because the experimentally observed jets cannot distinguish between the parton types):

\begin{equation}
  \sigma = \sum_{ij}{\int{dx_1dx_2f_i(x_1,\mu_F^2)f_j(x_2,\mu_F^2)\hat{\sigma}_{ij}\left(\alpha_s(\mu_R^2),\frac{Q^2}{\mu_F^2},\frac{Q^2}{\mu_R^2}\right)}}
\end{equation}
In the equation above, $Q$ is the characteristic hard scale of the interaction (e.g. the dijet invariant mass in a two-to-two parton scattering), $\mu_F$ is the factorization scale, which is of the same order as $Q$ and separates the long-distance, non-perturbative interactions from the hard scattering, and $\mu_R$ is the renormalization scale. Both the $\mu_{F,R}$ scales are arbitrary parameters of a fixed-order calculation. At all orders of the perturbative expansion, the cross section should be independent of them $\left(\partial\sigma/\partial\mu_R=\partial\sigma/\partial\mu_F=0\right)$. In all practical calculations of cross sections at a fixed order, it is assumed that $\mu_R=\mu_F=Q$. It should be noted, that the higher the order of the calculation, the weaker is the dependence on $\mu_{R,F}$.

It is often helpful in hadron collisions to quantify the effect of the parton distribution functions by introducing the parton luminosity factor. This is defined as:
\begin{equation}
  \frac{dL_{ij}}{d\tau} = \int_0^1\int_0^1{dx_1dx_2f_i(x_1)f_j(x_2)\delta(x_1x_2-\tau)},
\end{equation}
where
\begin{equation}
  \tau = x_1x_2 = \frac{\hat{s}}{s}.
\end{equation}

In practice, experimental constraints are imposed on the rapidities of the outgoing partons, observed as hadronic jets. It is therefore more convenient to express the parton luminosity as a functions of the variables $\tau$ and $\bar{y}$, rather than $x_{1,2}$: $dx_1dx_2 = \frac{\partial(\tau,\bar{y})}{\partial{x_1,x_2}}d\tau d\bar{y} = d\tau d\bar{y}$. The parton luminosity then is:
\begin{equation}
  \frac{dL_{ij}(\bar{y}_{min},\bar{y}_{max})}{d\tau} = \int_{\bar{y}_{min}}^{\bar{y}_{max}}{f_i\left(\sqrt{\tau}e^{\bar{y}}\right)f_j\left(\sqrt{\tau}e^{-\bar{y}}\right)d\bar{y}}
\end{equation}
The hadronic cross section of any process, can be expressed generally as a function of the parton luminosity factor and the partonic cross section~\cite{QCDColliderPhysics}:
\begin{equation}
  \sigma_{had} = \sum_{i,j}{\int{\frac{d\tau}{\tau}\left[\frac{1}{s}\frac{dL_{ij}}{d\tau}\right]\left[\hat{s}\,\hat{\sigma}_{ij}\right]}}.
\end{equation}

In the specific case of a two-to-two scattering, resulting in the production of two jets, the differential cross section can be expressed as a function of the di-parton invariant mass and the scattering angle at the center-of-mass frame. To first approximation, this cross section is equal to the observed dijet cross section. The matrix elements presented in table~\ref{tab:LOXSEC} are folded with the parton distribution functions, giving:

\begin{equation}\label{massXsection}
  \frac{d^2\sigma_{had}}{dm\,d\cos\theta^*} = \frac{\pi\alpha_s^2}{m}\sum_{ij}{\left[\frac{1}{s}\frac{dL_{ij}}{d\tau}\right]_{\tau=m^2/s}\hat{\mathcal{F}}_{ij}(\cos\theta^*)},
\end{equation}
with  
\begin{equation}
  \hat{\mathcal{F}}_{ij}(\cos\theta^*) = \sum_{kl}{\mathcal{S}(ij\rightarrow kl)\frac{1}{1+\delta_{kl}}}.
\end{equation}
In the equations above $m\approx\sqrt{\hat{s}}=\sqrt{\tau s}$ is the dijet invariant mass.

\subsection{Models of partonic resonances}\label{sec:models}

\subsubsection{Excited quarks}

In various theoretical models, ordinary quarks can be composite objects~\cite{Qstar1}, with a characteristic compositeness scale $\Lambda$. As a natural consequence, excited states are expected, called simply \textit{excited quarks} and denoted by $q^*$. Depending on the details of the composite models, the excited quarks can have various values of spin and weak isospin. In the simplest case, they take the value of $1/2$. The interaction of excited quarks with the Standard-Model fields is of a "magnetic" type, and the Lagrangian takes the form~\cite{Qstar2}:  
\begin{equation}
  \mathcal{L} = \frac{1}{2\Lambda}\bar{q}^*_R\sigma^{\mu\nu}\left(g_s f_s t_a G^a_{\mu\nu}+gf\frac{\vec{\tau}\cdot\vec{W}_{\mu\nu}}{2}+g^\prime f^\prime\frac{Y}{2}B_{\mu\nu}\right)q_L+\text{h.c},
\end{equation}
where $t_a$ and $\vec{\tau}$ are the generators of the color $SU(3)$ and isospin $SU(2)$, $Y$ is the hypercharge, $G^{a}_{\mu\nu},\,\vec{W}_{\mu\nu},\,B_{\mu\nu}$ are the field tensors, $g_s,g,g^\prime$ are the gauge couplings, and $f_s,f,f^\prime$ are dimensionless constants, accounting for possible deviations from the Standard-Model couplings.  

In hadron collisions, the production of an excited quark happens through the quark-gluon fusion. Subsequently, the excited quark decays to an ordinary quark and a gauge boson. The dominant decay channel is $q^*\rightarrow qg$, leading to a dijet signature. The partial width for the decay of an excited quark with mass $m^* $ is given by:
\begin{equation}
  \Gamma(q^*\rightarrow qg) = \frac{1}{3}\alpha_sf_s^2\frac{{m^{*}}^3}{\Lambda^2}
\end{equation}

\subsubsection{Randall-Sundrum gravitons}

The gravity model from Randal and Sundrum~\cite{RS1,RS2} (RS) was proposed as a solution to the electroweak vs Planck scale hierarchy problem. In this model the hierarchy is generated by an exponential function of the compactification radius of one extra dimension. The metric in the 5-dimensional space is given by: 

\begin{equation}
  ds^2 = e^{-2kr_c\phi}\eta_{\mu\nu}x^\mu x^\nu+r_c^2d\phi^2,
\end{equation}
where $\phi$ is the extra dimension with compactification radius $r_c$, $k$ is a constant of the same order and related to the 5-dimensional Planck scale $M$, and $x^\mu$ are the usual space-time dimensions. The reduced effective 4-D Planck scale $\bar{M}_{Pl}$ is given by:
\begin{equation}
  \bar{M}_{Pl}^2 = \frac{M^3}{k}\left(1-e^{-2kr_c\pi}\right).
\end{equation}

In this model, spin-2 gravitons appear as the Kaluza-Klein (KK) excitations of the gravitational field $h^{\mu\nu}$, whose coupling to the Standard-Model fields is given by the interaction Lagrangian~\cite{RS3}:
\begin{equation}
  \mathcal{L}_I=-\frac{1}{\Lambda_\pi}h^{\mu\nu}T_{\mu\nu},
\end{equation}
with $T_{\mu\nu}$ being the energy-momentum tensor of the matter fields. The scale $\Lambda_\pi$ and the mass $m_n$ of the KK excitations can be expressed as a function of the fundamental constants of the theory:
\begin{equation}
  \Lambda_\pi = \bar{M}_{Pl}e^{-kr_c\pi},\,\,\, m_n = kx_ne^{-kr_c\pi}.
\end{equation}
The coupling constant of the graviton-matter interaction is the inverse of the scale $\Lambda_\pi$:
\begin{equation}
  g = \frac{1}{\Lambda_\pi} = x_n\frac{\left(k/\bar{M}_{Pl}\right)}{m_n},
\end{equation}
where $x_n$ is the n-th root of the the Bessel function of order 1. The phenomenological consequences of the RS-gravitons are essentially determined by their mass, and the ratio $k/\bar{M}_{Pl}$. If the fundamental constants of the model satisfy the relation $kr_c\sim 12$, then $\Lambda_\pi\sim \text{TeV}$, and RS gravitons can be produced in hadron collisions. Through the graviton coupling to the matter fields, RS-gravitons can decay to partons, leading to a dijet signature. The relevant partial widths~\cite{RS4} for the first KK excitation are given by: 
\begin{equation}
  \Gamma(G\rightarrow gg) = \frac{x_1^2}{10\pi}\left(\frac{k}{\bar{M}_{Pl}}\right)^2m_1,
\end{equation}
and
\begin{equation}
  \Gamma(G\rightarrow q\bar{q}) = \frac{3x_1^2}{160\pi}\left(\frac{k}{\bar{M}_{Pl}}\right)^2m_1.
\end{equation}

\subsubsection{Axigluons}

In the axigluon models~\cite{Axigluon}, the symmetry group of the strong sector is expanded to a chiral color group $\text{SU}(3)_L\times\text{SU}(3)_R$ which, at some energy, breaks to the diagonal $\text{SU}(3)$.
Under such a symmetry group, the left-handed and right-handed fermions $\psi_{L,R}=\frac{1}{2}(1\mp\gamma_5)\psi$ transform differently and the transformations are generated by the $T^a_{L,R}$ generators. Equivalently, the group can be described by a linear transformation of the generators, divided into vectorial $T_V^a=T_L^a+T_R^a$ and axial $T_A^a=T_L^a-T_R^a$ ones. The associated gauge field to the vectorial generators is identified as the usual color field of QCD, while the gauge field associates to the axial generators is called the \textit{axigluon} field. While the exact implementation of the chiral color group is model dependent, there are two universal features: the existence of a massive color octet axigluon field (corresponding to the broken symmetry), and the existence of new particles which are needed to cancel the triangular anomalies.

Axigluons can decay to quark-antiquark pairs, which leads to a dijet experimental signature. Note that due to parity conservation, the axigluon cannot decay to a gluon-gluon pair (all gluon-axigluon vertices must have an even number of axigluons). The axigluon decay to fermions is described by the Lagrangian:
\begin{equation}
  \mathcal{L}_A=-ig_s\left(\sum_{ij}{\bar{q}_i\gamma_5\gamma_\mu t_a q_j}\right)\mathcal{A}^{a\mu},
\end{equation}  
where $g_s=\sqrt{4\pi\alpha_s}$, $\mathcal{A}$ is the axigluon field, and $t^a$
are the usual color group generators. The width of the axigluon decay can be 
shown to be~\cite{UA1ptLimit}:
\begin{equation}
  \Gamma_A = \frac{N\alpha_sM_A}{6},
\end{equation}
where $N$ refers to the open decay channels, and $M_A$ is the axigluon mass. The latter is a free parameter of the theory, determined by the chiral color symmetry breaking scale and the details of the model.

\subsubsection{Colorons}

Similar to the axial color models, their exist other possibilities to enrich the group structure of the strong sector. Such a model is the flavor-universal coloron~\cite{Coloron1}, where the gauge group is extended to $\text{SU}(3)_1\times\text{SU}(3)_2$. The corresponding gauge couplings are denoted as $\xi_{1,2}$. Additionally, the model includes a scalar boson $\Phi$, which develops a non-zero vacuum expectation value and breaks spontaneously the symmetry of the two groups. The diagonal subgroup remains unbroken and is identified as the familiar color group of QCD. In the rotated phase of the physical gauge fields, the initial gauge bosons are mixed, forming an octet of massless gluons and an octet of massive \textit{colorons}. The mass of the colorons is expressed as a function of the fundamental parameters~\cite{Coloron2}: 
\begin{equation}
  M_C = \left(\frac{g_s}{\sin\theta\cos\theta}\right)f,
\end{equation}
where $\theta$ is the gauge boson mixing angle with $\cot\theta = \frac{\xi_2}{\xi_1}$, and $f$ is the vacuum expectation value of the scalar field. The Lagrangian of the interaction between the colorons field $C^{\mu a}$ and the quarks is similar to QCD:
\begin{equation}
  \mathcal{L} = -g_s\cot\theta\left(\sum_{ij}{\bar{q}_i\gamma_\mu t_a q_j}\right)C^{\mu a}.
\end{equation}
The above interaction predicts the decay of colorons to quarks with kinematically allowed masses. It can be shown that the decay width is:
\begin{equation}
  \Gamma_C \approx \frac{N}{6}\alpha_s\cot^2\theta M_C,
\end{equation}
where $N$ is the number of quark flavors with mass less than $M_C/2$.

\subsubsection{Color octet scalars}

In various theoretical models, bosonic states can arise from gluon-gluon fusion. The \textit{color octet scalar} model
$(S_8)$ is one example of exotic color resonances~\cite{S8}. The coupling of the color octet scalar field with gluons is expressed with the Lagrangian:
\begin{equation}
  \mathcal{L} = g_s\frac{\kappa}{\Lambda}d^{abc}S_8^aG_{\mu\nu}^bG^{c,\mu\nu},
\end{equation}
where $g_s$ is the strong coupling constant, $\kappa$ is the scalar coupling, $\Lambda$ is the characteristic scale of the interaction, $d^{abc}$ are structure constants of the $\text{SU}(3)$ group defined by the relation $\left\{t^a,t^b\right\} = \frac{1}{3}\delta^{ab}+d^{abc}t^c$, and $S_8,\,G_{\mu\nu}$ are the color octet scalar field, and the gluon field tensor, respectively. The width of the color octet scalar resonance is given by:
\begin{equation}
  \Gamma = \frac{5}{6}\alpha_s\kappa^2\frac{M^3}{\Lambda^2}.
\end{equation}

\subsubsection{$Z^\prime$ and $W^\prime$}

New gauge bosons arise in models where the symmetry $\text{SU}(2)_L\otimes
\text{U}(1)_Y$ of the electroweak Standard-Model sector is enlarged. Common
features in these models are the new gauge coupling constants, which are of the
same order as the $\text{SU}(2)_L$ coupling of the Standard Model, and the
existence of new gauge bosons, namely $W^\prime$ and $Z^\prime$. Under the
assumption that the new gauge bosons couple to ordinary quarks and leptons
similar to their Standard-Model counterparts, the cross sections of these
particles are calculated by scaling the the corresponding Standard-Model cross
section. In particular, the Fermi constant $G_F$ becomes~\cite{WZprime}:
\begin{equation}
  G^\prime_F = G_F\left(\frac{M}{M^\prime}\right)^2,
\end{equation}
where $M$ and $M^\prime$ are the masses of $W$ or $Z$, and $W^\prime$ or $Z^\prime$, respectively.

\subsubsection{String resonances}

According to the string theory, particles are created by vibrations of relativistic strings, with mass $M_s$, and they populate Regge trajectories, which relate their spins and masses. In principle, the mass of the fundamental strings is of the order of the Planck scale. However, in some theories with large extra dimensions, it is plausible that $M_s$ lies in the $TeV$ scale. In this case, Regge excited states of quarks and gluons occur in hadron collisions. If the string coupling is small, the basic properties of the Regge excitations (production cross section and width) are model independent (the details of the compactification are irrelevant). This statement is true for parton scattering involving gluons, but only approximately true in the four-quark scattering.

The effect of the Regge excitations can be quantified~\cite{String1,String2}
through the presence of a common form factor in the two-to-two parton scattering
amplitudes, which is called the \textit{Veneziano form factor} and is written in terms of the $\Gamma$-function:
\begin{equation}
  V(\hat{s},\hat{t},\hat{u})=\frac{\Gamma(1-\hat{s}/M_s^2)\Gamma(1-\hat{u}/M_s^2)}{\Gamma(1-\hat{t}/M_s^2)},
\end{equation}
where $\hat{s},\hat{t},\hat{u}$ are the usual Mandelstam variables. The physical content of the Veneziano form factor is revealed by an expansion in terms of s-channel poles. Each such pole represents a virtual Regge resonance, with mass $\sqrt{n}M_s$. For the purpose of resonances in the dijet spectrum, only the first-level ($n=1$) excitation is relevant, while the string mass $M_s$ is the only free parameter. The exact values of the cross sections depend also on the color factors and spin values of the excited states.

\subsubsection{$E_6$ Diquarks}

In the context of superstring theory in 10 dimensions, anomaly cancellation requires that the gauge group is $E_8\times E_8$. Certain models for the compactification of the additional 6 dimensions, predict that the grand unification symmetry group is $E_6$~\cite{E6}. The $E_6$ models~\cite{E6Diquarks1} contain color-triplet scalar diquarks, $D$ and $D^c$ with charges $-\frac{1}{3}$ and $\frac{1}{3}$ respectively, which couple to the light quarks $u,d$.

The interaction Lagrangian between the $E6$ diquarks and the light quarks is given by~\cite{E6Diquarks2}: 
\begin{equation}
  \mathcal{L} = \lambda\epsilon_{ijk}\bar{u}^{ci}\frac{1}{2}(1-\gamma_5)d^jD^k+\frac{\lambda_c}{2}\epsilon_{ijk}\bar{u}^i\frac{1}{2}(1+
\gamma_5)d^{cj}D^{ck}+h.c,
\end{equation}
where $i,j,k$ are color indices, and $\lambda,\lambda_c$ are parameters of the hyper-potential of the general $E_6$ model. The squared amplitudes for the diquark decays to light quarks are given by~\cite{E6Diquarks3}:
\begin{equation}
  |\mathcal{M}(D\rightarrow\bar{u}\bar{d})|^2 = 24\lambda^2m_D^2,\,\,\,|\mathcal{M}(D^c\rightarrow ud)|^2 = 6\lambda_c^2m_{D_c}^2.
\end{equation}
The corresponding widths are:
\begin{equation}
  \Gamma_D = \alpha_\lambda M_D,\,\,\,\Gamma_{D^c} = \frac{1}{4}\alpha_{\lambda_c} M_{D^c},
\end{equation}
with $\alpha_\lambda = \lambda^2/4\pi$, $\alpha_{\lambda_c} = \lambda^2_c/4\pi$.

\subsubsection{Color octet technirho}

Technicolor models predict a rich spectrum of technirho vector mesons $(\rho_T)$ that decay predominantly into technipion $(\pi_T)$ states. In most of the models, $\rho_T$ is either heavy, leading to small production cross section, or its decay width to $\pi_T$ is very large, leading to broad peak structures on the dijet mass spectrum. However, in the context of the "walking technicolor" models~\cite{WalkingTC}, the technifermion chiral condensate acquires large values, resulting in $\pi_T$ with large mass, such that the $\rho_T\rightarrow\pi_T\pi_T$ decay is kinematically suppressed. Furthermore, if the technirho is a color multiplet, its production cross section in hadron collisions through the $q\bar{q},gg\rightarrow\rho_T$ subprocesses is enhanced. In retrospect, giving the kinematically suppressed decay to technipions, $\rho_T$ decays almost $100\%$ to partons, appearing as a narrow resonance in the dijet mass spectrum. 

It should be noted that the phenomenology of the walking technicolor models can be fairly complex~\cite{Technirho1}, with one reason being the fact that the $\rho_T$ states are mixed with ordinary gluons. Because of the matrix form of the corresponding propagator, the predicted resonance on the dijet mass spectrum appears higher than the pole mass. Also, the resonance cross section cannot be calculated independently of QCD, to which it is coupled. For the experimental searches, a simple benchmark model is considered~\cite{Technirho2}, which predicts a single "bump" on the dijet mass spectrum. The corresponding parameters are listed below:
\begin{itemize}
  \item standard topcolor-assisted-technicolor (TC2) couplings
  \item degenerate technirhos: $M(\rho_{11})=M(\rho_{12})=M(\rho_{21})=M(\rho_{22})=M(\rho)$ (the pole mass)
  \item mixing among the technirhos: $M^\prime_8=0$, which reduces the mass shift from the pole mass,
  \item octet technipion mass: $M(\pi_{22}^8)=\frac{5M(\rho)}{6}$, which suppresses the decay $\rho_T\rightarrow\pi_T\pi_T$,
  \item singlet technipion mass: $M(\pi_{22}^1)=M(\pi_{22}^1)/2$,
  \item coloron mass: $M(V8)\rightarrow\infty$ , so that the coloron does not affect the cross section,
  \item mass parameter $M_8=\frac{5M(\rho)}{6}$, which keeps the branching fraction of the process $\rho_T\rightarrow g\pi_T$ small and leads to narrow resonance 
\end{itemize}

\subsubsection{Benchmark Models}
\label{benchmarks}

The parton-parton resonance models presented in the previous sections involve limited number of free parameters each. The experimental searches traditionally consider benchmark models, with certain parameter assumptions, which are then used to set limits on the masses of the corresponding resonances. Below is a summary of the benchmark models:
\begin{itemize}
  \item {\it Axigluons:} the number of quark flavors to which the axigluon can decay is set to $N=6$, corresponding to the known quarks.
  \item {\it Colorons:} the number of quark flavors to which the coloron can decay is set to $N=6$, and the gauge boson mixing angle is set to $\cot\theta = 1$.
  \item {\it Excited quarks:} standard model couplings are assumed ($f_s=f=f^\prime=1$) and the compositeness scale is set equal to the excited quark mass $\Lambda = M^*$.
  \item {\it RS graviton:} the ratio $k/\bar{M}_{Pl}$ is set to $k/\bar{M}_{Pl} = 1$.
  \item {\it $W^\prime,\,Z^\prime$:} standard model couplings are assumed.
  \item {\it $E_6$ diquark:} electromagnetic coupling constants are assumed $\alpha_\lambda = \alpha_{\lambda_c} = \alpha_e$.
  \item {\it Color octet scalars:} the gauge coupling is set equal to the QCD coupling $(\kappa=1)$, and the characteristic scale of the interaction is set equal to the resonance mass $\Lambda=M$. 
\end{itemize}
Table~\ref{tab:MODELS} summarizes the basic properties of the resonances discussed in this review. In particular, the 
decay widths are approximate values for the benchmark models, since they also depend on the running of $\alpha_s$ which
should be evaluated at a scale equal to the resonance mass. For string resonances the decay width varies significantly, depending on the spin and color quantum numbers of the resonances. 

\begin{table*}[htbH]
  \tbl{Summary of resonances considered in this review.}
  {
  \begin{tabular}{|c|c|c|c|c|c|}
    \hline
    &&&&&\\
    Resonance & Symbol & $J^P$ & Color & $\Gamma/(2m_R)$ & Decay\\
    & & & Multiplet & & Channel \\
    &&&&&\\
    \hline
    excited quark   & $q^*$      & $1/2^+$ & triplet & $0.02$          & $qg$                 \\
    axigluon        & $A$        & $1^+$   & octet   & $0.05$          & $q\bar{q}$           \\
    coloron         & $C$        & $1^-$   & octet   & $0.05$          & $q\bar{q}$           \\
    RS graviton     & $G$        & $2^-$   & singlet & $0.01$          & $q\bar{q}$, $gg$     \\
    $E_6$ diquark      & $D_6$      & $0^+$   & triplet & $0.004$         & $ud$                 \\
    color octet scalar    & $S_8$      & $0^+$   & octet   & $0.04$          & $gg$                 \\
    color octet technirho & $\rho_T$   & $1^-$   & octet   & $0.01$          & $q\bar{q}$,$gg$      \\
    heavy W         & $W^\prime$ & $1^-$   & singlet & $0.01$          & $q_1\bar{q}_2$       \\
    heavy Z         & $Z^\prime$ & $1^-$   & singlet & $0.01$          & $q\bar{q}$           \\
    string          & $S$        & various & various & $0.003 - 0.037$ & $q\bar{q}$,$qg$,$gg$ \\
    \hline
  \end{tabular}\label{tab:MODELS}}
\end{table*}

\subsubsection{Model Calculations}\label{secCalc}

The exact LO calculations of the cross sections and the decay widths of the various resonances involve all the relevant Feynman diagrams associated with each Lagrangian. In practice, the experimental searches presented in this review are focused on narrow resonances, which appear as "bumps" on a steeply falling dijet mass spectrum. In all the cases, it is the $s-\text{channel}$ decay mode of the resonances which produces a "bump". 

The cross section of a resonance decaying through the $s-\text{channel}$ is given by the Breit-Wigner expression:
\begin{equation}
  \hat{\sigma}(m)\left(1+2\rightarrow R\rightarrow 3+4\right) = 16\pi\mathcal{N}\times\frac{\Gamma(1+2\rightarrow R)\times\Gamma(R\rightarrow 3+4)}{\left(m^2-m_R^2\right)^2+m_R^2\Gamma^2_R},
\end{equation}
where $m_R$ and $\Gamma_R$ are the mass and the total width of the resonance, respectively, $\Gamma(1+2\rightarrow R)$ and $\Gamma(R\rightarrow 3+4)$ are the partial widths for the creation and the decay of the resonance to the specific final state. The spin and color multiplicity factor $\mathcal{N}$ is
\begin{equation}
  \mathcal{N} = \frac{N_{S_R}}{N_{S_1}N_{S_2}}\frac{C_R}{C_1C_2},
\end{equation}
where $N_{S_R}$, $N_{S_{1,2}}$ are the spin multiplicities of the resonance and the initial state particles, while $C_R$ and
$C_{1,2}$ are the corresponding color factors. The cross section above arises after integrating over $\cos\theta^*$. The
angular dependence of the cross section, in the $s-\text{channel}$ decay mode, is determined by the spin of the resonance and
the spin of the final state particles. It should be noted, that in all resonance cases decaying to two partons, the angular
dependence is expressed as a polynomial of $\cos\theta^*$, as opposed to the dominant QCD background, which exhibits a $t-\text{channel}$ pole at $\cos\theta^*\rightarrow 1$. More specifically, the angular distributions of the various resonances, in the $s-\text{channel}$ decay mode, are listed below:
\begin{itemize}
  \item {\it $E_6$ diquark, color octet scalars:} $F(\cos\theta^*) = \text{const.}$
  \item {\it excited quark:} $F(\cos\theta^*)\sim 1+\cos\theta^*$, which becomes $F(|\cos\theta^*|) = \text{const.}$ (odd in $\cos\theta^*$).
  \item {\it axigluon, coloron, $\textit{W}\,^\prime$, $\textit{Z}\,^\prime$:} $F(\cos\theta^*)\sim 1+\cos^2\theta^*$.
  \item {\it RS gravitons:} $F(gg\rightarrow G\rightarrow q\bar{q}) = F(q\bar{q}\rightarrow G\rightarrow gg)\sim 1-\cos^4\theta^*$, $F(gg\rightarrow G\rightarrow gg)\sim 1+6\cos^2\theta^*+\cos^4\theta^*$, and $F(q\bar{q}\rightarrow G\rightarrow q\bar{q})\sim 1-3\cos^2\theta^*+4\cos^4\theta^*$.
\end{itemize}
where $F(\cos\theta^*) \equiv d\hat{\sigma}/d\cos\theta^*$.

In practice, experimental searches impose kinematic constraints on the scattering angle $\theta^*$, such that the QCD background is suppressed. In this case, the Breit-Wigner partonic cross section is written as:
\begin{equation}
  \hat{\sigma}(m)=\frac{16\pi\times\mathcal{N}\times\mathcal{A}_{\cos\theta^*}\times BR \times \Gamma_R^2}{\left(m^2-m_R^2\right)^2+m_R^2\Gamma^2_R},
\end{equation}
where $BR$ is the branching fraction of the subprocess, and $\mathcal{A}_{\cos\theta^*}$ is the acceptance after the $\cos\theta^*$ cut. If the resonance is sufficiently narrow ($\Gamma_R << m_R$), the {\it narrow-width approximation} holds:
\begin{equation}
  \frac{1}{\left(m^2-m_R^2\right)^2+m_R^2\Gamma^2_R}\approx\frac{\pi}{m_R\Gamma_R}\delta(m^2-m_R^2).
\end{equation}
Finally, the hadronic cross section in the narrow-width approximation is derived: 
\begin{equation}
  \sigma_{had}(m_R) = 16\pi^2\times\mathcal{N}\times\mathcal{A}_{\cos\theta^*}\times BR\times\left[\frac{1}{s}\frac{dL(\bar{y}_{min},\bar{y}_{max})}{d\tau}\right]_{\tau=m^2_R/s}\times\frac{\Gamma_R}{m_R},
\end{equation}
where the parton luminosity $\frac{dL}{d\tau}$ is calculated at $\tau=m_R^2/s$, and constrained in the kinematic range $[\bar{y}_{min},\bar{y}_{max}]$.

%% file: Experiment.tex
\section{Experiment}
\label{secExperiment}

In this review paper we consider all searches that used the dijet mass or $p_T$ spectra
to search for dijet resonances. The searches considered are listed in table~\ref{tabSearches}
in chronological order, along with a summary of the energy and luminosity of the dataset, and the 
techniques of the search. The searches by the UA1 and UA2 experiments used data from the 
proton anti-proton collisions at the CERN $S\bar{p}pS$ collider at a center-of-mass 
energy of $0.63$ TeV. The searches by the CDF and D0 experiments used data from proton anti-proton 
collisions at the Fermilab Tevatron at a center-of-mass energy of $1.8$ and $1.96$ TeV. 
The searches by the ATLAS and CMS experiments used data from proton proton collisions 
at the CERN Large Hadron Collider at a center-of-mass energy of $7$ TeV. The search techniques
listed in table~\ref{tabSearches} are discussed in the following sections.

\begin{table}[htbp]
\tbl{Searches for dijet resonances at hadron colliders. For each search we list
the experiment, year of data publication, center-of-mass energy, integrated luminosity, techniques used
to define the resonance shape and the background, dijet mass range of the data, the cut applied
on the center of mass scattering angle, and the primary reference for the search.}
{\begin{tabular}{@{}llccccccc@{}} \toprule
Expt. & Yr. & $\sqrt{s}$ & $\int L \ dt$ & Resonance & Background & $m_{JJ}$ & $\cos\theta^*$ & Ref. \\
      &      & (TeV)      & ($pb^{-1}$) & Shape & Shape & (TeV) & Cut  & \# \\ \colrule
UA1   & 86  & 0.63 & 0.26 & BW $\oplus$ Gaussian & LO QCD & $.07 - 0.3$ & - & \refcite{UA1ptLimit}\\
UA1 & 88 & 0.63 & 0.49 & BW $\oplus$ Gaussian & LO QCD & $.11 - 0.3$ & bins  & \refcite{UA1mass}\\
CDF & 90 & 1.8 & 0.026  & BW $\oplus$ Gaussian & LO QCD & $.06 - 0.5$ & -  & \refcite{CDF1990}\\
UA2 & 90 & 0.63 & 4.7 & Gaussian & Fit Func. & $.05 - 0.3$ & - & \refcite{UA21990} \\
CDF & 93 & 1.8 & 4.2   & BW $\oplus$ Resolution & LO QCD & $.14 - 1.0$ & - & \refcite{CDF1993}  \\
UA2 & 93 & 0.63 & 11  & Gaussian & Fit Func. & $.05 - 0.3$ & .60 &  \refcite{UA21993}\\
CDF & 95 & 1.8 & 19  & \textsc{Pythia} $\oplus$ Sim. & Fit Func. & $.15 - 0.9$ & .67  & \refcite{CDF1995}\\
CDF & 97 & 1.8 & 106  & \textsc{Pythia} $\oplus$ Sim. & Fit Func. & $.18 - 1.0$ & .67 & \refcite{CDF1997}\\
D0 & 04  & 1.8 & 109  &\textsc{Pythia} $\oplus$ Sim. & NLO QCD & $.18-1.2$ & .67 & \refcite{D02004} \\
CDF & 09 & 1.96 & 1130 & \textsc{Pythia} $\oplus$ Sim. & Fit Func. & $.18 -1.3$ & -  & \refcite{CDF2009}  \\
ATLAS & 10 & 7 & 0.32 & \textsc{Pythia} $\oplus$ Sim. & Fit Func. & $.20-1.7$ & .57 & \refcite{ATLAS2010} \\
CMS & 10 & 7 & 2.9 & \textsc{Pythia} $\oplus$ Sim. & Fit Func. & $.22-2.1$ & .57 & \refcite{CMS2010} \\
ATLAS & 11w & 7 &  36 &\textsc{Pythia} $\oplus$ Sim. & Fit Func. & $.50-2.8$ & .57 & \refcite{ATLAS2011} \\
CMS & 11 & 7 & 1000 & \textsc{Pythia} $\oplus$ Sim. & Fit Func. & $.84-3.7$  & .57 & \refcite{CMS2011}\\ 
ATLAS & 11s & 7& 1000 & \textsc{Pythia} $\oplus$ Sim. & Fit Func. & $.72-4.1$  & .54 & \refcite{ATLAS2011summer} \\ 
 \botrule
\end{tabular} \label{tabSearches}}
\end{table}

\subsection{Resonance Shapes}
\label{secResonanceShapes}

Narrow resonances are those whose observed width is dominated 
by experimental resolution, where the natural width of the resonance
is small in comparison.  Searches for narrow resonances are prevalent
in high energy physics, because they in principle only require the use of
one shape for the signal, dominated by the experimental resolution. This
is an approximation, as we will see explicitly in this section.

Searches for dijet resonances have modeled the resonance line 
shape using either an analytic or a Monte Carlo technique, as
shown in Table~\ref{tabSearches}.

 The analytic technique used by the older searches 
started with a natural shape for the resonance which was then 
smeared with the experimental resolution function. That natural shape 
was either a Breit-Wigner (BW) for wide resonances, where the natural
width could not be ignored, or a delta-function for narrow resonances. 
The principle of the technique is the idea that the resolution can be understood
separately from the natural line shape, and the two can then be convolved.  In practice
things were often done this way in older searches because of limitations in the MC modeling 
or the detector simulation. 
The search~\cite{UA1ptLimit} using 1986 data from UA1~\cite{UA1ptData} considered the entire spectrum,
with QCD and resonances determined analytically and summed together, and then smeared with
the experimental resolution, $\sigma(p_T)/p_T = 0.53/\sqrt{p_T} + 0.05$.  
The UA1 search in 1988~\cite{UA1mass} smeared a BW
shape in a toy MC with the Gaussian experimental dijet mass resolution, $\sigma/m=0.11$, and checked 
the process with the \textsc{Isajet} MC~\cite{ISAJET} and a full UA1 simulation. The UA2 search in
1990~\cite{UA21990},
and again in 1993~\cite{UA21993}, considered only narrow resonances and used the Gaussian experimental 
resolution for the resonance shapes. The resolution varied from 10.7\% at 80 GeV to 8.4\% at 300 GeV for
$W^\prime\rightarrow q\bar{q}$ resonances, and UA2 noted that the resolution was worse for 
a $gg$ resonance, 13.2\% at 80 GeV~\cite{UA21990}. The CDF searches in 1990~\cite{CDF1990} and 1993~\cite{CDF1993} 
coherently summed 
an axigluon resonance shape to the QCD background and then smeared with the experimental dijet mass resolution.
The resolution was 
$\sigma/m=0.68/\sqrt{m} + 0.065$ in 1990, 
and in 1993 it was parameterized in 
a more complex way and also included the effects of QCD radiation from \textsc{Herwig}~\cite{HERWIG}. 
The 1990 search was for 
narrower axigluons, $N=5$ and $10$, and the 1993 search was for wider axigluons, $N=10$ and $20$.
The 1993 paper also considered a more generic search for three widths, one 
narrow ($\Gamma=0.02M$) and two wide ($\Gamma=0.1M$ and $0.2M$).

\begin{figure}[htbp]
\centerline{
\psfig{file=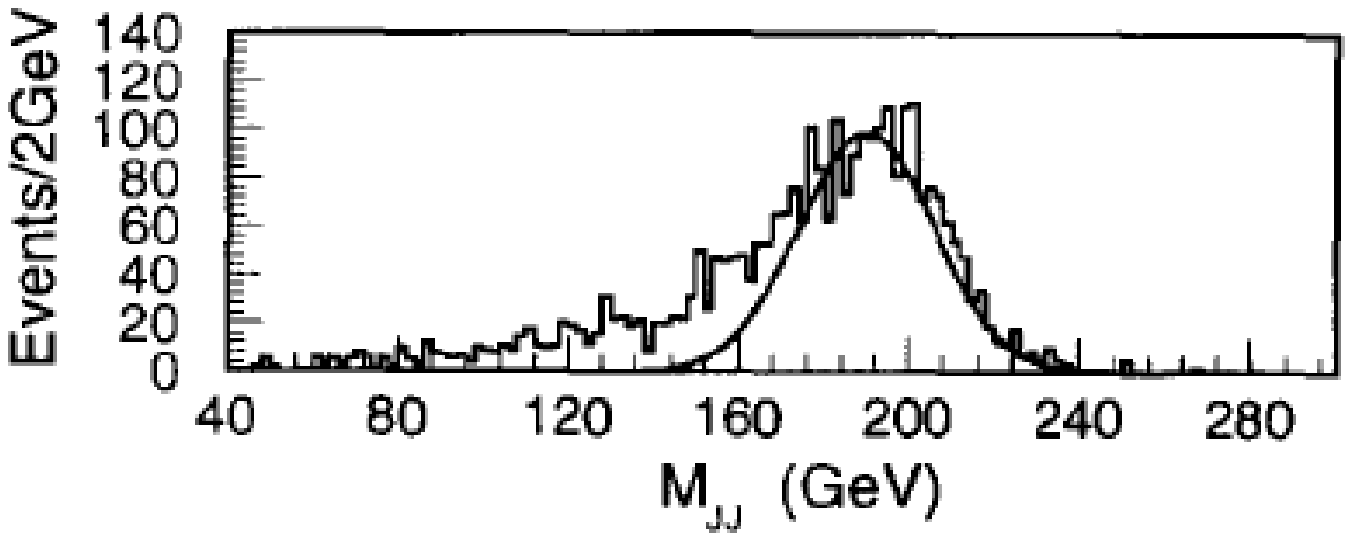,width=2.5in}
\psfig{file=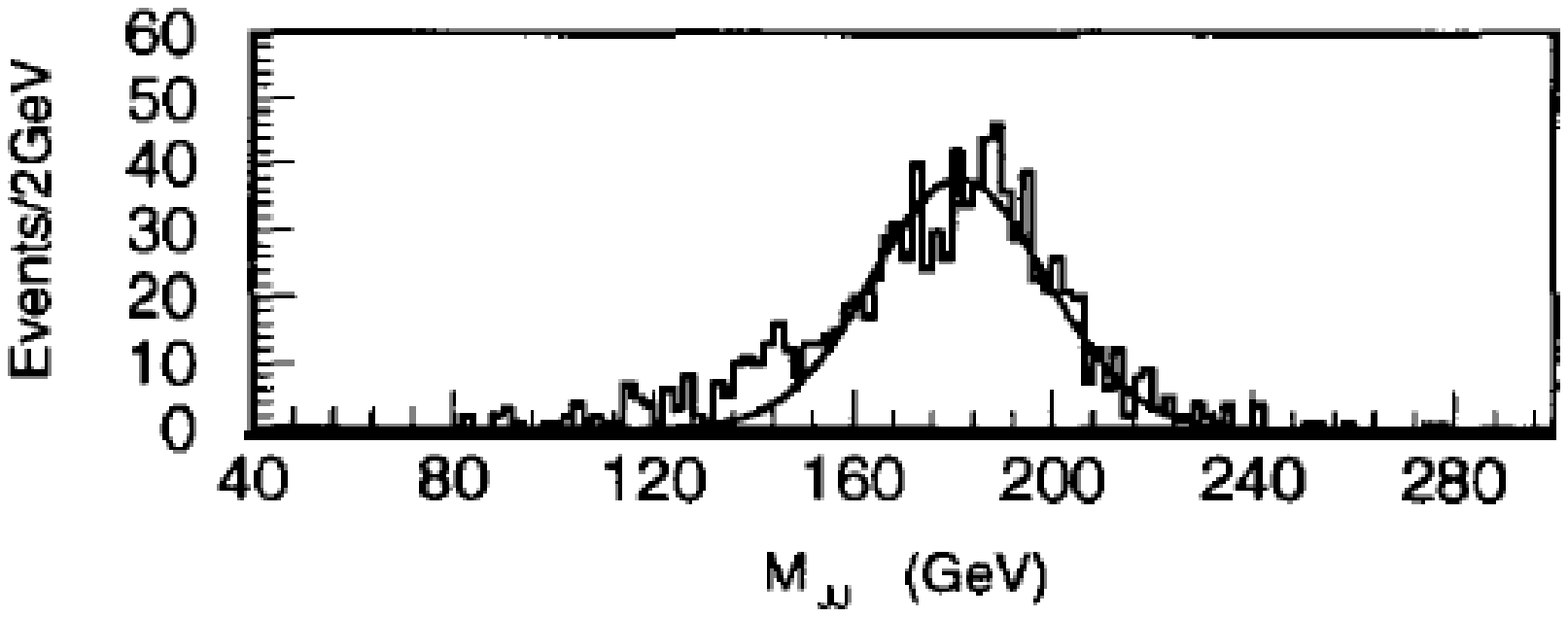,width=2.5in}
}
\vspace*{8pt}
\caption{ Resonance shapes from UA2 in 1993.
(left) $Z^{\prime}$ and (right) excited quark simulations at a mass of 200 GeV 
with Gaussian fits overlaid. 
Reprinted from Ref.~\protect\refcite{UA21993}, Copyright 1993, 
with permission from Elsevier.
\label{UA21993shapes}
}
\end{figure}

A nice early illustration of narrow resonance shapes was provided by UA2 in 1993~\cite{UA21993} and 
is reproduced in Fig.~\ref{UA21993shapes}.  The Gaussian shapes used to model
the signal represent reasonably well the central core of the simulated
distribution using \textsc{Pythia}~\cite{PYTHIA}, but
there is also a long tail at low mass due to final state radiation. UA2 accounted for
the tail using a signal efficiency, and stated that {\it ``this tail is re-absorbed in the parameterization 
of the QCD background''}. The tail at low mass is generally negligible compared to the 
QCD background, so a Gaussian was an adequate approximation. Nevertheless, all  
subsequent searches used the full resonance shape including the radiative tail to fit
for resonances and set limits.

\begin{figure}[htbp]
\centerline{
\psfig{file=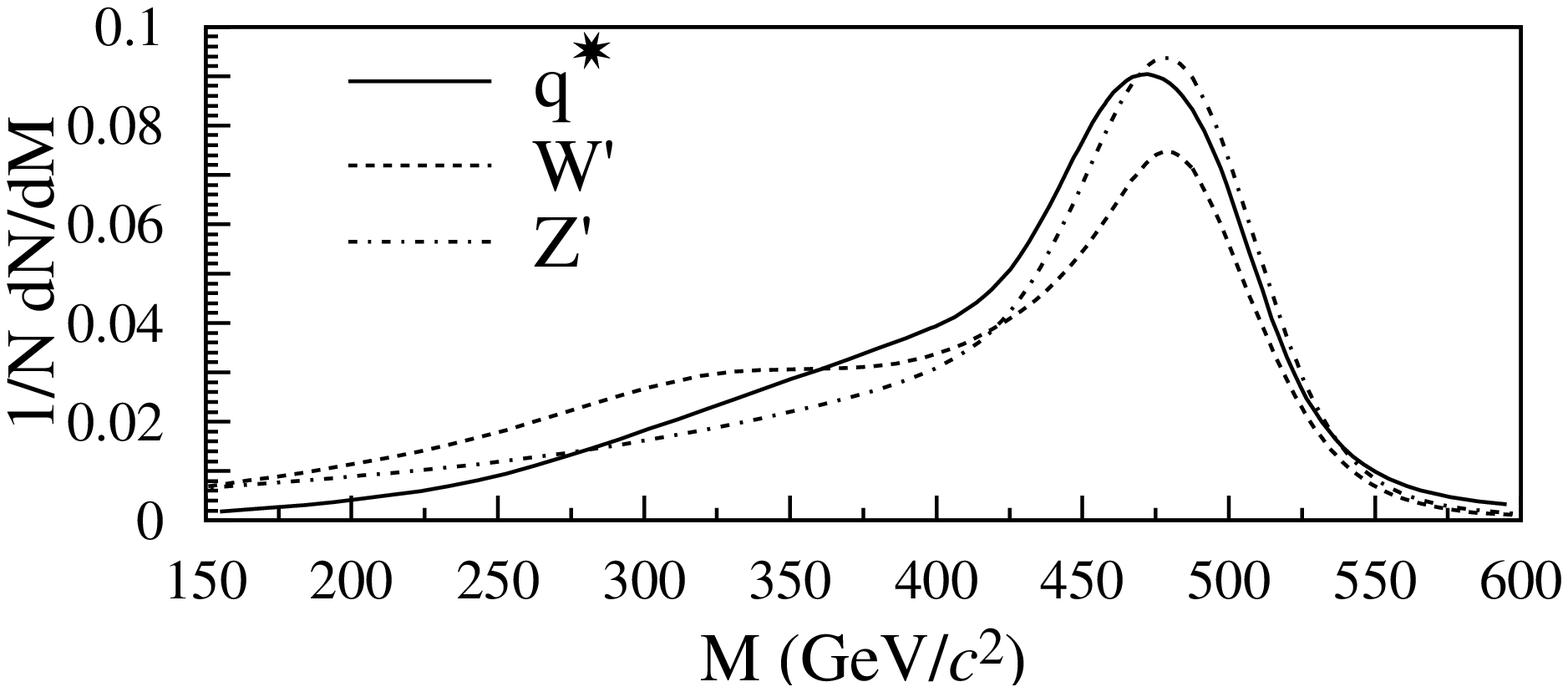,width=2.5in}
\psfig{file=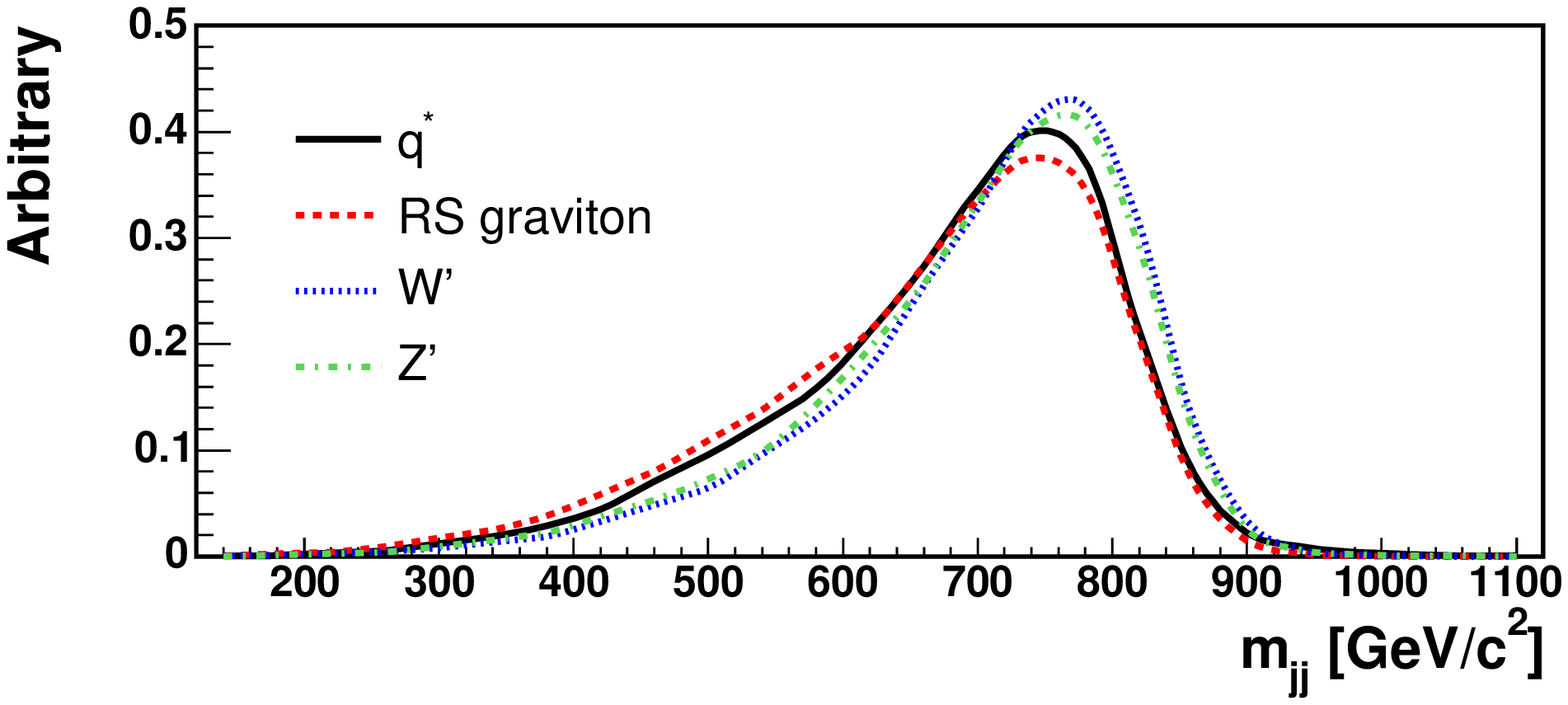,width=2.5in}
}
\vspace*{8pt}
\caption{Resonance shapes from D0 in 2004 and CDF in 2009.
(left) D0 simulation of three models of a 500 GeV resonance from Ref.~\protect\refcite{D02004}, Copyright 2004, 
and (right) CDF simulation of four models of an 800 GeV resonance from Ref.~\protect\refcite{CDF2009},
 Copyright 2009 by the American Physical Society. 
\label{D0CDFshapes}}
\end{figure}

The MC technique used by the more recent searches modeled the resonance line shape using 
\textsc{Pythia} with either a fast or full detector simulation. The CDF searches in 1995~\cite{CDF1995}
and 1997~\cite{CDF1997} used an excited quark decaying to $qg$ to model the resonance shape
for all dijet resonances. Uncertainties on the amount of final state radiation were included 
by changing the amount of the ``radiation tail'' at low mass by $\pm$50\%. The Gaussian core 
of the dijet mass resolution was reported as approximately 10\% at CDF~\cite{CDF1995,CDF1997,CDF2009}, 
but the resolution did depend on dijet mass. The D0 search in 2004~\cite{D02004},
and the CDF search in 2009~\cite{CDF2009}, used independent shapes for different models of narrow
resonances available in \textsc{Pythia}, as shown in Fig.~\ref{D0CDFshapes}. The shapes varied primarily 
due to the different amounts of final state radiation, with the low mass tail and the width usually increasing 
with the number of gluons in the final state.

\begin{figure}[htbp]
\centerline{
\psfig{file=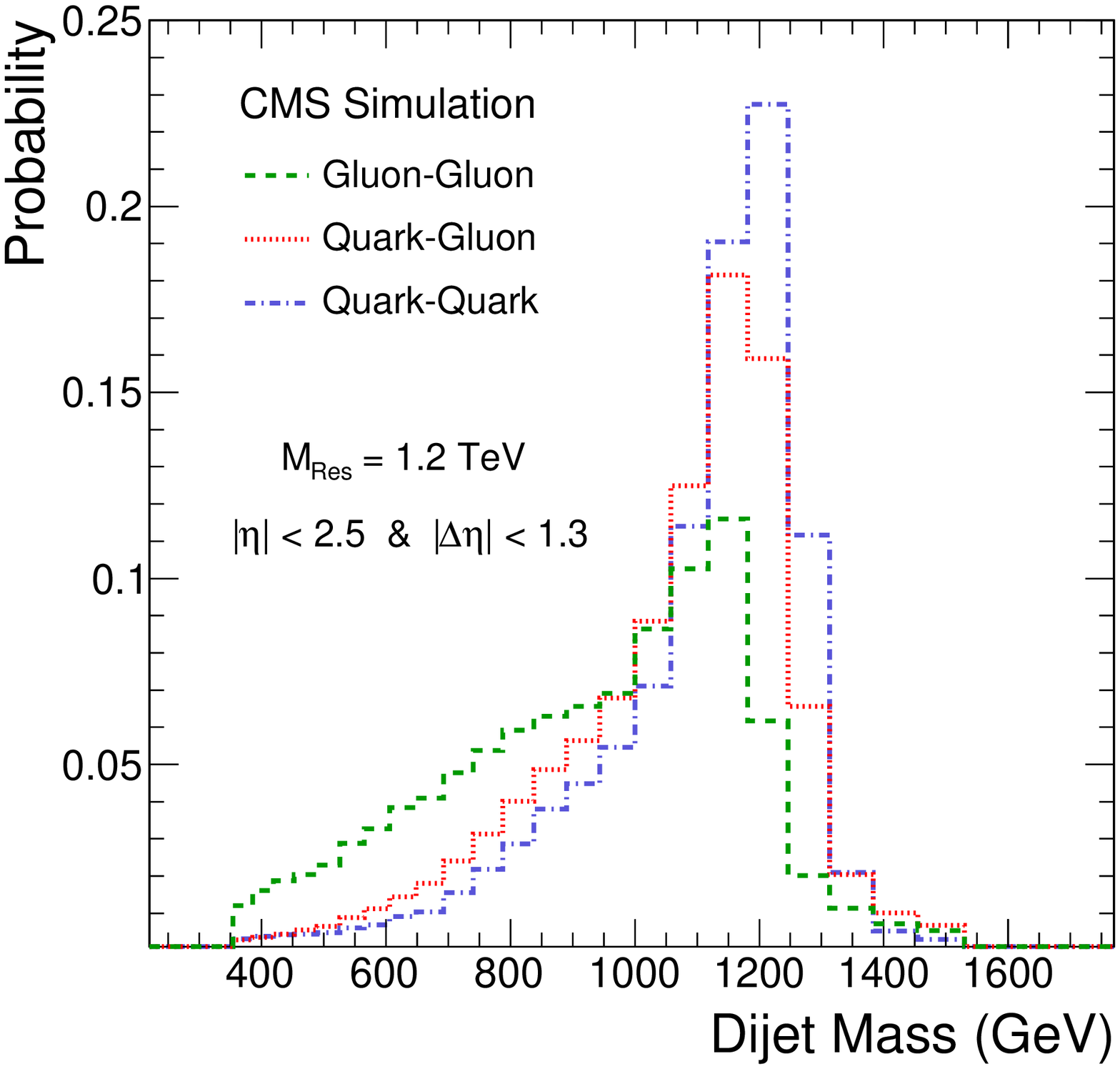,width=2.5in}
\psfig{file=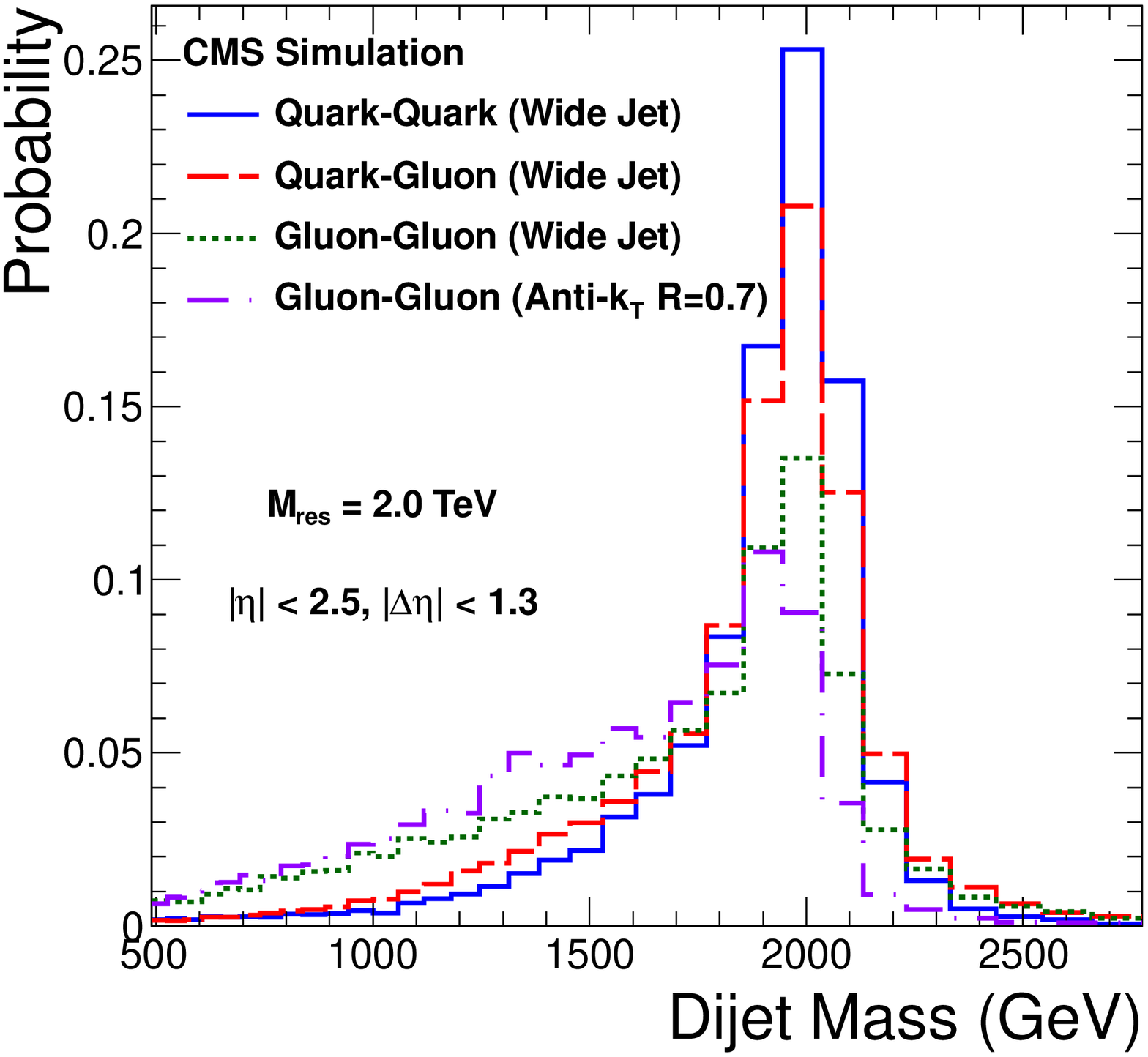,width=2.5in}
}
\vspace*{8pt}
\caption{Resonance shapes from CMS.
(left) Simulations of 1.2 TeV resonances shown for the different possible final state partons, from
Ref. ~\protect\refcite{CMS2010}, Copyright 2010 by the American Physical Society.
(right) Simulation of 2 TeV resonances reconstructed with two different types of jets, from
Ref. ~\protect\refcite{CMS2011}, Copyright 2011, with permission from Elsevier.
\label{CMSshapes}}
\end{figure}

The CMS search in 2010~\cite{CMS2010} 
introduced independent narrow resonance shapes for each generic type 
of final state: quark-quark, quark-gluon, and gluon-gluon shown in Fig.~\ref{CMSshapes}.  
They reported in Ref.~\refcite{CMS2010} that for resonance masses between $0.5$ and $2.5$ TeV the Gaussian resolution 
of the core of the distribution
{\it ``varies from 8\% to 5\% for $qq$, 10\% to 6\% for $qg$, and 16\% to 10\% for $gg$''}.
These were estimated by fitting mainly the peak and high-mass edge to a Gaussian~\cite{resonanceCmsNote}.
To reduce the radiation tail the CMS search in 2011~\cite{CMS2011} introduced wide-jets 
with an effective radius of $\Delta R=1.1$. 
Wide-jets also improved the Gaussian component of the resolution which in 2011 was 
$\sigma/M=1.31/\sqrt{M} + 0.018$ for $qq$, $\sigma/M=1.56/\sqrt{M} + 0.027$ for $qq$, 
and $\sigma/M=2.09/\sqrt{M} + 0.015$ for $gg$ resonances, with resonance mass $M$ given
in GeV, and between $500$ and $3500$ GeV~\cite{refMaxime}.
The improvement is largest for gluon-gluon resonances, as shown in Fig.~\ref{CMSshapes},
where the shape using wide-jets is compared with the shape using narrower jets from the 
anti-$k_T$ algorithm~\cite{anti-kt} with a distance parameter $R=0.7$. 

The ATLAS search in 2010~\cite{ATLAS2010} considered the explicit dijet resonance shape of the excited quark 
model from \textsc{Pythia}, {\it ``including all possible SM final states, which were dominantly $qg$ 
but also $qW$, $qZ$, and $q\gamma$''}. The reported Gaussian resolution for an excited quark
ranged from 11\% to 7\% for $q^*$ masses between $0.3$ and $1.7$ TeV~\cite{ATLAS2010}, 
and the reported detector level resolution $\sigma_{m_{jj}}/m_{jj}$ was ``5\% at 1 TeV, 4.5\% 
at 2 TeV, and asymptotically approached 4\% at masses of 5 TeV and above''~\cite{ATLAS2011summer}. 
In 2011~\cite{ATLAS2011,ATLAS2011summer} ATLAS 
reused the $q^*$ shapes to set limits on axigluons, after noting that {\it ``the axigluon and
$q^*$ signal templates result in very similar limits''}. In summer 2011~\cite{ATLAS2011summer} 
ATLAS introduced a shape for $gg$ resonances. The ATLAS searches in 2011~\cite{ATLAS2011,ATLAS2011summer} 
also considered generic limits for Gaussian shaped resonances of varying width, not just narrow resonances.

The shapes of narrow resonances presented in the literature are all well described by 
a Gaussian core and a radiation tail, but the different resonance shapes in a single 
experiment also demonstrates that the concept of narrow resonances is often approximate.

In addition to radiation, another parton-level physical mechanism produces a significant tail 
at low mass, further perturbing the narrow resonance search technique: the parton luminosities 
at low mass are larger than at the resonance pole mass, and significantly lift the natural Breit-Wigner 
tail of the resonances, in the cases where the PDFs are falling off rapidly with increasing dijet mass.
This frequently happens at resonance masses approaching the kinematic bound for both sea quarks and gluons,
and the size of the effect increases with the resonance intrinsic width, even for resonances normally 
considered narrow. The extreme end of this tail due to the PDFs is 
sometimes suppressed in the searches by requiring the partons to be have mass close to the pole mass, within a few standard
deviations on the dijet mass resolution. This is generally a reasonable solution for the shapes, as
the QCD background overwhelms the signal at low dijet mass. However, the way that this tail from PDFs is handled 
can significantly affect the total resonance cross section quoted for specific models, as we discuss in
~\ref{sectionCompare}.

The narrow resonance search technique, where a single resonance shape dominated by experimental resolution 
is used to model the effect of all narrow resonances, is only strictly applicable when the 
half-width of the resonance, $\Gamma/2$ is significantly less than the experimental Gaussian 
resolution $\sigma$. The model half-widths are listed in table~\ref{tab:MODELS}, and vary from about
$0.4\%$ to as large as $\alpha_s/2$ for axigluons and colorons. For the searches at UA1, UA2, CDF, and D0, 
$\alpha_s/2$ is about a factor of 2 less than the dijet mass resolution of roughly 10\%, and  
the half-widths of the other models are significantly less than 10\%, so the narrow resonance technique should 
be applicable. However, the searches at CMS and ATLAS are now looking for very massive resonances, 
producing very energetic jets measured with better resolution. 
For an axigluon or coloron of mass 3 TeV the half-width is $\alpha_s(3\ \mbox{TeV})/2=3.9\%$.
For a resonance mass of 3 TeV, the Gaussian dijet mass resolution at CMS in the $qq$ channel 
in 2011 is 4.2\%, and at ATLAS at 3 TeV the dijet mass resolution quoted above is roughly the same. 
In addition, both CMS and ATLAS have also included a 
long tail to low mass due to radiation, which is a part of the modification of the natural line width into
an observed line width, so the experimental resolution is somewhat worse than 4.2\%. Nevertheless,
the natural half-width of 3.9\% is comparable to the experimental resolution, and so the widest resonances ATLAS
and CMS have considered, like axigluons and colorons, are beginning to push the boundaries of the narrow 
resonance classification.

\subsection{Dijet Data and QCD Background}
\label{secData}

The heart of the search for dijet resonances is the measurement of the dijet mass distribution 
and the estimation of the background. Unlike many other searches in high energy physics, the 
search for dijet resonances is completely dominated by a single background. The observed dijet 
mass distribution comes from the dominant process in hadronic collisions: $2\rightarrow2$ scattering
of partons predicted by perturbative QCD. 

\subsubsection{Angular Requirement}

The event selection requirements of each search can only be understood by examining the dijet production in QCD.
Most experiments chose a fiducial region in the experiment to measure dijet production, limiting
the pseudorapidity, $\eta$, of each jet to a central region.  In addition, as shown in 
table~\ref{tabSearches}, many of the searches place a cut on the center of momentum frame scattering
angle $|\cos\theta^*|$.\footnote{For some experiments we have translated their cut on dijet $\Delta\eta$ or 
$\Delta y$ to the equivalent cut on $\cos\theta^*\approx \tanh(\Delta\eta/2)$, which follows from 
Eq.~\ref{costheta} and $\Delta\eta \approx \Delta y$.}
  Both of these selections, but the $|\cos\theta^*|$ one in particular, are
designed to enhance the dijet resonance signal and suppress the QCD background.  This is because
QCD production of dijets at high mass is dominated by $t-channel$ production, with an angular distribution
that is approximately Rutherford scattering (Section~\ref{QCDxsections}):
\begin{equation}
\frac{dN}{d \cos\theta^*} \sim \frac{1}{\hat{t}^2} \sim \frac{1}{(1 - \cos\theta^*)^2},
\label{eqnBack}
\end{equation}
peaking strongly at $\cos\theta^*=1$. In practice, only $|\cos\theta^*|$ is measured, because the partons in the final state emerge as jets and are experimentally 
indistinguishable. In the $s-\text{channel}$ mode, dijet resonances lead to angular distributions that are
much flatter in $\cos\theta^*$ than QCD, with the exact angular distribution
depending on the spin of the resonance and the final state partons (Section~\ref{secCalc}). 
The optimal $|\cos\theta^*|$ cut is determined by maximizing the integrated signal, S, over the 
square root of the integrated background, $\sqrt{B}$, often by using a Monte Carlo. 
However, one can estimate the optimal $\cos\theta^*$ cut\footnote{Integrate both signal and background
from $-\cos\theta^*_{MAX}$ to $\cos\theta^*_{MAX}$ and then solve $d(S/\sqrt{B})/d\cos\theta^*_{MAX}=0$.}, 
using equation~\ref{eqnBack} for the background and a signal flat in $|\cos\theta^*|$, resulting in 
the optimal cut $|\cos\theta^*_{MAX}|=1/\sqrt{3}=0.58$. Table~\ref{tabSearches} shows that all experiments which 
employed a cut on $|\cos\theta^*|$ used a similar value in order to optimize their searches for dijet resonances.

\subsubsection{Background Models}
\label{secBackgound}

Searches for dijet resonances have chosen to model the background with either a QCD calculation 
or with a background parameterization, as shown in table~\ref{tabSearches}. No search has used 
a Monte Carlo simulation to model the QCD background, 
even if they used a Monte Carlo to model the resonance signal. This is because Monte Carlo may 
not model the spectrum shape accurately enough. QCD calculations
are also challenged by high statistics jet data. In the cases
where leading order (LO) QCD was used to model the spectrum shape, the normalization of LO QCD was always 
adjusted to obtain agreement with the data. Despite the considerable progress in 
next-to-leading order (NLO) QCD calculations
over the last 20 years, only one experiment has successfully used NLO QCD to 
model the dijet background~\cite{D02004}, and remarkably was able to use the absolute normalization of the 
calculation. Even when normalization and shape agree between QCD and the data, there are still considerable
theoretical uncertainties (PDFs, renormalization scale, non-perturbative effects, ...) and experimental uncertainties 
(jet energy scale, resolution smearing, ...) which can produce significant differences between data and 
QCD that can potentially mimic a resonance signal. For these reasons, 
the majority of searches simply used a fit function to parametrize the QCD background.
This is a traditional method to search for a large resonance signal. It also takes 
advantage of the fact that the QCD background always produces a smooth and monotonically decreasing spectrum
with high statistics that can simply be fit. This methodology makes use of Occam's razor~\cite{OccamsRazor}, noting that
if the data can be well fit with the background fit function alone, then there is no significant evidence
for signal, and the data is compatible with background alone. Hence it is simplest to use the data itself for the 
background. Nevertheless, most of these searches are aware that this argument has its limitations, for 
example on the low statistics tail at high dijet mass, and have 
also compared the data with a QCD Monte Carlo just to confirm that dijet mass data are at least roughly 
compatible with QCD alone.
Experimenters have also noted other potential biases when using a parameterization to model the 
background~\cite{bertram}.

\subsubsection{Data from the CERN S$\bar{p}p$S Collider Experiments}

\begin{figure}[bhtp]
\centerline{
\psfig{file=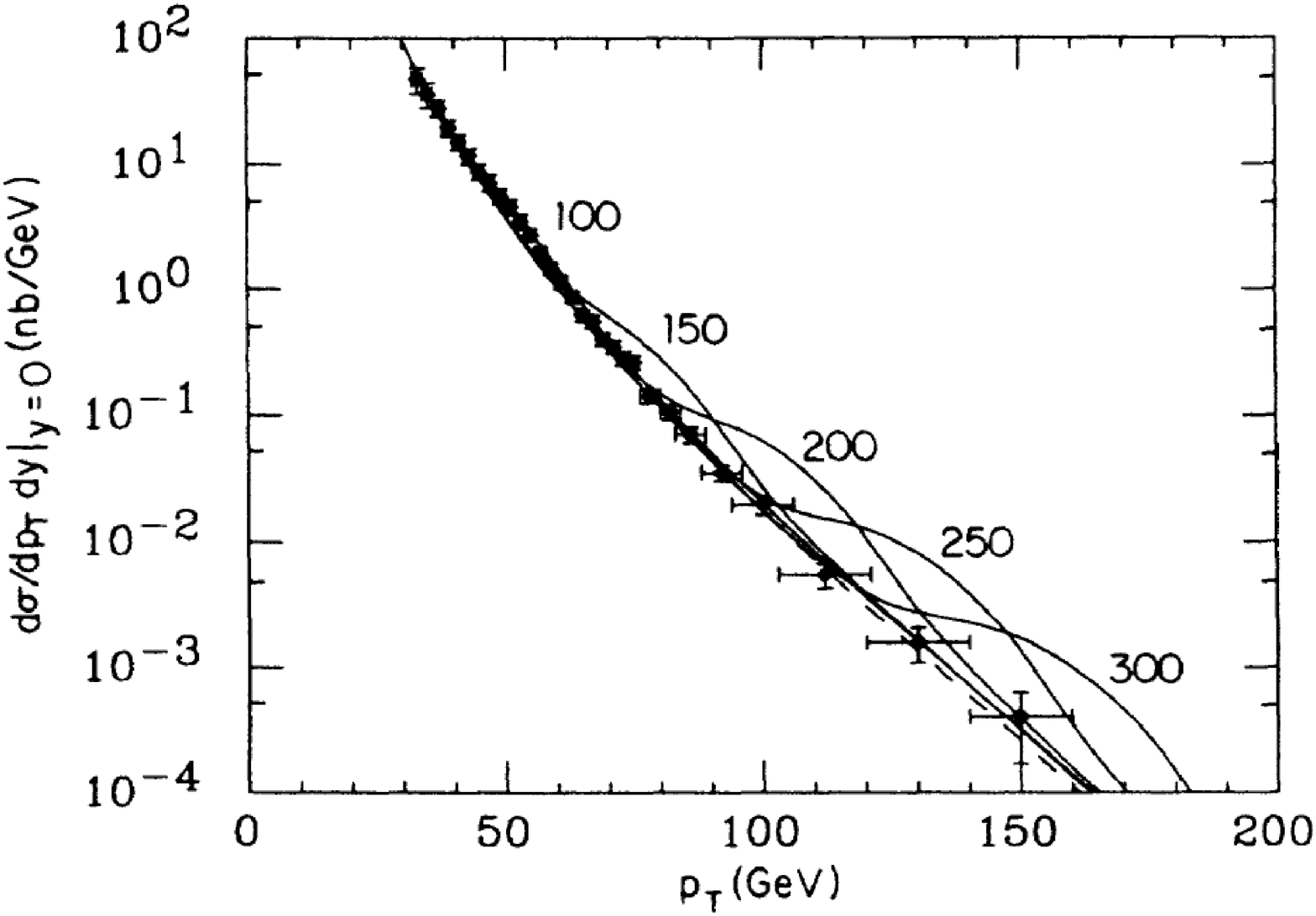,width=2.5in}
\psfig{file=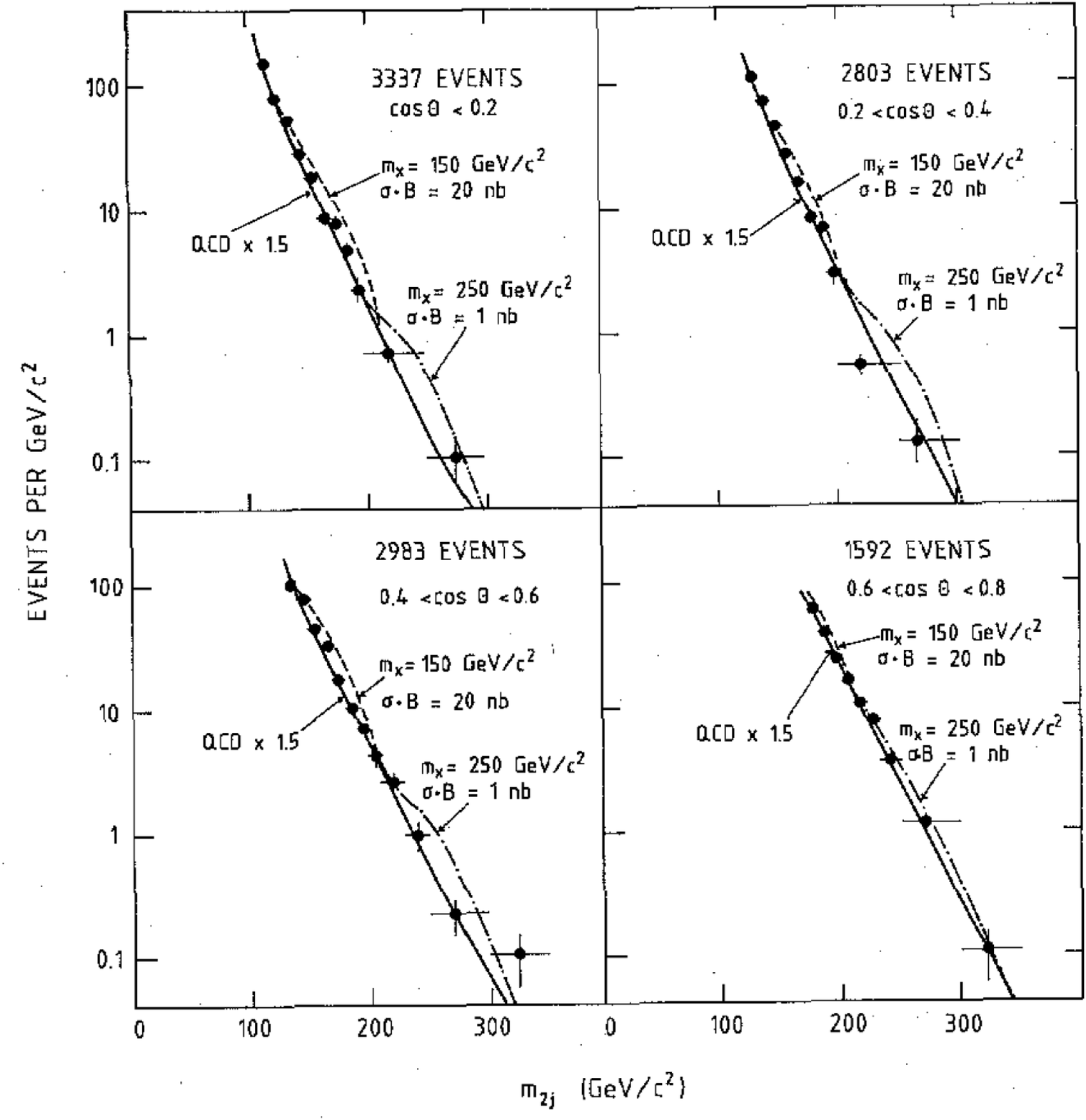,width=2.5in}
}
\caption{Data from UA1 in 1986 and 1988. (left) Inclusive jet $p_T$ spectrum using 260 nb$^{-1}$, originally from 
Ref.~\protect\refcite{UA1ptData} in 1986, compared to predictions of 
QCD and axigluons 
decaying to $N=5$ flavors of quarks from Ref.~\protect\refcite{UA1ptLimit}, 
Copyright 1988 by the American Physical Society. (right) Dijet mass spectra using 
490 nb$^{-1}$ in four bins of $|\cos\theta^*|$ compared to predictions of QCD and a vector
resonance with width $\Gamma=0.1M$ from  Ref.~\protect\refcite{UA1mass},
Copyright 1988, with permission from Elsevier.
\label{UA1data}}
\end{figure}

The earliest search for dijet resonances at hadron colliders~\cite{UA1ptData,UA1ptLimit} was atypical in many 
respects. In Fig.~\ref{UA1data} we reproduce the comparison of UA1 1986 data\cite{UA1ptData} from the inclusive 
jet $p_T$ spectrum for $|\eta|<0.7$ with a leading order calculation of QCD plus the axigluon model of a dijet 
resonance. This comparison was published by theorists~\cite{UA1ptLimit} in the paper which introduced
the theoretical cross section for axigluon production. The analysis was challenging, as the jet $p_T$ data 
had been unsmeared for the effects of jet resolution by the UA1 collaboration, and had to be resmeared by 
the theorists in order to make a valid comparison with the expected smeared shapes of axigluon
resonances in the UA1 detector. Figure~\ref{UA1data} 
also shows a search from UA1 in 1988~\cite{UA1ptLimit}, 
which used the dijet mass spectrum, and again compared with LO QCD normalized to the
data. Both searches using UA1 data required a multiplicative factor of 1.5 to normalize LO QCD to the data. 
The UA1 1988 data extended to a dijet mass of $0.3$ TeV and agreed with the QCD background.
The UA1 analysis of the dijet mass spectrum was unique in that it analyzed the data in four bins of 
$|\cos\theta^*|$ instead of simply applying a single cut on $|\cos\theta^*|$.

\begin{figure}[htbp]
\centerline{
\psfig{file=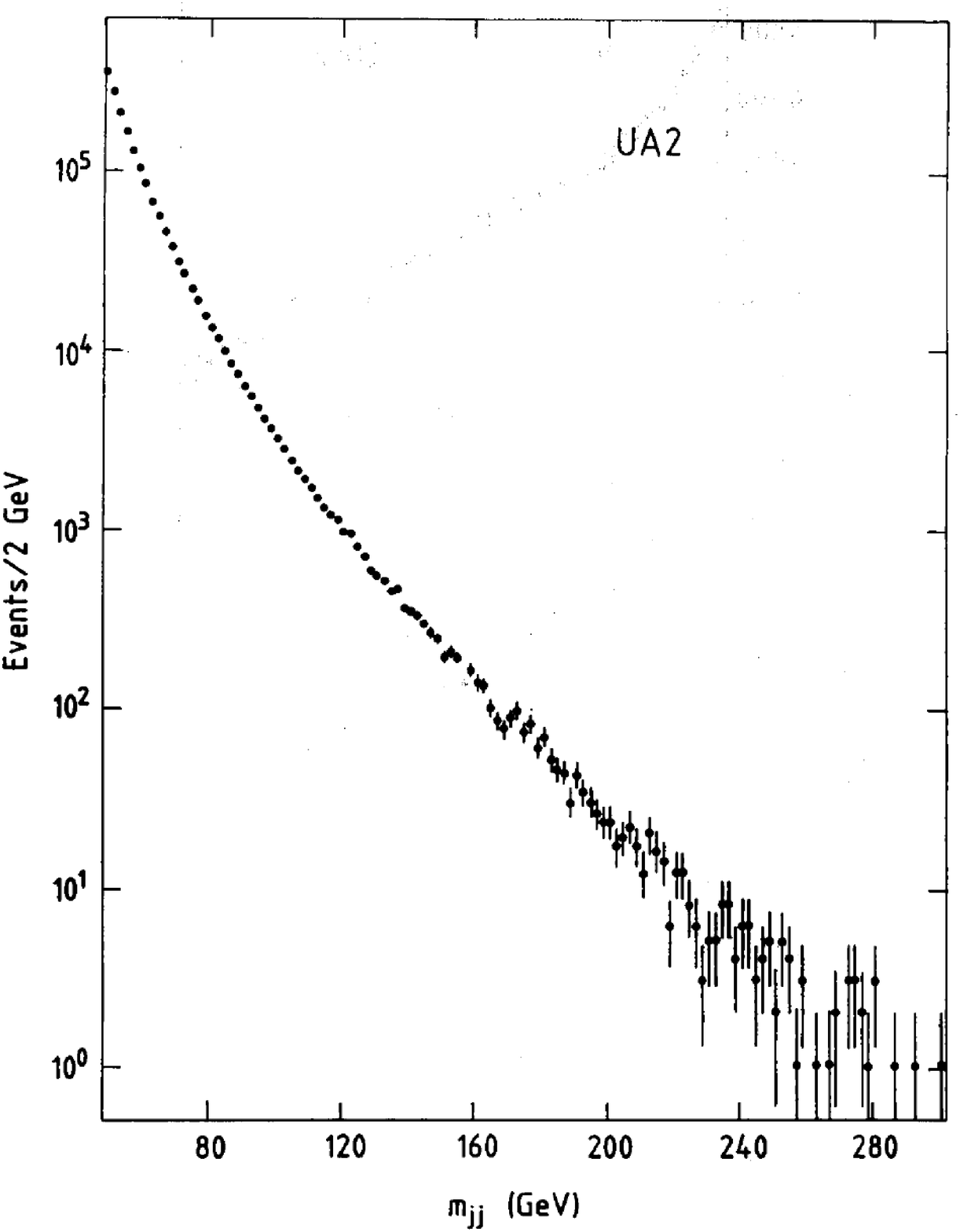,width=2.5in}
\psfig{file=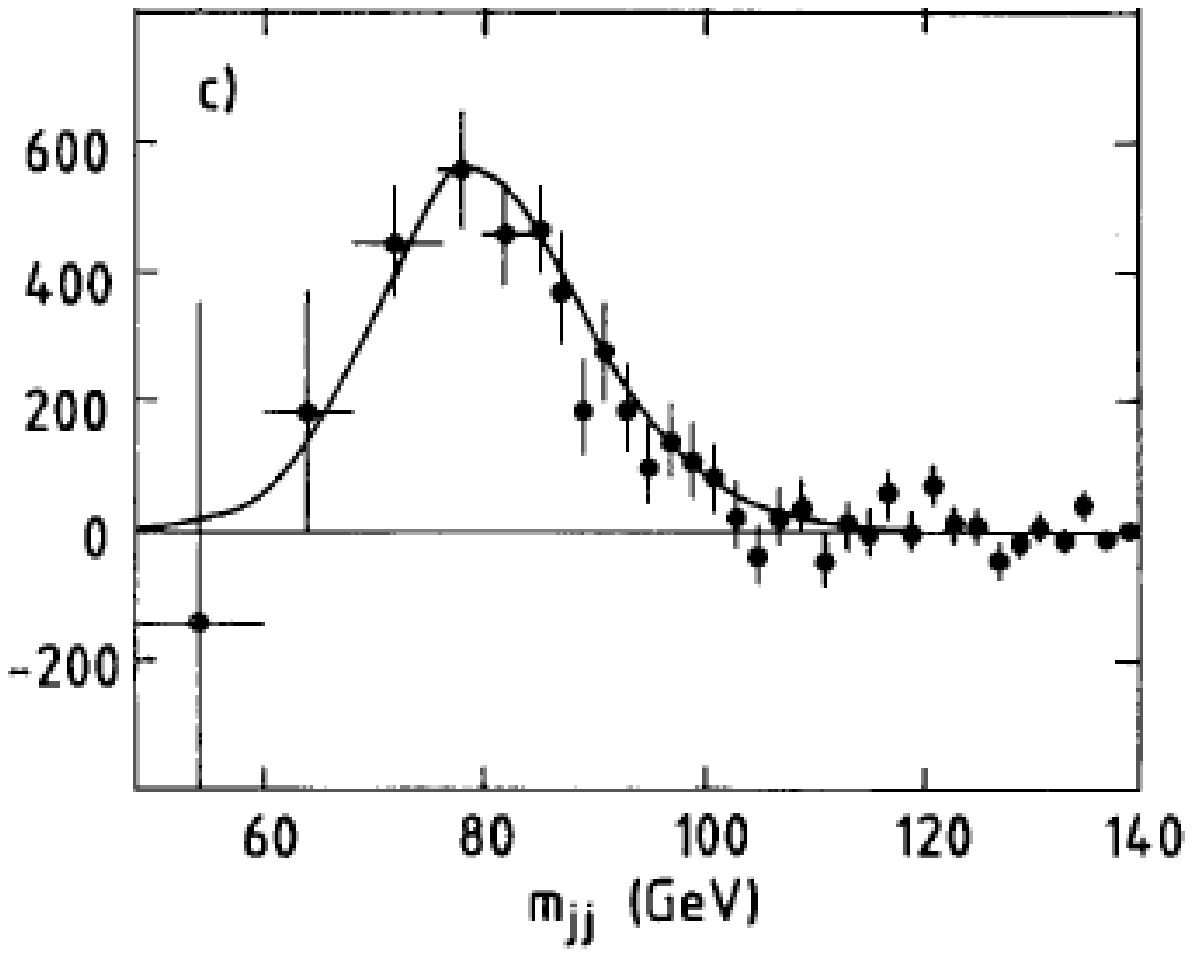,width=2.5in}
}
\vspace*{8pt}
\caption{Data from UA2 in 1990. 
(left) The dijet mass spectrum using 4.7 pb$^{-1}$ and (right) the difference between the data and the fitted background.
The fitted signal from W and Z decays to dijets is shown by the solid curve. From 
Ref.~\protect\refcite{UA21990} with kind permission from Springer Science+Business Media.
\label{UA21990data}}
\end{figure}

The UA2 collaboration in 1990 published the only observation\footnote{Later observations of $W$ decaying to dijets always involved more complicated processes, such as top quark decay.} in hadron collisions~\cite{UA21990} of an $s-\textit{channel}$ dijet resonance:
the W and Z boson, whose mass was already known from its previous discovery in decays to leptons. In 
Fig.~\ref{UA21990data} we reproduce the dijet mass spectrum before and after the subtraction of the QCD background. 
The data after the subtraction is well fit by a single peak with shape, accounting for both the W resonance and the 
smaller rate for the Z resonance. The QCD background came from a parameterization of the differential rate as a 
function of dijet mass, $m$,
\begin{equation}
\frac{dN}{dm}=\frac{p_0}{m^{p_1}} e^{-(p_2 m + p_3 m^2)},
\label{UA2param}
\end{equation}
with four parameters $p_0$, $p_1$, $p_2$ and $p_3$. The fit of this parameterization to the data in the full dijet mass region 
$48<m<300$~GeV had a probability of only 1\%, but excluding the dijet mass region 
$70<m<100$ that contained the W and Z boson, and redoing the fit, gave a probability of 78\%. The clear peak 
after subtracting this second background fit is quite significant, as 
shown in Fig.~\ref{UA21990data}. The search for the dijet decays of the W and Z benefited from 
apriori knowledge of the existence and the mass of the resonance. Otherwise, the 1\% probability of the first background 
fit would not be sufficient disagreement with the background to claim a discovery. Nevertheless, this provided 
an important confirmation of the W and Z resonances in the dijet channel. The background 
parameterization was also used outside the mass region of the W and Z to search for 
higher mass resonances in 1990~\cite{UA21990}, and again in 1993~\cite{UA21993}. These two papers led 
the field by introducing searches with parametrized background shapes. It is unfortunate that no figures 
exist comparing either UA2 analysis for $m>0.12$ TeV to the background parameterization\footnote{Also, no figure 
with all the 1993 UA2 dijet mass data was ever published: UA2 in 1993 reused the data of the 1990 search and added a 
similar amount of new data}. Both UA2 analyses used dijet mass data in the region $0.05 < m < 0.3$~TeV to search
for resonances, with the 1990 search requiring $|\eta|<0.7$ and the 1993 search requiring $|\cos\theta^*|<0.6$.

\subsubsection{Data from the Fermilab Tevatron Collider Experiments}

\begin{figure}[hbtp]
\centerline{
\psfig{file=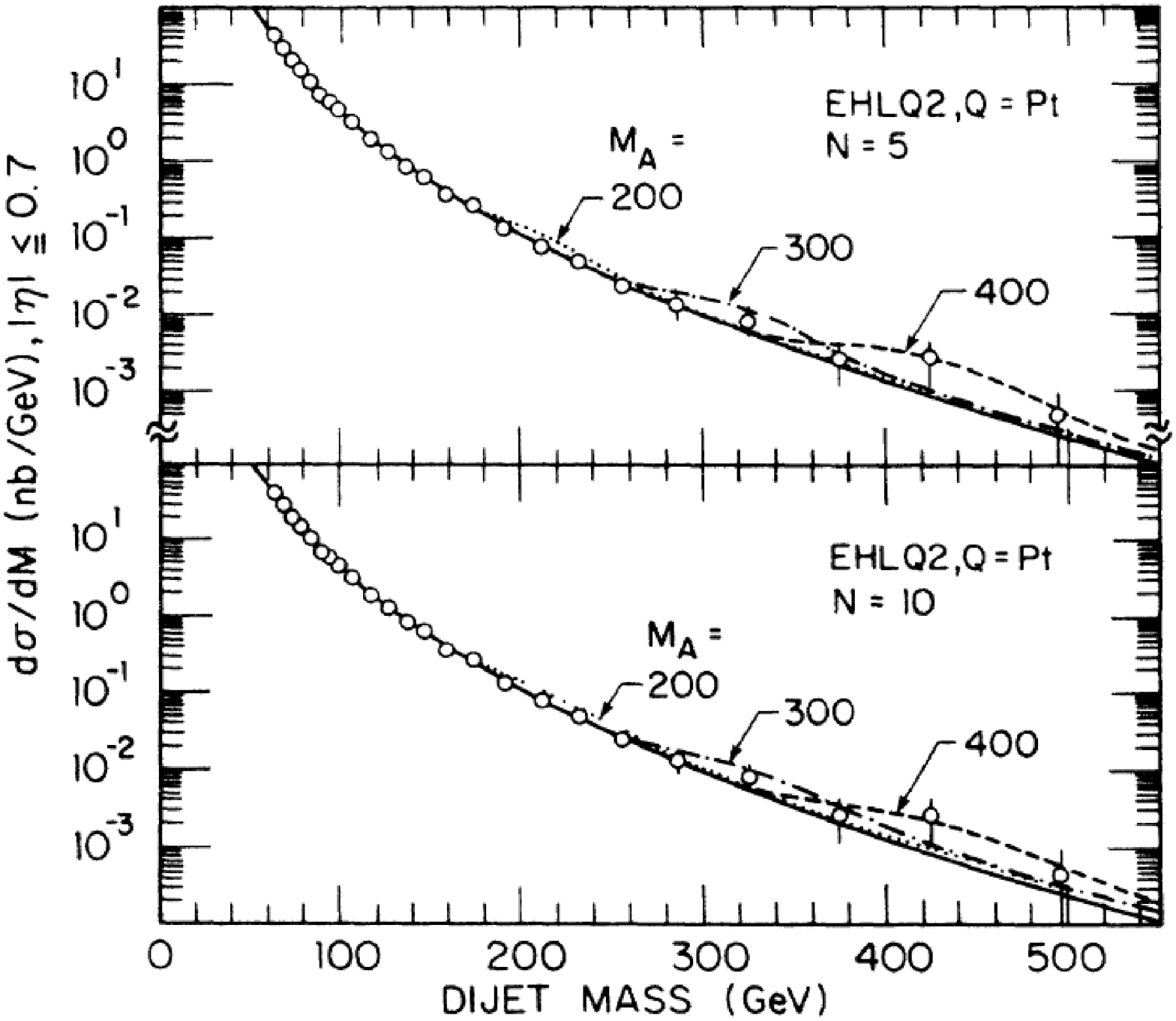,width=2.7in}
\psfig{file=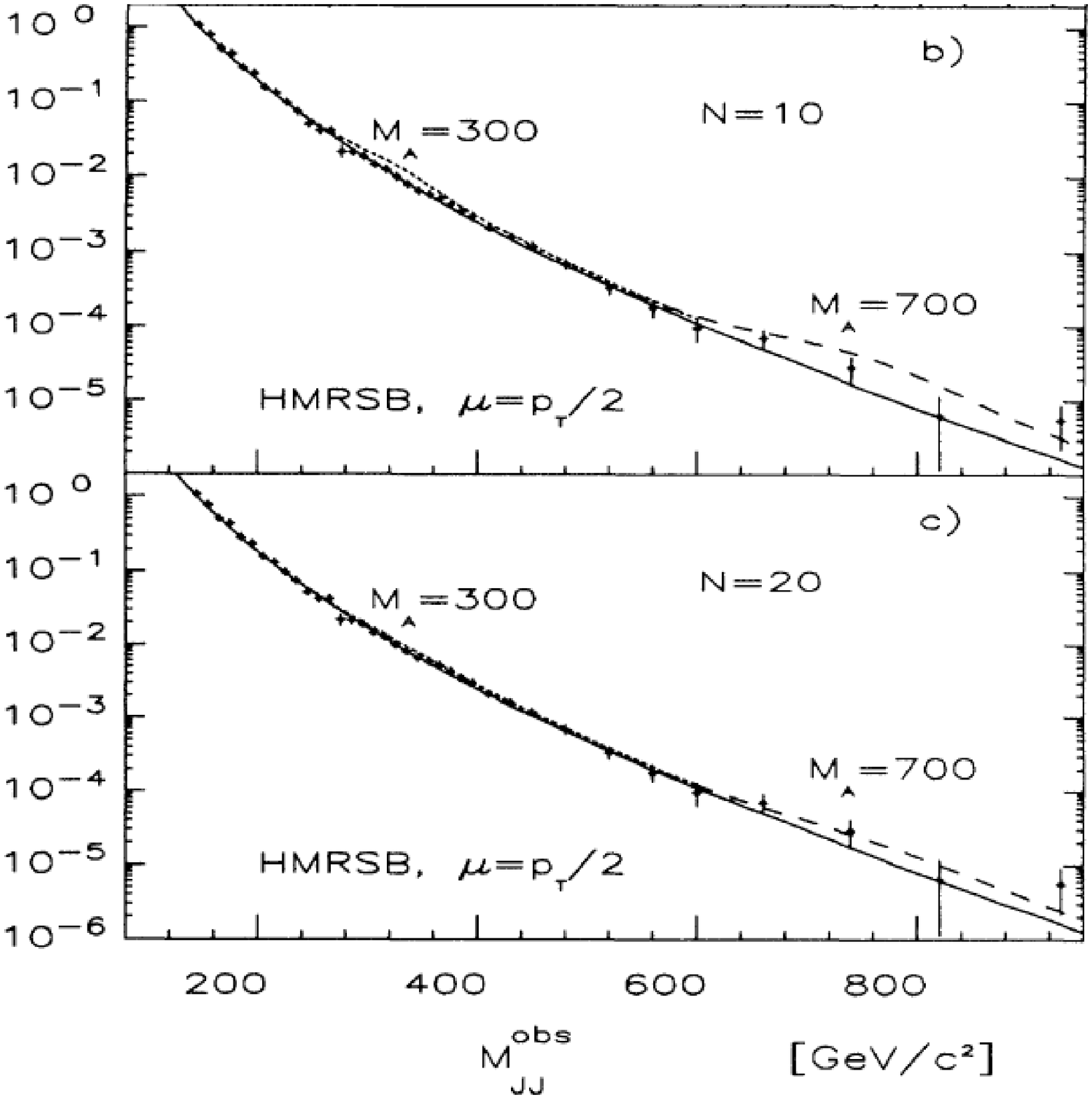,width=2.3in}
}
\vspace*{8pt}
\caption{Data from CDF in 1990 and 1993. (left) Dijet mass spectrum using 26 nb$^{-1}$ 
compared to predictions of QCD and axigluons decaying to $N=5$ (upper) and $N=10$ (lower) 
flavors of quarks from Ref.~\protect\refcite{CDF1990}, Copyright 1990, and 
(right) using 4.2 pb$^{-1}$ for $N=10$ (upper) and $N=20$ (lower) from 
Ref.~\protect\refcite{CDF1993}, Copyright 1993 by the American Physical Society.
\label{CDFearly}}
\end{figure}

The two earliest CDF searches in 1990~\cite{CDF1990} and 1993~\cite{CDF1993} are very similar, and 
model the background roughly like the UA1 search~\cite{UA1mass}.
Dijet mass data with $|\eta|<0.7$ from these CDF searches are shown in Fig.~\ref{CDFearly} compared 
to predictions from the QCD background and axigluon signals.
In 1990 CDF searched for dijet resonances, using 26 nb$^{-1}$, with data in the mass interval $60 < m < 500$ GeV. In 1993 CDF used 4.2 pb$^{-1}$ of data to search in the dijet mass interval 
$140<m<1000$ GeV. Both CDF analyses modeled the background using LO QCD calculations smeared 
with the dijet mass resolution and normalized to the data at low dijet mass. Both datasets were compatible
with the QCD prediction, however, CDF reported in the 1993 paper
that {\it ``a small excess of events is observed in the data between 350 and 400 GeV''} to explain a mass gap in its
axigluon exclusions for $N=20$.

\begin{figure}[htbp]
\centerline{
\psfig{file=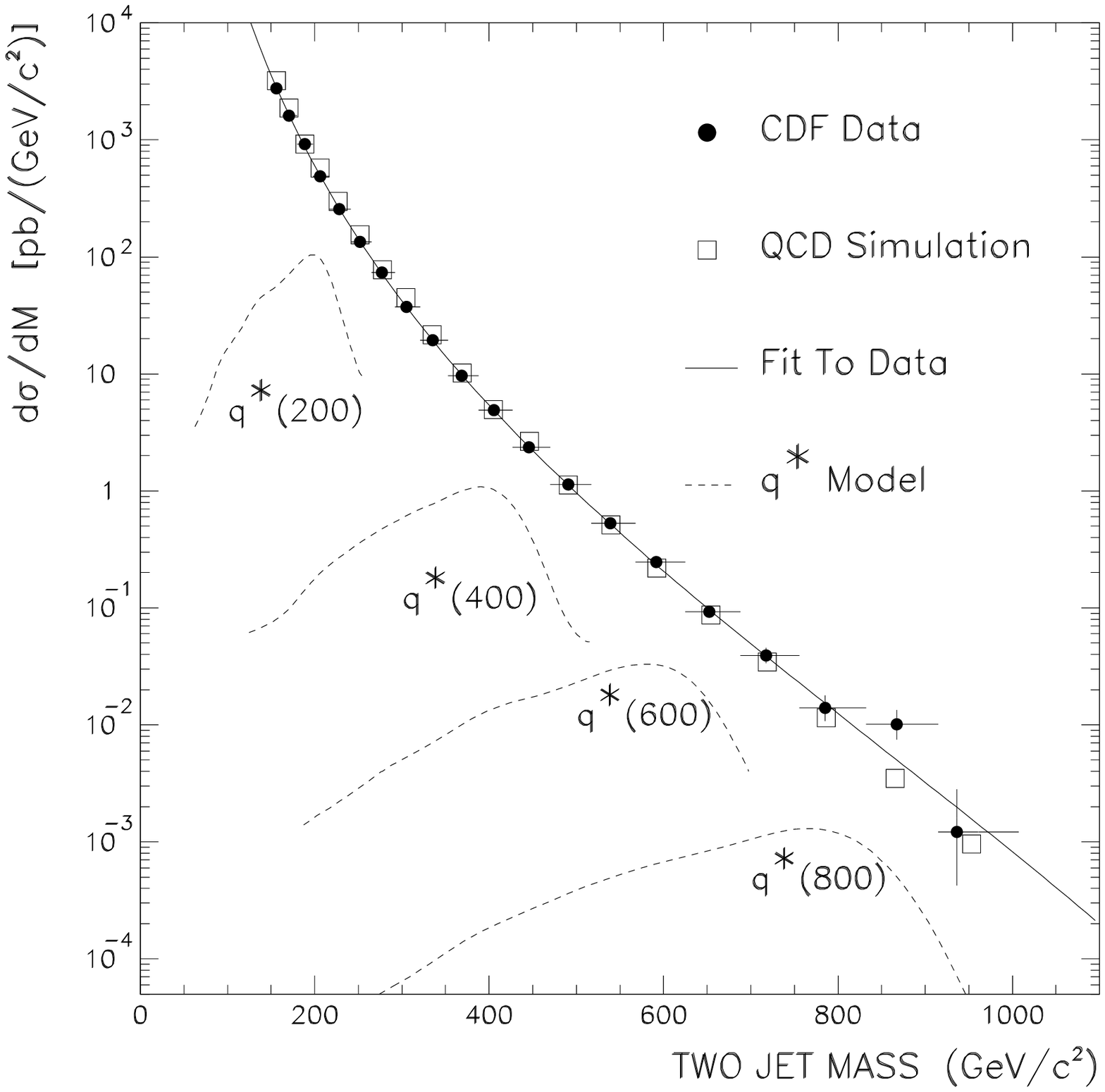,width=2.5in}
\psfig{file=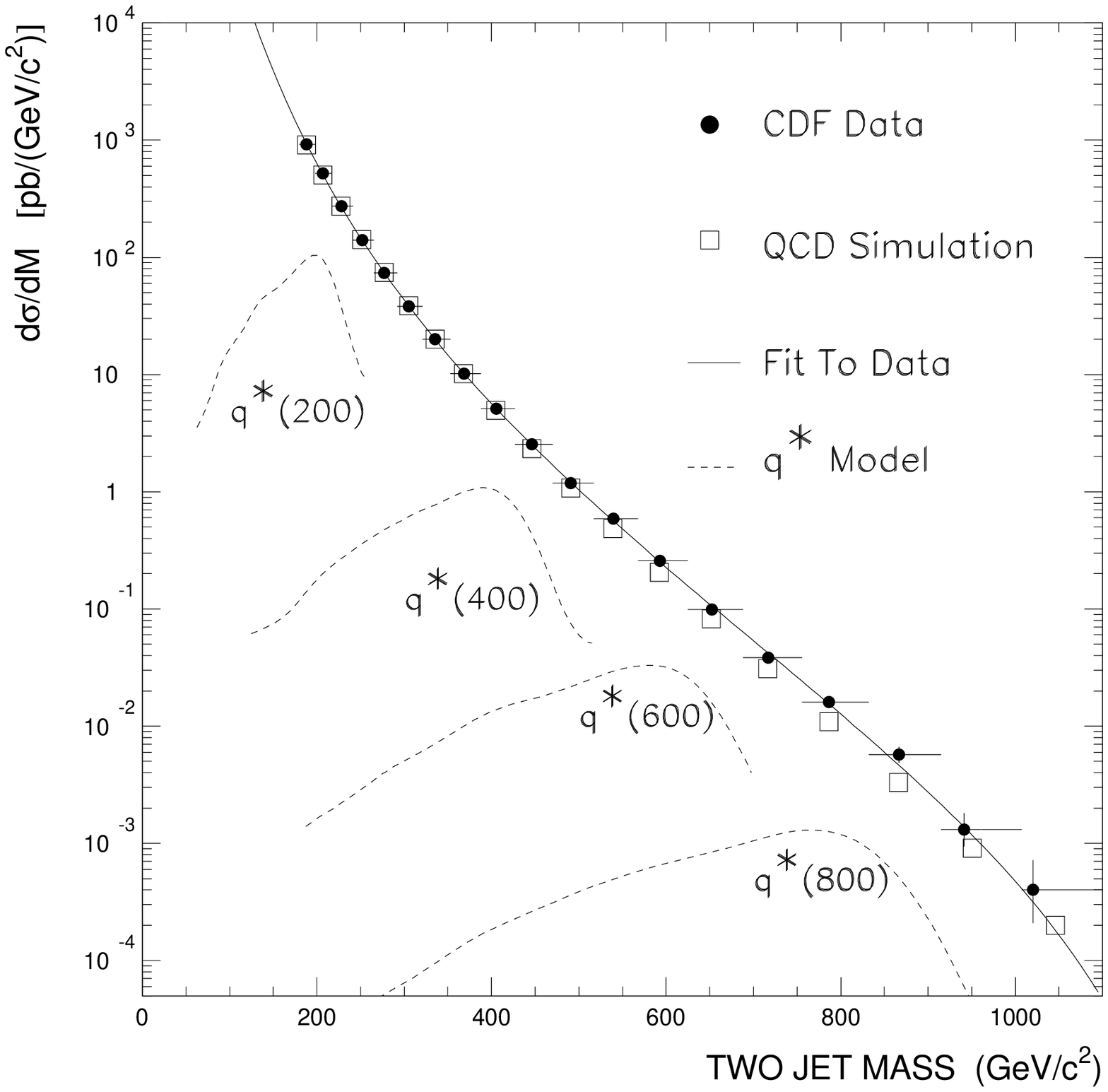,width=2.5in}
}
\vspace*{8pt}
\caption{Data from CDF in 1995 and 1997.
(left) Dijet mass spectrum using 19 pb$^{-1}$ compared to a background fit, and simulations of QCD 
and excited quark signals from Ref.~\protect\refcite{CDF1995}, Copyright 1995, and 
(right) using 106 pb$^{-1}$ from Ref.~\protect\refcite{CDF1997}, Copyright 1997 by the
American Physical Society.
\label{CDFmiddleData}}
\end{figure}

\begin{figure}[htbp]
\centerline{
\psfig{file=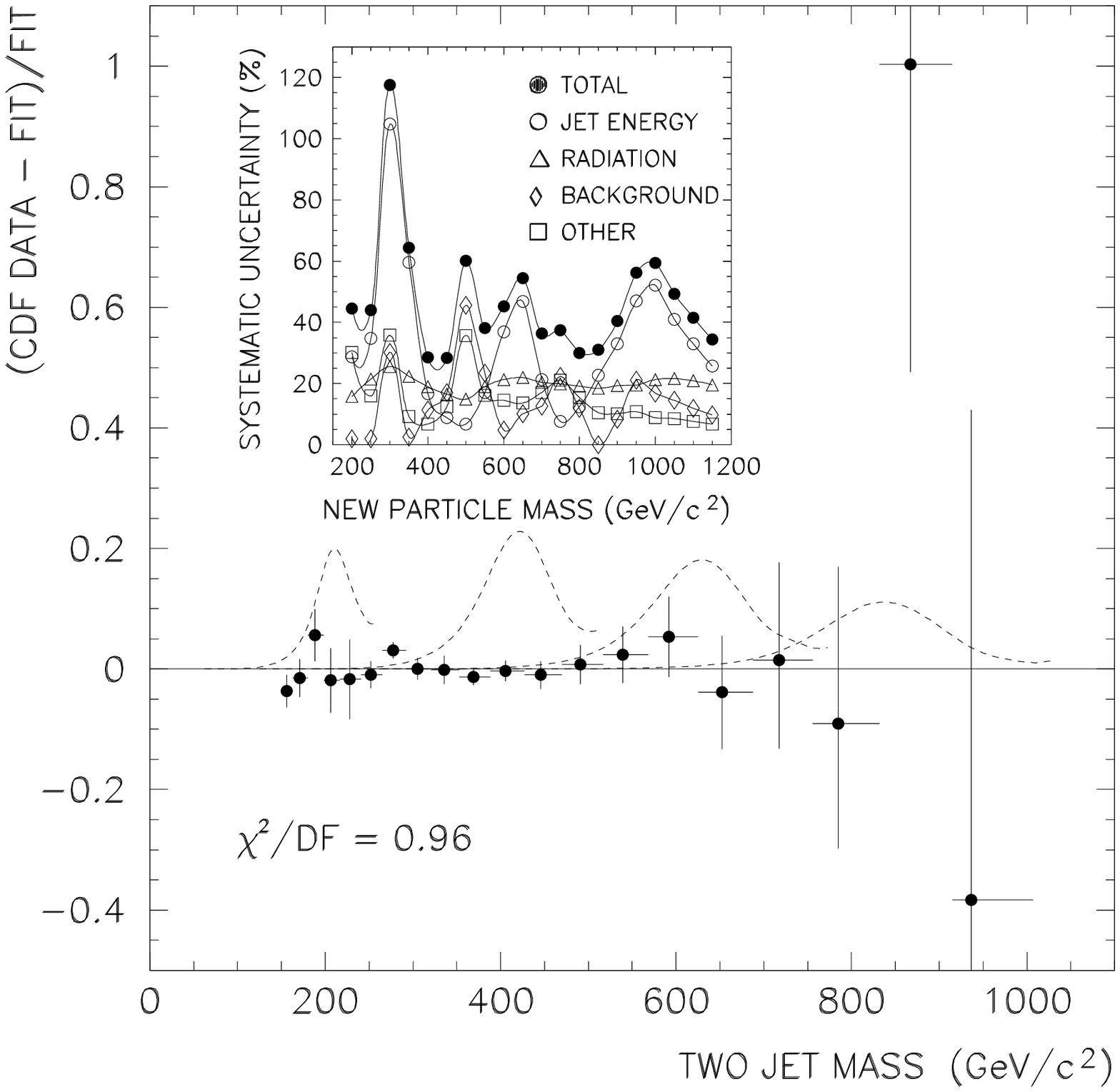,width=2.5in}
\psfig{file=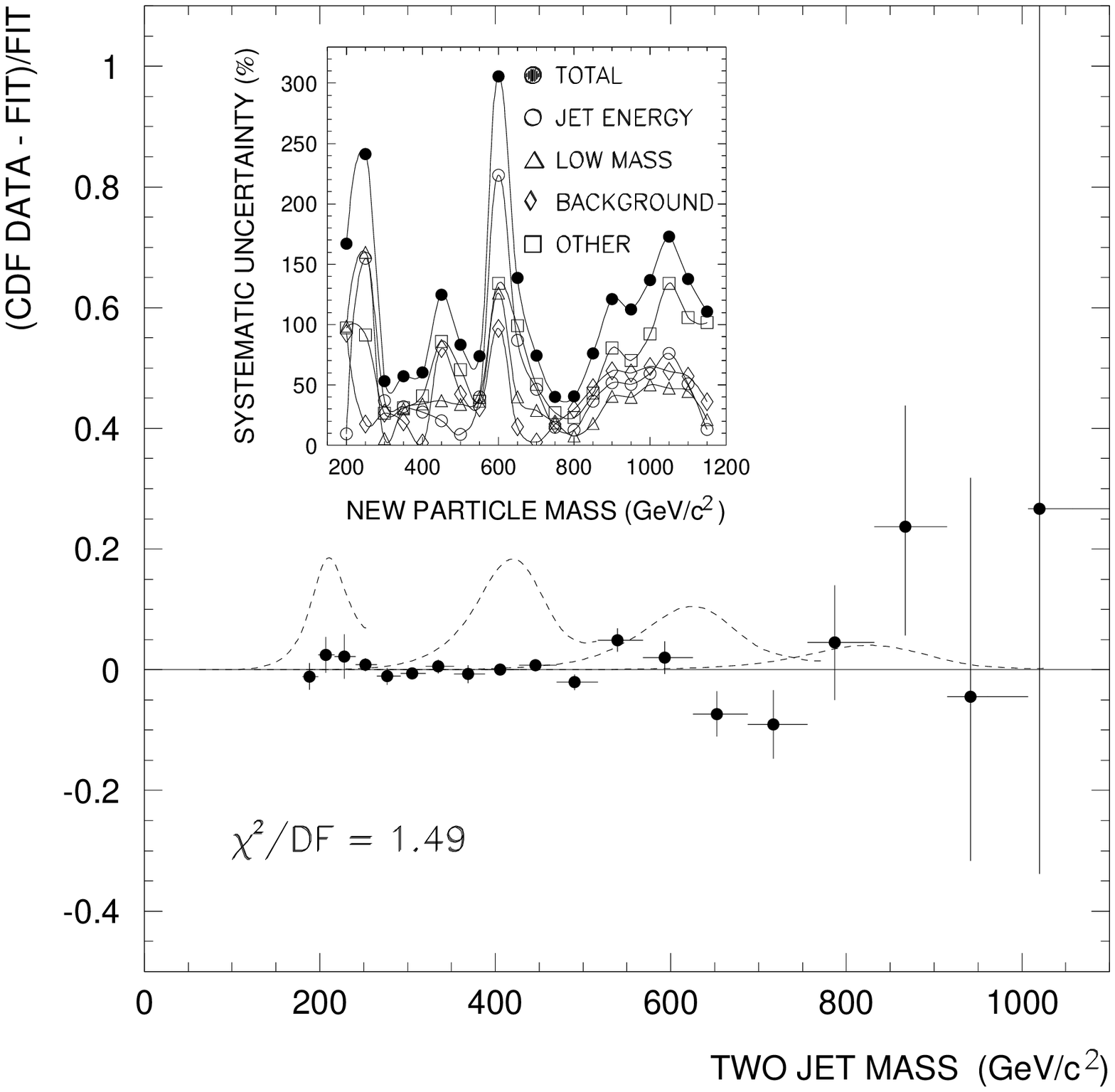,width=2.5in}
}
\vspace*{8pt}
\caption{Fit and systematics from CDF in 1995 and 1997. 
(left) Fractional difference between data using 19 pb$^{-1}$ and a background fit, compared to 
simulations of excited quark 
signals.  Inset shows systematic uncertainty for signal. From Ref.~\protect\refcite{CDF1995},
Copyright 1995, and (right) using 106 pb$^{-1}$ from Ref.~\protect\refcite{CDF1997}, Copyright 1997 
by the American Physical Society.
\label{CDFmiddleFit}}
\end{figure}

The CDF searches in 1995~\cite{CDF1995} and 1997~\cite{CDF1997} used similar techniques, 
and modeled the background like the UA2 searches~\cite{UA21990,UA21993}.  
Dijet mass data with $|\eta|<2.0$ and $|\cos\theta^*|<2/3$ from these CDF searches are shown in 
Fig.~\ref{CDFmiddleData}. Despite having $5-20$ times more luminosity than the previous
search published in 1993, the 1995 and 1997 data also extended to about $m=1$ TeV because the 
$|\cos\theta^*|<2/3$ cut suppressed the QCD background. The 1995 search fit the background 
with a functional form containing only three parameters:
\begin{equation}
\frac{d\sigma}{dm}=\frac{p_0}{m^{p_1}}(1 - m/\sqrt{s})^{p_2},
\label{CDFparam1}
\end{equation}
while the 1997 search used a similar form with four parameters:
\begin{equation}
\frac{d\sigma}{dm}=\frac{p_0}{m^{p_1}}(1 - m/\sqrt{s} +p_3 m^2/s)^{p_2},
\label{CDFparam2}
\end{equation}
where $\sqrt{s}$ is the $p\bar{p}$ collision energy. These parameterizations were motivated
by leading order QCD. The term $m^{p_1}$ mimics the mass dependence of the QCD matrix elements (Eq.~\ref{massXsection}), 
and was borrowed from the UA2 parameterization in Eq.~\ref{UA2param}.
The term $(1-m/\sqrt{s})^{p_2}$ was introduced by CDF and mimics the mass dependence of 
the parton distributions at an average fractional momentum $x=m/\sqrt{s}$. With increased 
statistics in 1997 the additional $p_3$ term was needed in order to model the data at high dijet mass.
In addition to the background fit used to search for new physics, CDF also compared their data 
to a QCD prediction from \textsc{Pythia}, with the prediction normalized to the data, on average in 1995, and
normalized to the low mass data in 1997. In the 1997 paper CDF noted that the data was {\it ``above the 
QCD simulation at high mass''}, similar to the excess they reported in the inclusive jet cross section 
at high $p_T$~\cite{CDFpt}. The source of the difference was not understood at the time, but it has 
since been attributed to the proton PDFs within the QCD prediction. By using a
parameterization for the background, CDF was able to minimize the effects of considerable experimental
and theoretical uncertainties on the determination of the background shape. In Fig.~\ref{CDFmiddleFit} 
CDF showed the difference between their data and the background fit, and concluded there was no
evidence for new physics. In 1995 they quoted the local 
significance of fluctuations interpreted as a resonance mass of 250 GeV (2.3$\sigma$), 
550 GeV (1.3$\sigma$) and 850 GeV (1.8$\sigma$), and in 1997 pointed out that the single bin near 
dijet mass 550 GeV was 2.6 $\sigma$ above the fit. It was not common practice in those years to estimate 
the "look elsewhere effect"~\cite{LEE}, used to quote a global significance of observing local 
fluctuations by also taking into account the probability of observing the fluctuation anywhere in 
the mass range considered.
In Fig.~\ref{CDFmiddleFit} CDF also compared the data with the shape and normalization of excited 
quark signals, allowing the excluded mass to roughly be estimated by eye directly from the data.

\begin{figure}[htbp]
\centerline{
\psfig{file=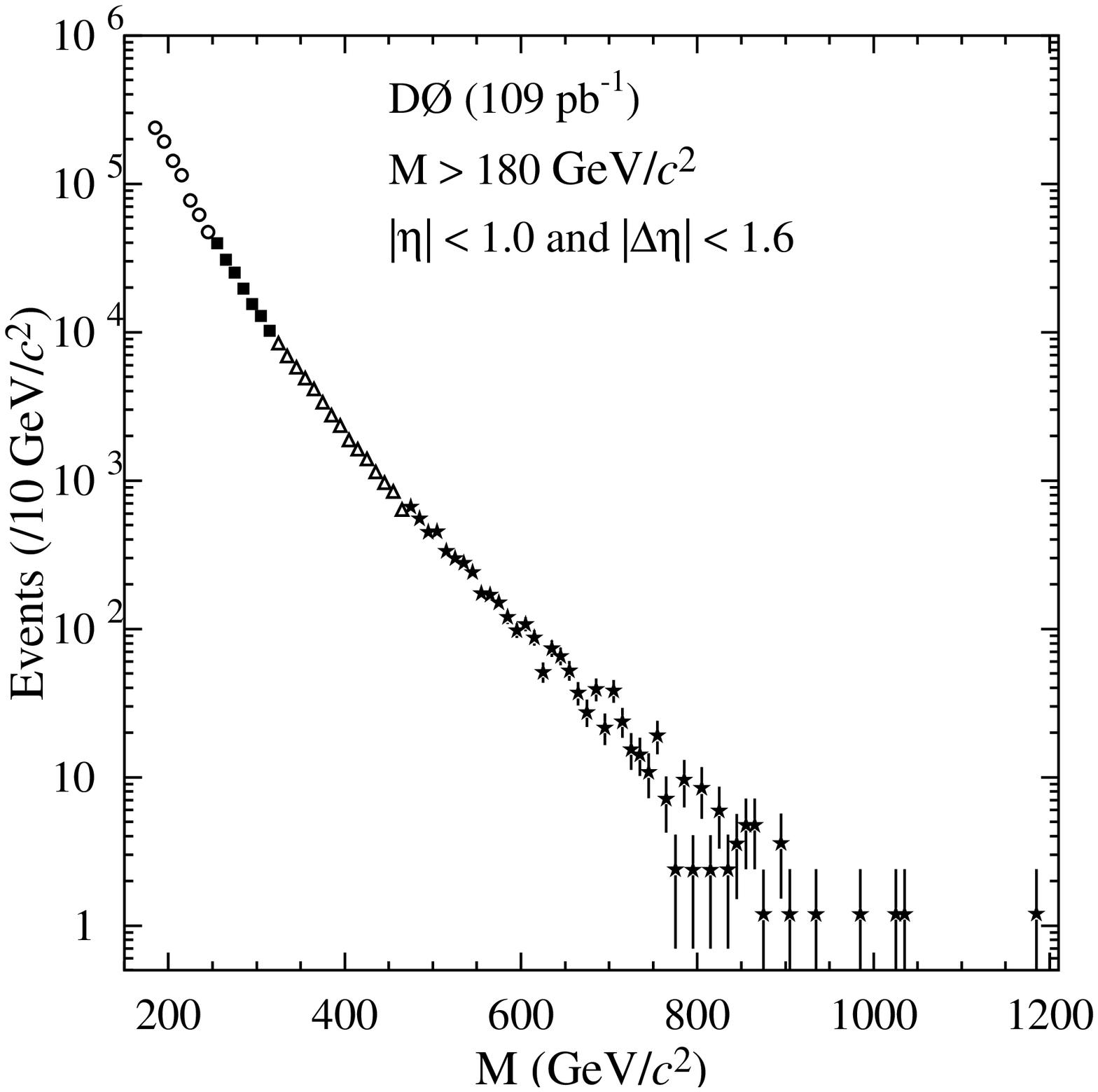,width=2.1in}
\psfig{file=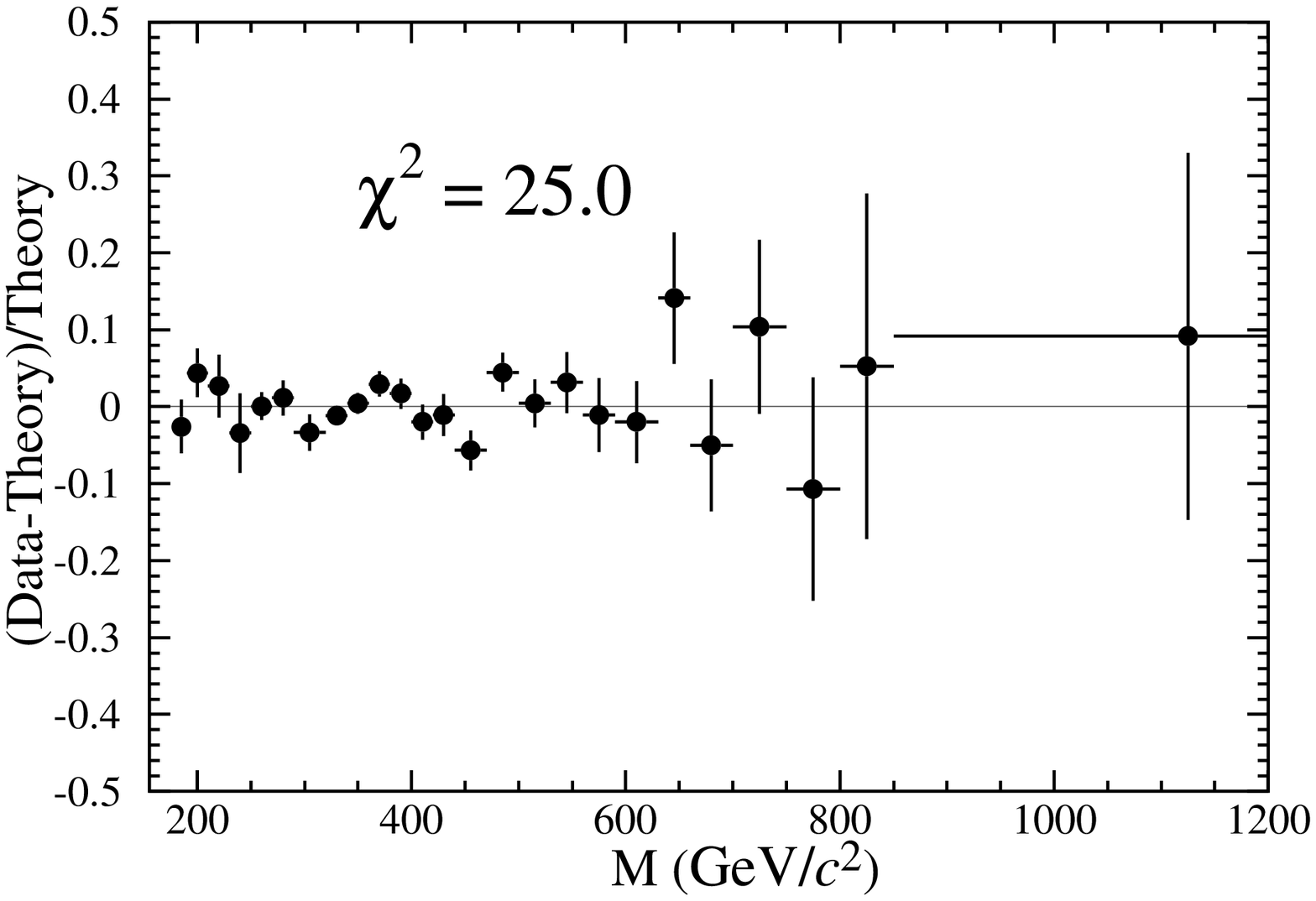,width=2.9in}
}
\vspace*{8pt}
\caption{Data from D0 in 2004. (left)  The dijet mass spectrum using 109 pb$^{-1}$ and (right) 
the fractional difference between the same data and NLO QCD from 
Ref.~\protect\refcite{D02004}, Copyright 2004 by the American Physical Society.
\label{D02004Data}}
\end{figure}

Figure~\ref{D02004Data} shows D0 data from a search published in 2004~\cite{D02004}. The analysis required 
$|\eta|<1$ and $|\cos\theta^*|<0.67$ and measured dijet masses in the range $0.18<m<1.2$ TeV.
The D0 analysis was unique in using NLO QCD to model the background. A QCD calculation from
JETRAD~\cite{JETRAD} with CTEQ6M~\cite{CTEQ6} PDFs and renormalization scale $\mu=0.5 E_T^{MAX}$ was smeared with
the measured dijet mass resolution. It agreed remarkably
well with the data. The $\chi^2/DF$ was $25/25$, considering only statistical uncertainties and without 
any change to the normalization of the QCD 
prediction. Considering the experimental and theoretical uncertainties, the agreement was impressive. This
level of agreement between experiment and QCD calculation for the dijet mass distribution, both in shape
and in normalization, has not been seen by any of the other experiments.

\begin{figure}[htbp]
\centerline{
\psfig{file=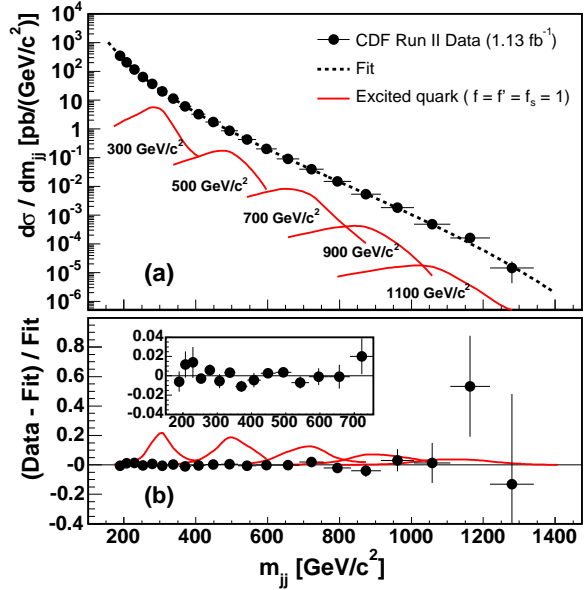,width=3.3in}
}
\vspace*{8pt}
\caption{Data from CDF in 2009.
(top) Dijet mass spectrum using 1.13 fb$^{-1}$ compared to a background fit,  
and simulations of excited quark signals, and (bottom) the fractional difference between the same data and 
fit and (inset) a zoom at low mass.  From Ref.~\protect\refcite{CDF2009}, Copyright 2009 
by the American Physical Society.
\label{CDF2009Data}}
\end{figure}

Dijet mass data shown in Fig.~\ref{CDF2009Data} was published
by CDF in 2009~\cite{CDF2009}, the last search from the Tevatron.  
To use the same dataset for both a QCD measurement and a resonance search 
only $|\eta|<1$ was required, with no cut on $|\cos\theta^*|$, and
the mass distribution was measured in the range $0.18<m<1.3$ TeV.
The technique for modeling
the background is the same as the 1995 and 1997 CDF searches, however, the following parameterization
was developed for an improved fit to the high statistics data over a wide mass range:
\begin{equation}
\frac{d\sigma}{dm}=\frac{p_0(1 - m/\sqrt{s})^{p_1}}{(m/\sqrt{s})^{({p_2}+{p_3}\ln{[m/\sqrt{s}]})}}.
\label{CDFparam3}
\end{equation}
This parameterization gave a good fit to the data with $\chi^2/DF=16/17$, and CDF used it as the background
in the search for resonances. In the same paper CDF also compared the data with a full NLO QCD 
calculation from fastNLO~\cite{fastNLO}, using CTEQ6.1M~\cite{CTEQ6.1}, and with renormalization scale $\mu=p_T^{AVG}$, and found 
$\chi^2/DF=21/21$ after taking into account the systematic
uncertainties. However, unlike the D0 search, the shape agreement between data and 
NLO QCD at CDF without including systematic uncertainties was not sufficient
to use NLO QCD for the background in a search for resonances.

\subsubsection{Data from the CERN Large Hadron Collider Experiments}

\begin{figure}[htbp]
\centerline{
\psfig{file=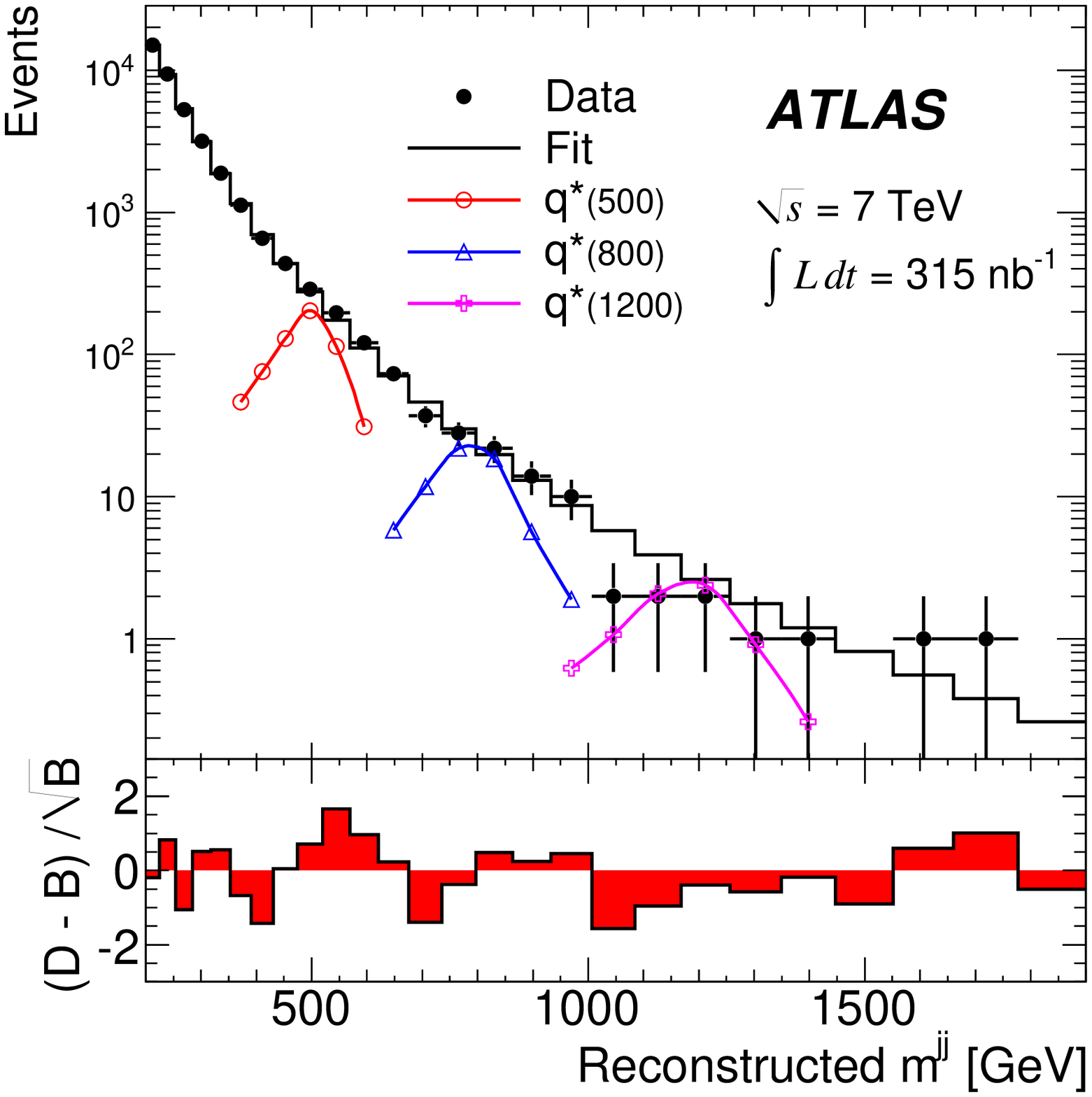,width=2.5in}
\psfig{file=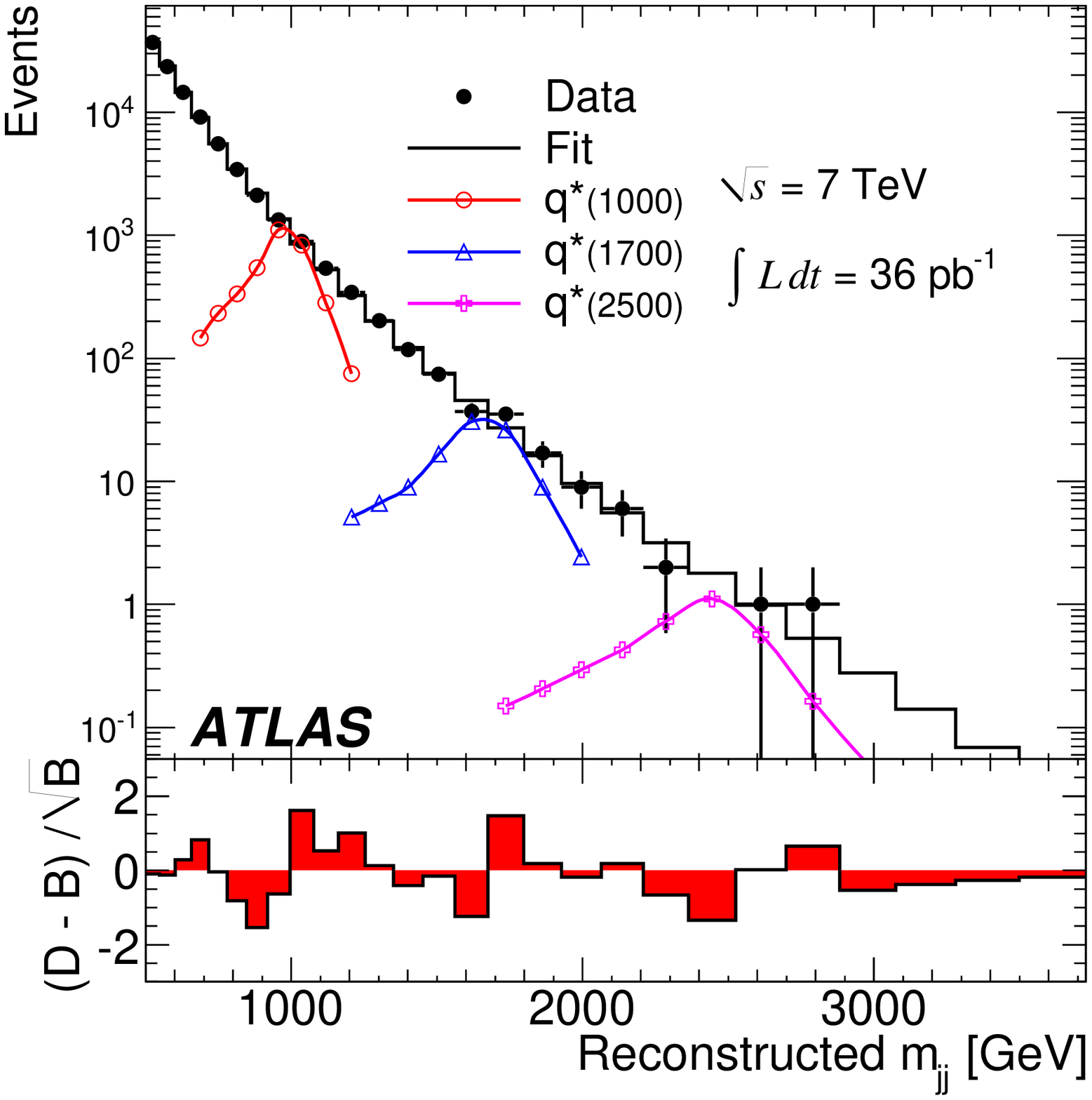,width=2.5in}
}
\centerline{
\psfig{file=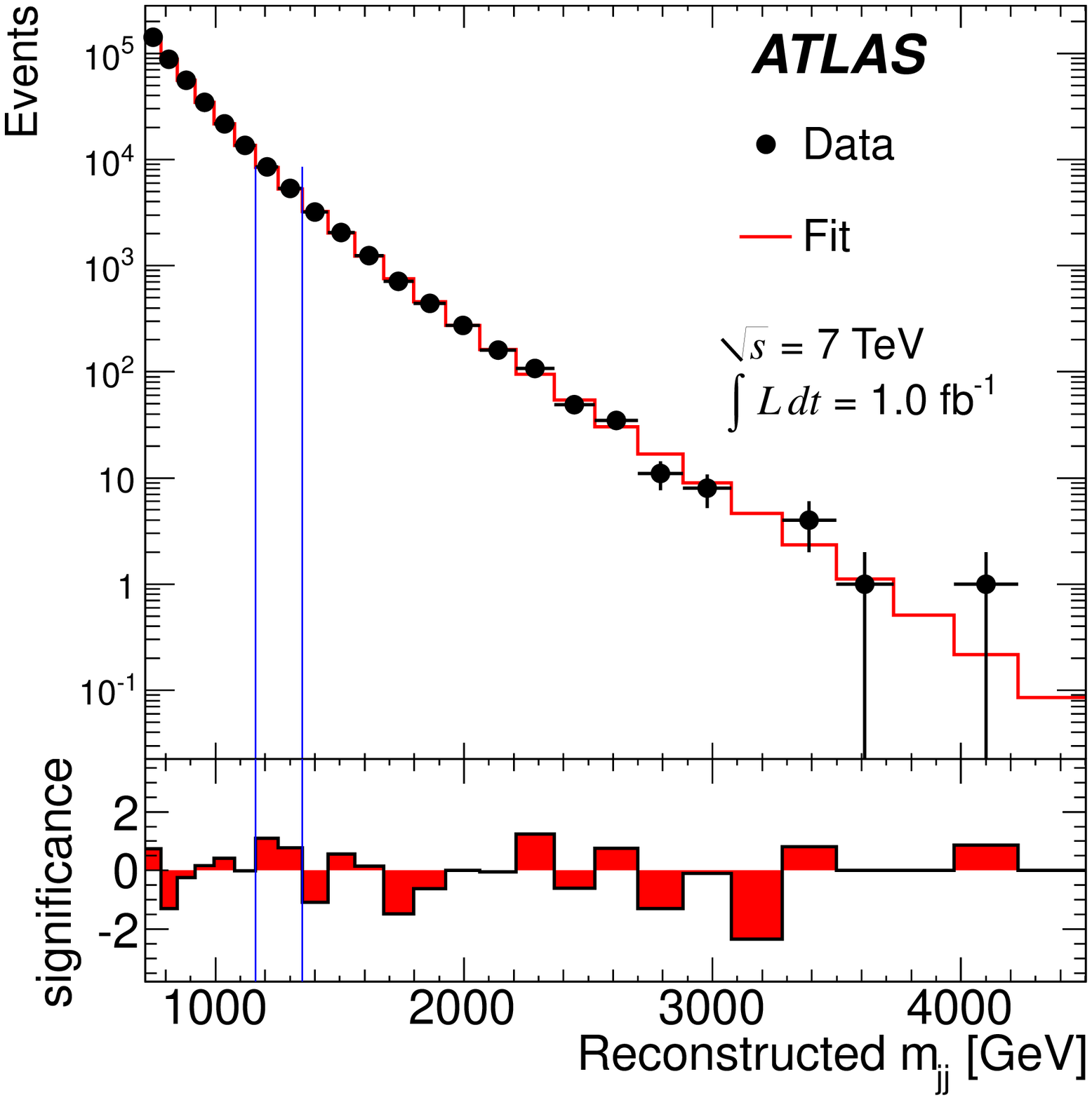,width=2.5in}
}
\vspace*{8pt}
\caption{Data from ATLAS in 2010 and 2011.
(top left) Dijet mass spectrum using 315 nb$^{-1}$ compared to a background fit,  
and simulations of excited quark signals, and the bin-by-bin significance of the difference between
the data and the fit from Ref.~\protect\refcite{ATLAS2010} Copyright 2010 
by the American Physical Society. (top right) Same using 36 pb$^{-1}$ from
from Ref.~\protect\refcite{ATLAS2011}. (bottom)  Same using $1.0$ fb$^{-1}$, and without
simulations, from Ref.~\protect\refcite{ATLAS2011summer}.
\label{ATLASdata}}
\end{figure}

The introduction of the Large Hadron Collider greatly accelerated the pace of dijet 
resonance searches. Data from three ATLAS searches are shown in Fig.~\ref{ATLASdata}, all 
published within roughly a year. The three searches required 
$|\eta|<2.5$ and $|\cos\theta^*|<0.57$ for the search in 2010~\cite{ATLAS2010} and winter 
2011~\cite{ATLAS2011}, and $|\eta|<2.8$ and $|\cos\theta^*|<0.54$ in summer 2011~\cite{ATLAS2011summer}. The measured 
dijet mass range for the three searches was $0.2<m<1.7$ TeV, $0.5<m<2.8$ TeV, and 
$0.72 < m < 4.1$ TeV respectively. The lowest dijet mass used increased with each search because the 
jet trigger threshold used for the search increased. All ATLAS searches used the last CDF
parameterization in Eq.~\ref{CDFparam3} to model the background. As shown in Fig.~\ref{ATLASdata},
the ATLAS searches all reported the bin-by-bin statistical significance of the difference between
the data and the background fit, clearly showing that all fits were good and there was no evidence 
for new physics. The ATLAS searches introduced the \textsc{BumpHunter}~\cite{bumphunter} statistical method for finding the 
most significant upward fluctuation, and for quantifying its complete global probability, including the 
``look elsewhere effect'', of coming from the background. The largest upward fluctuations and the corresponding 
background probabilities were $0.55$ TeV and at least 51\% in 2010, $1.1$ TeV and 39\% in winter 2011, 
and $1.25$ TeV and 82\% in summer 2011.  

\begin{figure}[htbp]
\centerline{
\psfig{file=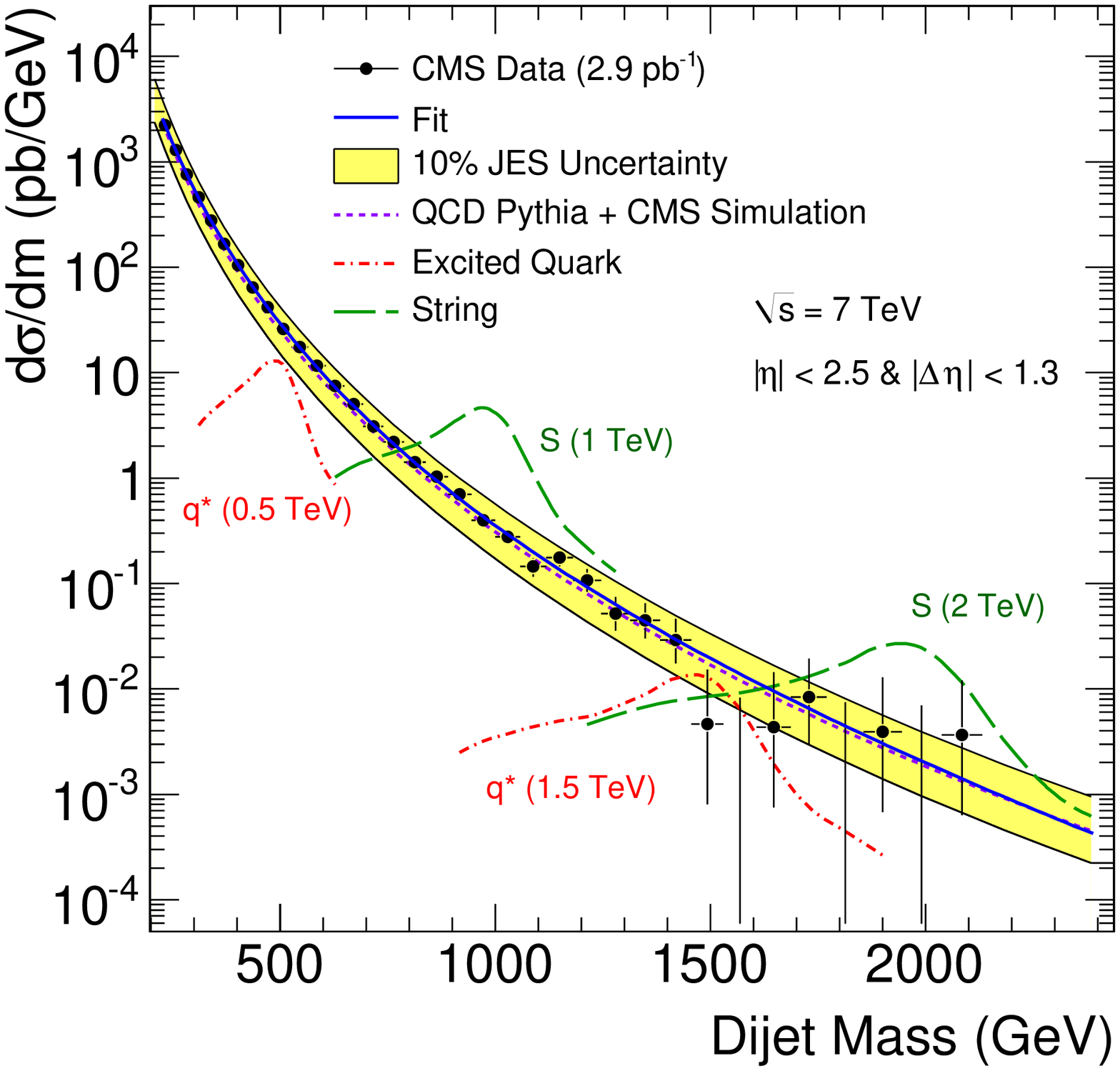,width=2.5in}
\psfig{file=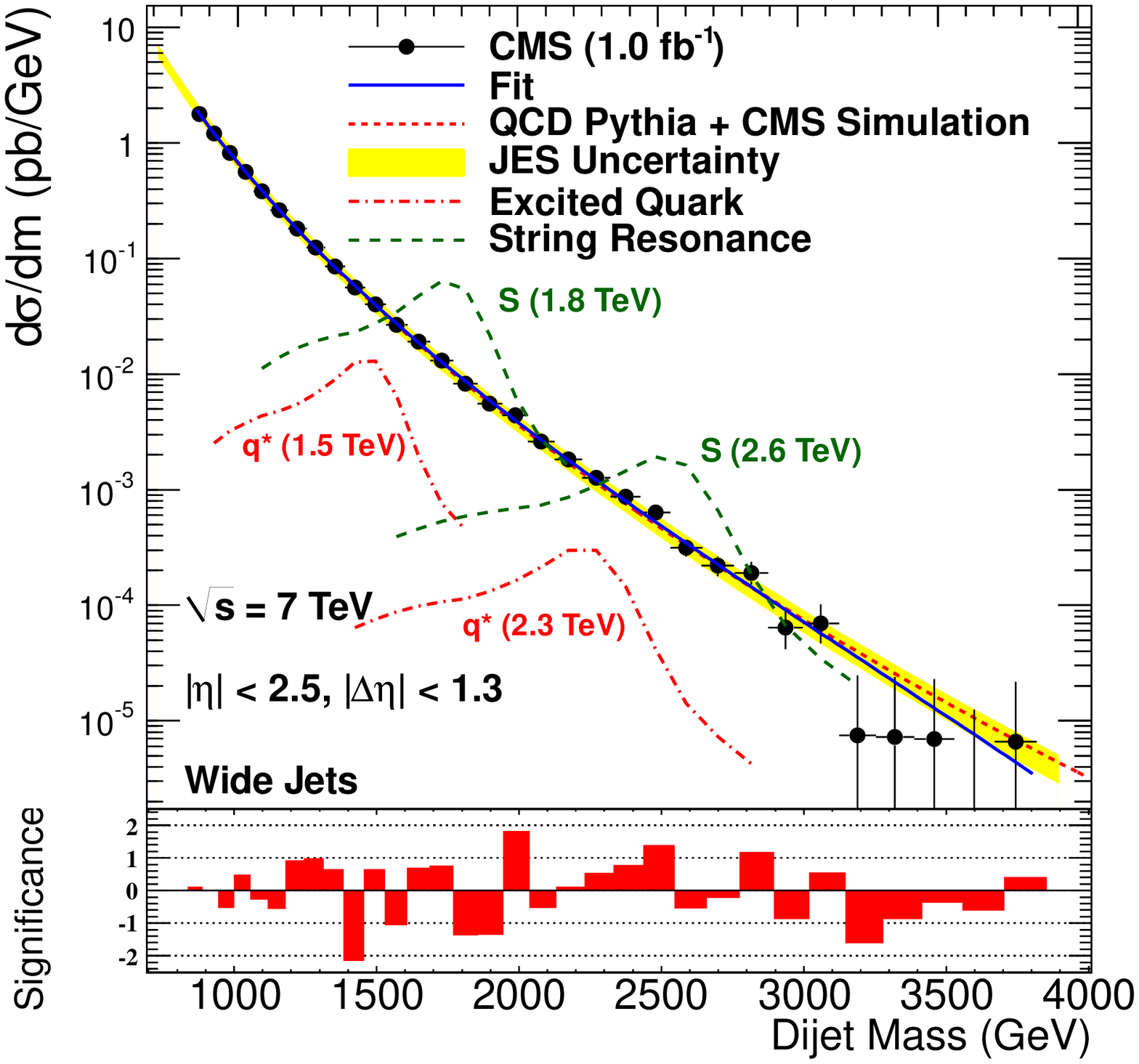,width=2.5in}
}
\centerline{
\psfig{file=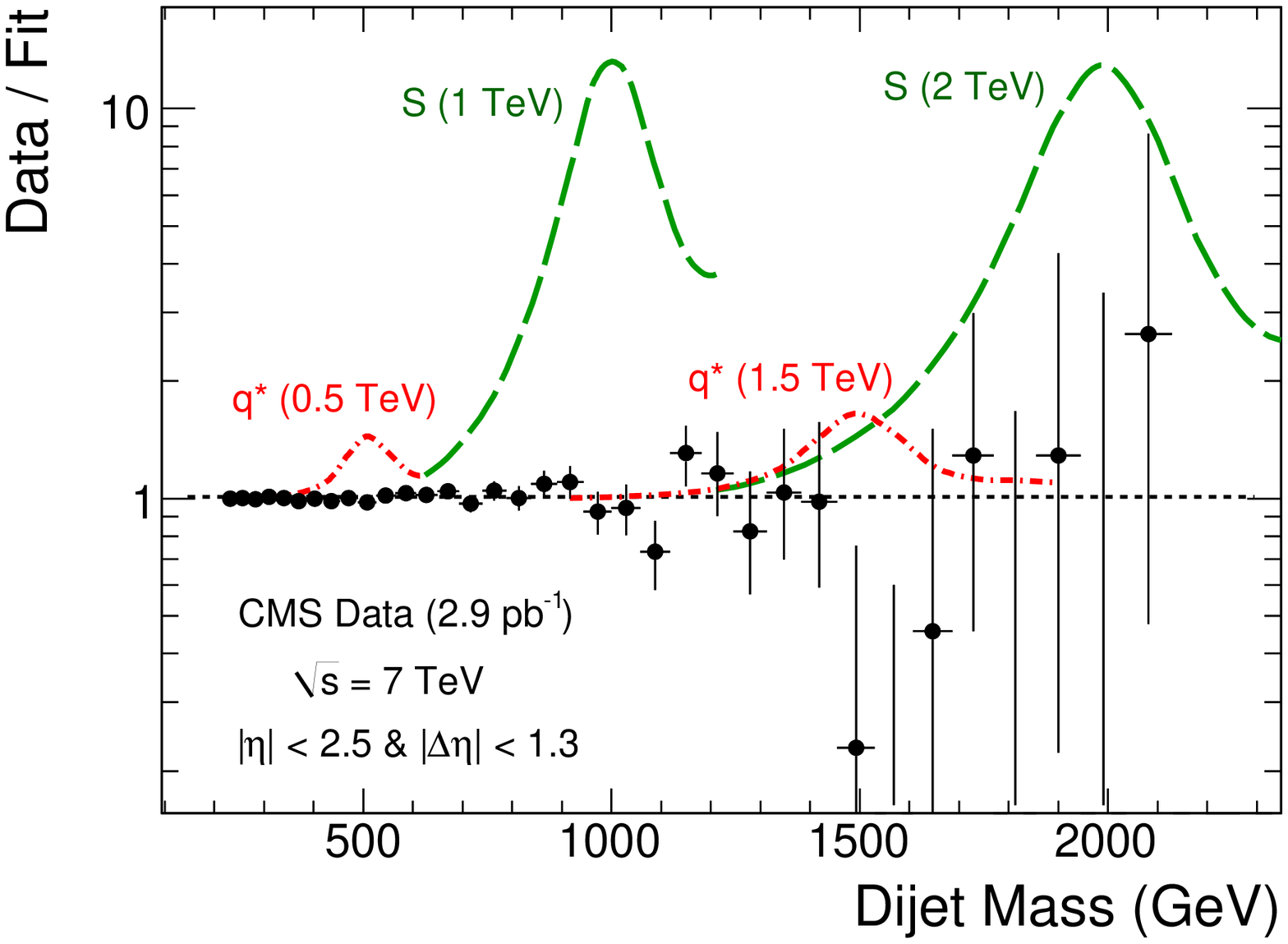,width=2.5in}
\psfig{file=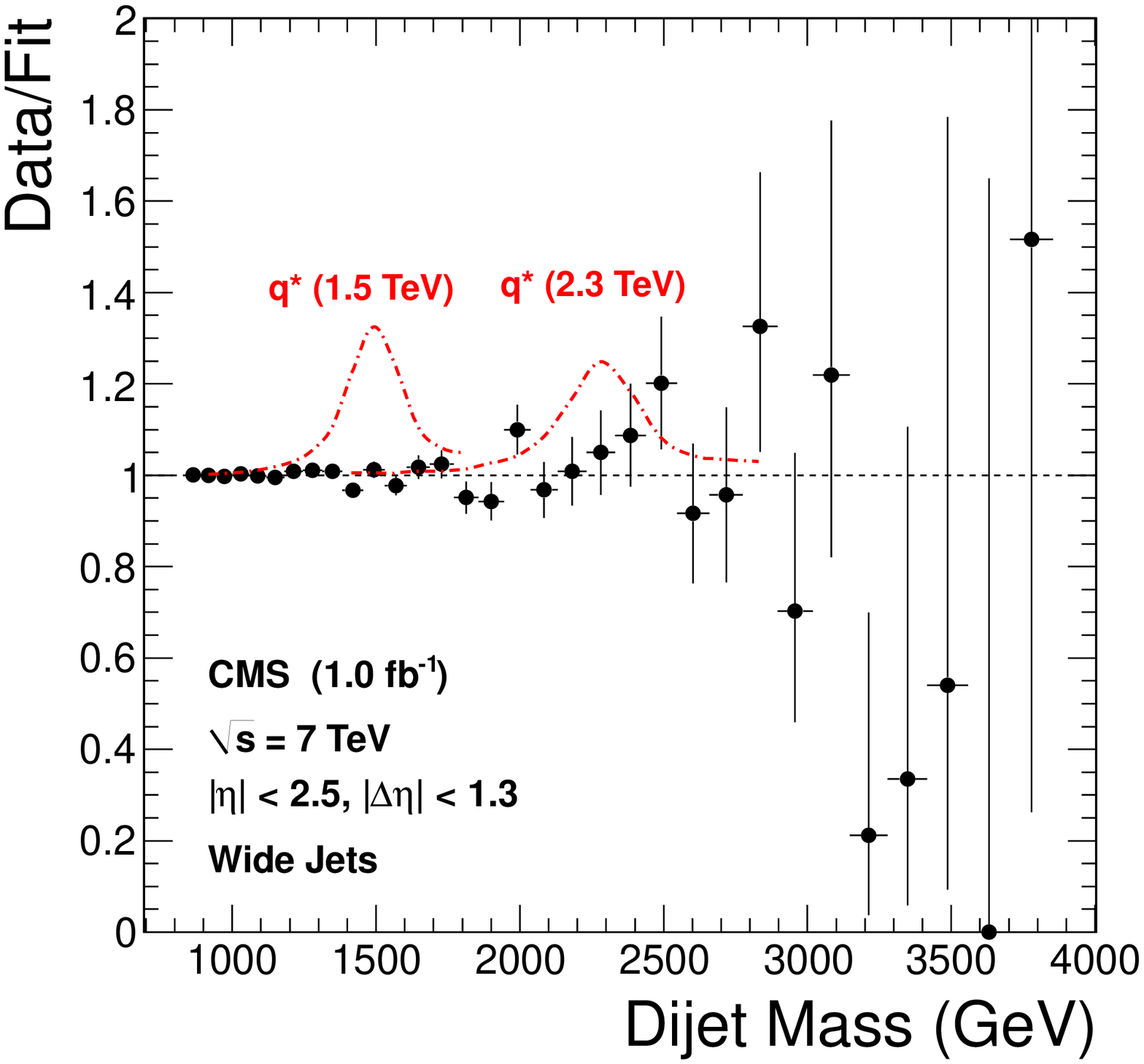,width=2.5in}
}
\vspace*{8pt}
\caption{Data from CMS in 2010 and 2011.
(top left) Dijet mass spectrum using 2.9 pb$^{-1}$ compared to a background fit,  
and simulations of QCD and signals from excited quarks and string resonances. 
The Band shows the jet energy scale uncertainty. (bottom left) Ratio of the same data
to the fit, and compared to the same signals. Left figures from Ref.~\protect\refcite{CMS2010} 
Copyright 2010 by the American Physical Society. (top right) Same as top left using
1.0 fb$^{-1}$ and including the bin-by-bin significance of the difference between
the data and the fit. (bottom right) Ratio of the data and the fit compared to excited
quark signals. Right figures from Ref.~\protect\refcite{CMS2011} 
Copyright 2011, with permission from Elsevier.
\label{CMSdata}}
\end{figure}

 The CMS collaboration has published searches for 
dijet resonances in 2010~\cite{CMS2010} and in 2011~\cite{CMS2011}.  CMS required $|\eta|<2.5$ 
and $|\cos\theta^*|<0.57$, the same as the first two ATLAS searches,
and measured the dijet mass spectrum in the interval $0.22<m<2.1$ TeV in 2010, and in $0.84<m<3.7$ TeV in 2011.
In Fig.~\ref{CMSdata} CMS compared data to both a QCD simulation and to a background fit.
The uncertainty in the jet energy scale decreased from 10\% in the 2010 publication to 2.2\% in
2011, decreasing the uncertainty in the comparison with QCD as shown in Fig.~\ref{CMSdata}. As
previously discussed, in 2011 CMS introduced wide jets to improve the resolution for resonances 
decaying to gluons. Following the CDF and ATLAS searches, CMS used Eq.~\ref{CDFparam3} to fit 
the background in the search. The background parameterization gave a good fit: $\chi^2/DF=32/31$ in 2010 and 
$\chi^2/DF=27.5/28$ in 2011. In 2011 CMS adopted the ATLAS style of displaying the bin-by-bin
significance of the difference between data and fit, just below the data points, from which one 
can visually see that there were no significant fluctuations. Unlike CDF and ATLAS, CMS did not 
quote estimates of the significance of fluctuations in either paper.
Following the CDF searches, CMS compared their data with the size and shape of various dijet resonance
signals in a plot of the data divided by the fit, allowing a direct visual estimate of the 
signal mass values excluded.

\subsection{Limits and Models}
\label{secLimit}

Limits on dijet resonances are often performed in three steps.
The first step is setting limits on the cross section, which can be 
purely experimental numbers determined using a statistical procedure, 
and can be fairly independent of any model. The second step is to evaluate 
the theoretical cross section for the model as a function of resonance mass, which is a completely model 
dependent process. The third and final step is to compare the experimental upper
limit on the cross section, with the theoretical cross section of the model,
and find what resonance masses, if any, are excluded. In Table~\ref{tabLimits} we
summarize the mass intervals excluded for the majority of models considered 
in dijet resonance searches.

\begin{table}[hbt]
\tbl{Excluded mass intervals for resonance models from searches in the dijet mass spectrum.  The model parameters
used for these exclusions are presented in Section~\ref{benchmarks}. See the text about each exclusion for 
any exceptions in the model parameters, discussion of the calculations, and for the additional exclusions 
of color octet technirhos by CDF and color octet scalars by ATLAS. Continuation rows beginning with a '' mark 
are present when the search excluded multiple mass intervals for a given model.}
{\begin{tabular}{@{}llccccccc@{}} \toprule
Expt. & Year &  Axigluon      & Excited     & $W^{\prime}$ & $Z^{\prime}$ & $E_6$        & String    \\
      &      &  or Coloron    & Quark       &              &              &   Diquark    &            \\ 
      &      &  (TeV)         & (TeV)       &  (TeV)       &  (TeV)       &   (TeV)      &   (TeV)    \\ \colrule
UA1 & 1986 &     0.13-0.28    & --          &  --          &  --          &   --         &   --        \\
UA1 & 1988 &     0.15-0.31    & --          &  --          &  --          &   --         &   --       \\
CDF & 1990 &     0.12-0.21    & --          &  --          &  --          &   --         &   --           \\
UA2 & 1990 &   --             & --          & 0.10-0.16    &  --          &   --         &   --          \\
CDF & 1993 &     0.22-0.64    & --          & --           &  --          &   --         &   --          \\
UA2 & 1993 &   --             & 0.14-0.29   & 0.13-0.26    &  0.13-0.25   &   --         &   --           \\
CDF & 1995 &   0.20-0.87      & 0.20-0.56   &  --          &  --          &   --         &   --          \\
CDF & 1997 &   0.20-0.98      & 0.20-0.52   & 0.30-0.42    &  --          & 0.29-0.42    &   --           \\
\ \  ''   & \ \  ''    &      & 0.58-0.76   &              &  --          &   --         &   --           \\
D0 & 2004  &   --             & 0.20-0.78   & 0.30-0.80    &  0.40-0.64   &   --         &   --          \\
CDF & 2009 &   0.26-1.25      & 0.26-0.87   & 0.28-0.84    &  0.32-0.74   & 0.29-0.63    &  0.26-1.4     \\
ATLAS & 2010&  --             & 0.30-1.26   &  --          &  --          &   --         &   --           \\
CMS & 2010 &   0.50-1.17      & 0.50-1.58   &  --          &  --          & 0.50-0.58    &  0.50-2.50          \\
\ \ ''   &   \ \ ''   &   1.47-1.52    &    &              &              & 0.97-1.08    &      \\        
\ \ ''   &   \ \ ''   &                &    &              &              & 1.45-1.60    &     \\ 
ATLAS & 2011w&  0.60-2.10     & 0.60-2.15   &  --          &  --          &   --         &   --          \\
CMS & 2011 &   1.00-2.47      & 1.00-2.49   & 1.00-1.51    &  --          & 1.00-3.52    & 1.00-4.00        \\
ATLAS & 2011s& 0.80-3.32      & 0.80-2.99   &  --          &  --          &   --         &   --          \\ 
 \botrule
\end{tabular} \label{tabLimits}}
\end{table}

\subsubsection{Experimental Systematic Uncertainties}

There are common systematics faced by all the experiments when setting upper limits on the
cross section of dijet resonances. Here we give an overview of these systematic uncertainties
and their relative importance for the majority of the experiments, before we discuss
the limits from each experiment below.

The largest source of experimental systematic uncertainty on the cross-section limits 
is usually the jet energy scale.  This is because the QCD background falls steeply 
with increasing dijet mass, and the amount of background underneath a potential dijet
resonance, at a fixed dijet resonance mass, depends critically on the jet energy scale.
For a typical spectrum that falls with a large power of the mass, $dn/dm \sim 1/m^N$,
where $N \sim 10$ at high masses, a typical 5\% uncertainty in the jet energy scale 
leads to an order of 50\% uncertainty in the amount of QCD background underneath 
the dijet resonance, which increases the cross section limit by roughly 20\% in 
smooth data samples. When there are upward fluctuations in the data sample the uncertainty
of the limit at a nearby resonance mass can be much larger, allowing for the possibility
that the fluctuation is attributed to both a resonance and an uncertainty on the jet energy
scale. Nearly all experiments correct the jet energy to remove non-linearities and non-uniformities
in detector response. There are potentially more serious uncertainties if these jet energy 
corrections are not continuous, or contain shape deformities, as they create the possibility 
of manufacturing a bump, or hiding one. The experiments work hard to insure that this does
not happen, but they generally do not include any residual uncertainty in a quantitative fashion.

The next largest source of experimental systematic uncertainty is usually the estimation
of the QCD background. As discussed in Section~\ref{secBackgound}, assigning systematic uncertainties 
is challenging when a QCD calculation is used to model the background. When a simple 
background parameterization is fit to the data instead, two common methods of evaluating
the systematic on the background are: trying alternate parameterizations, or assigning the
statistical uncertainty on the fit parameters as a background shape systematic. Either 
method will frequently give large systematics on the parametrized background in the region 
of high dijet mass, where the small amounts of data are not sufficient to constrain the background fit.

Most experiments report that the uncertainty on the jet energy resolution does not have 
a significant effect on the limits, and in particular is often significantly smaller than
the effect of the jet energy scale uncertainty. While this may seem surprising at first, it can be understood in the same way that the effect of the jet
energy scale was quantified. On a steeply falling QCD dijet mass spectrum, a
percentage shift in the mass will have a much larger effect on the amount 
of background underneath the resonance, compared to roughly the same percentage widening of the Gaussian 
resolution. The counter example would be a constant background, 
flat as a function of dijet mass, where a percentage shift in the jet energy scale would have no 
effect on the amount of background underneath the peak, while the same percentage widening of the jet
energy resolution would be more important. As a conclusion, the resolution uncertainties are negligible 
because the QCD background is falling steeply.

The astute reader should note that the arguments above are relevant for the limits
on dijet resonances, which is what most of the searches have reported, and the 
effect of systematic uncertainties would be different if a signal were observed.

\subsubsection{Limits from the CERN S$\bar{p}p$S Collider Experiments}

The first published exclusion~\cite{UA1ptLimit} which used 
UA1 data~\cite{UA1ptData} did not use a statistical
method, or discuss systematic uncertainties. Instead a purely visual exclusion was 
performed. The left plot of 
Fig.~\ref{UA1data} was presented overlaying the observed jet $p_T$ distribution with 
that from QCD plus the axigluon model. The paper states {\it ``We see . . . that 
axigluons in the range $M_A=125-275$ GeV are ruled out''} and {\it ``... are in contradiction 
with the data.''} 

\begin{figure}[htbp]
\centerline{
\psfig{file=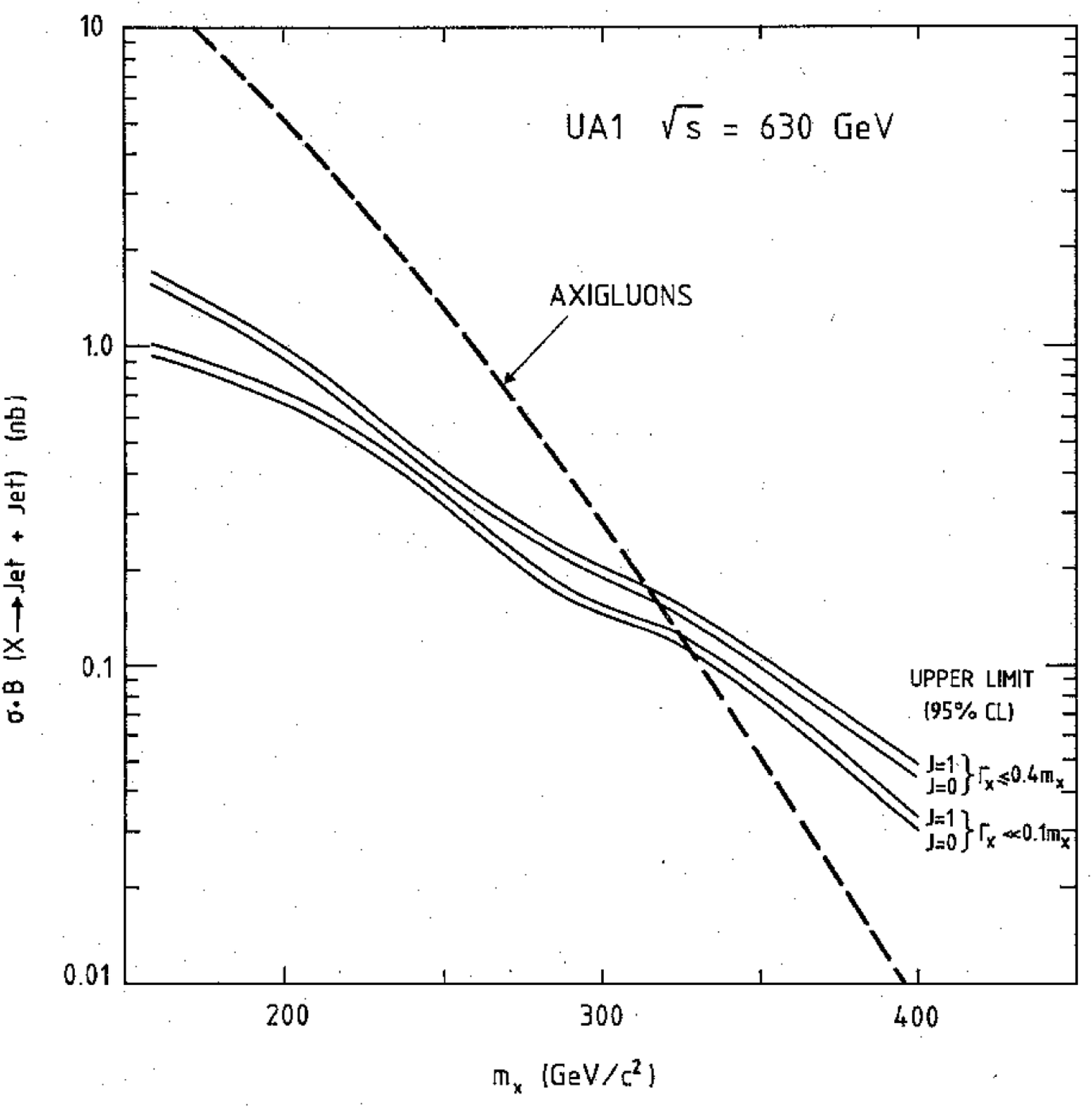,width=2.8in}
\psfig{file=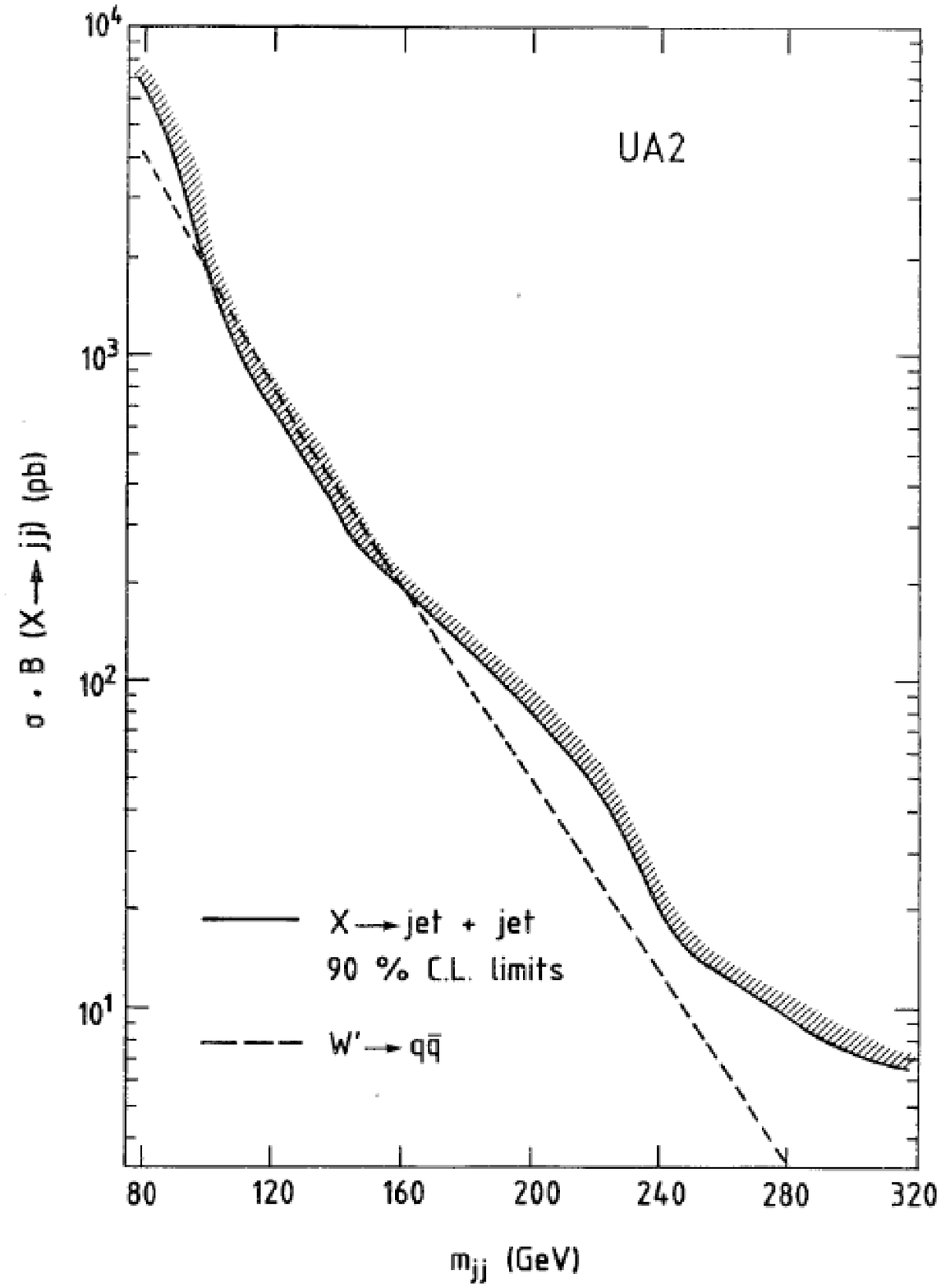,width=2.2in}
}
\vspace*{8pt}
\caption{Limits from UA1 in 1988 and UA2 in 1990: 
(left) Upper limits at 95\% CL on dijet resonance cross section times branching ratio from UA1 using 490 nb$^{-1}$, 
for both vector and scalar particles of two different widths, compared to an axigluon calculation. 
From Ref.~\protect\refcite{UA1mass}, Copyright 1988, with permission from Elsevier. 
(right)  Upper limits at 90\% CL on dijet resonance cross section times branching ratio from UA2 using
4.7 pb$^{-1}$, compared to a $W^{\prime}$ calculation. From Ref.~\protect\refcite{UA21990} with kind permission 
from Springer Science+Business Media.
 \label{sppsLimits}
 }
\end{figure}

The search from UA1 in 1988~\cite{UA1mass} was the first to exclude
cross sections for resonances at 95\% CL, and used the method of maximum likelihood in a fit 
of the QCD background prediction and the resonance shapes to the data. The UA1 jet energy scale 
uncertainty was reported as 9\%, however, UA1 used the agreement between
LO QCD and the shape of the UA1 data at low dijet mass to constrain the jet energy 
scale uncertainty for this search to less than 6\% at 95\% CL, and then included a 
possible 6\% increase in the jet energy scale into the limit. UA1 also noted that 
uncertainties in the background estimate from higher order QCD processes were not
included and could weaken the reported limits.
The upper limits shown in Fig.~\ref{sppsLimits} were found for narrow resonances
($\Gamma<0.1M$) and wide resonances ($\Gamma<0.4M$), with the limits corrected for the acceptance
of the decay's angular distribution, for the two cases of scalar and vector resonances. UA1 
then compared the wide vector resonance cross-section upper limit to the model cross section from
the previous search~\cite{UA1ptLimit} in Fig.~\ref{sppsLimits} to exclude axigluons in
the mass range $150<m_A<310$ GeV.

In 1990 UA2~\cite{UA21990} reported upper limits on the cross section shown in Fig.~\ref{sppsLimits}. 
Here limits at 90\% CL were obtained from a fit to the background parameterization and a signal, 
after subtracting the fitted signal for the W and Z resonances. A systematic
uncertainty on the cross-section upper limit, as much as 21.5\%, came from the dijet mass 
resolution uncertainty, which was the dominant systematic. UA2 reduced the jet energy
scale uncertainty to give only an 11\% uncertainty on the cross-section upper limit, 
by calibrating the jets with the observed W and Z peak.  
UA2 compared the upper limits on the cross section to an $O(\alpha_s^2)$ 
calculation of the $W^{\prime}$ resonance with standard model couplings to exclude 
the resonance decays in the mass interval $101<M<158$ GeV.

\begin{figure}[htbp]
\centerline{
\psfig{file=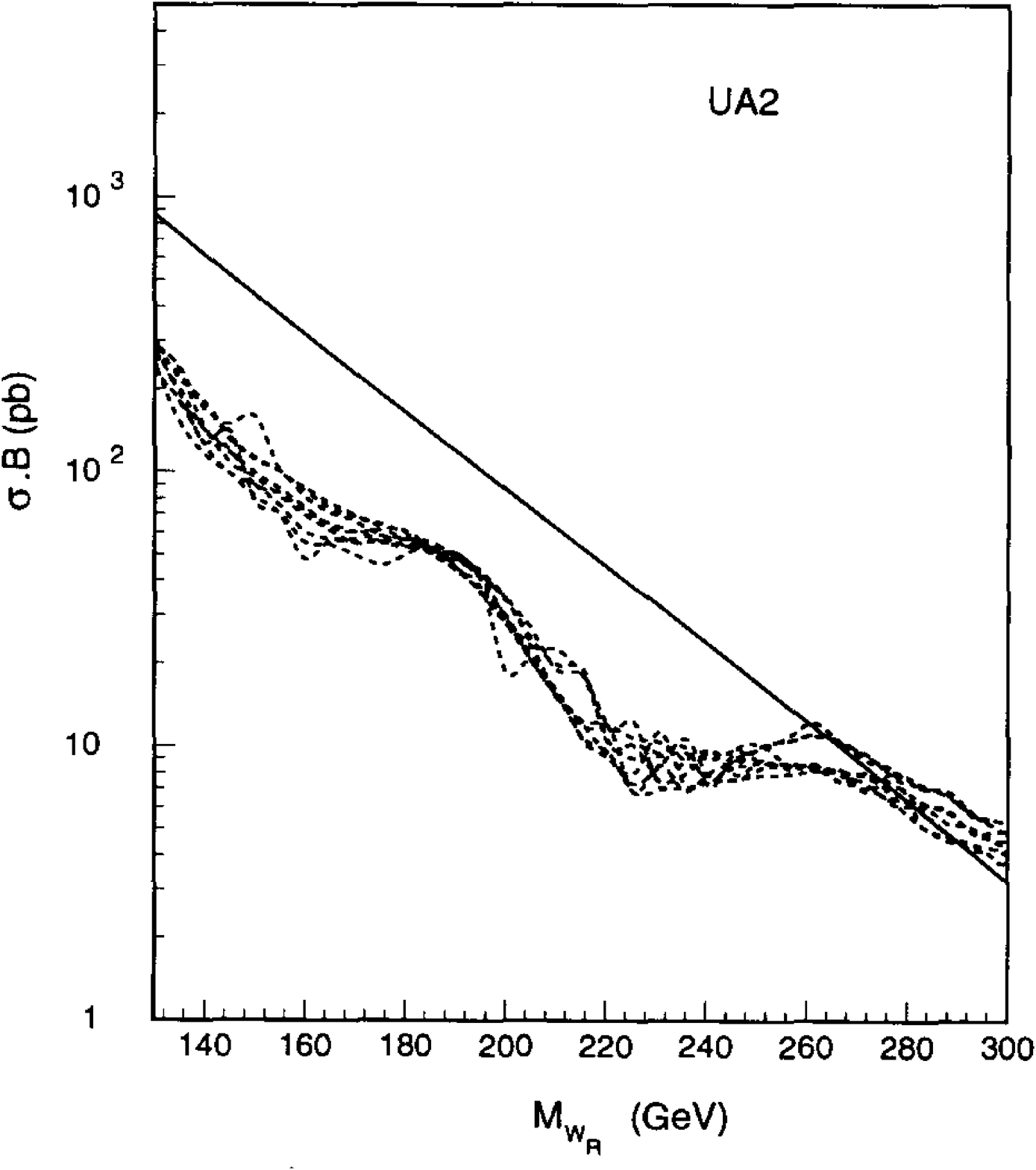,width=2.55in}
\psfig{file=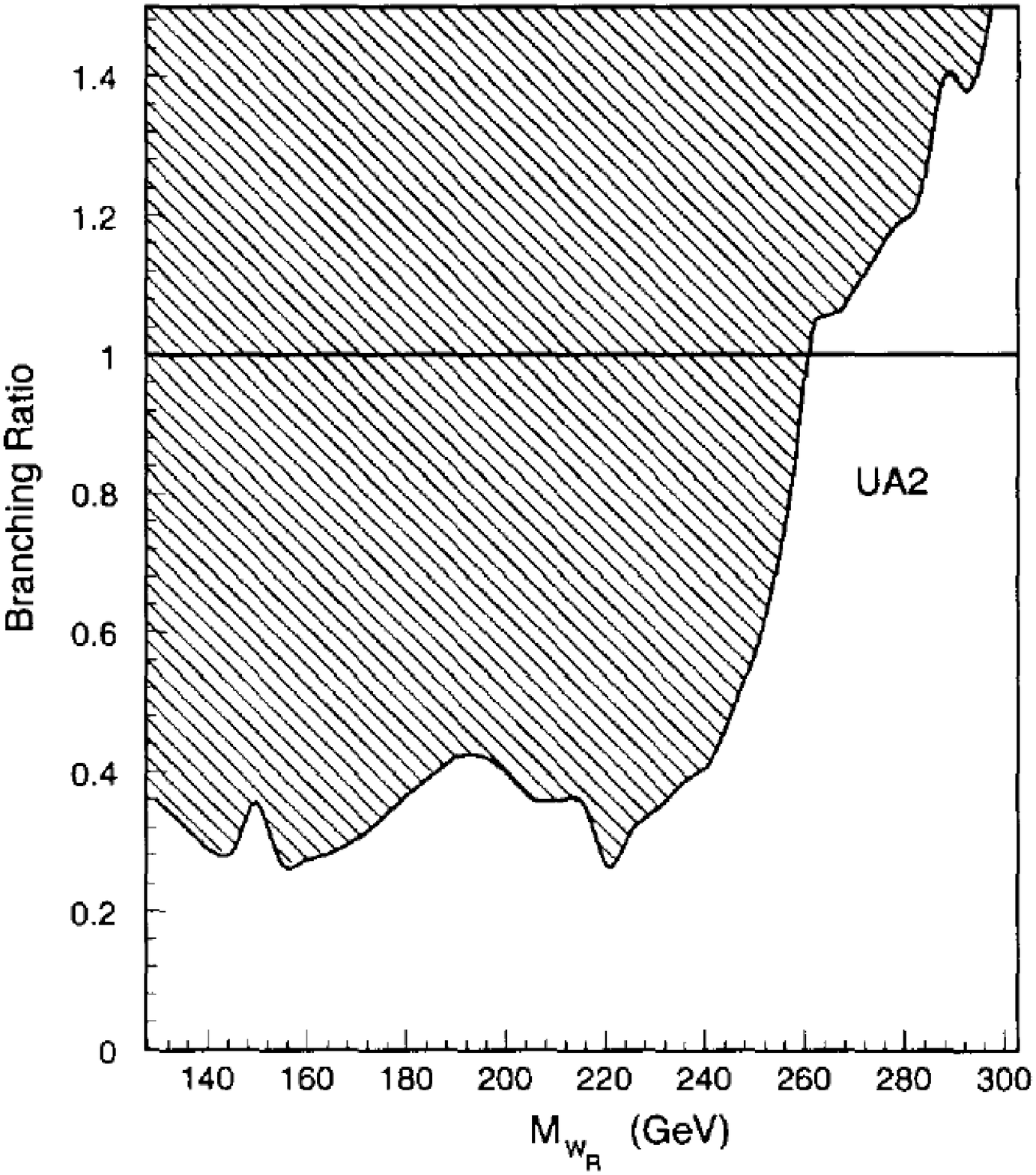,width=2.45in}
}
\vspace*{8pt}
\caption{Limits from UA2 in 1993.
(left) Upper limits at 90\% CL on dijet resonance cross section times branching ratio using 
11 pb$^{-1}$, with the nine curves showing systematic variations 
in the limit, compared to a $W^{\prime}$ cross section calculation,
and (right) the excluded mass region for a $W^{\prime}$ in the mass
vs. branching ratio plane, from Ref.~\protect\refcite{UA21993} 
Copyright 1993, with permission from Elsevier.
\label{UA21993}}
\end{figure}

The search from UA2 in 1993~\cite{UA21993} fit a dijet resonance signal and the background parameterization 
in Eq.~\ref{UA2param} to the data, obtaining the signal cross section and its error. From the best-fit signal 
cross section and its error UA2 calculated the 90\% CL upper limit including statistical uncertainties only. The systematics were
included in a second step and consisted of the uncertainty on the dijet resonances decay width, and the uncertainty 
on the stability of the dijet mass measurement between the two data samples used from the 1989 and 1990 running periods.
For each quantity, $x$, with uncertainty $\Delta x$, they performed the cross-section fit with the three different values
$x$, $x+\Delta x$ and $x - \Delta x$. Therefore, for the two systematics considered, there were nine independent fits and 
nine different values of the cross-section upper limit. For a $W^{\prime}$ signal, Fig.~\ref{UA21993} shows 
the cross-section upper limit from the nine fits. UA2 selected the largest value among the nine as the upper limit
on the cross section including systematics, and used that to exclude at 90\% CL the $W^{\prime}$ in the mass
region $130 < M < 261$ GeV. Also in Fig.~\ref{UA21993} UA2 presented their excluded region in the branching ratio vs. mass 
plane of the  $W^{\prime}$, where a branching ratio of 1 represents decays to $ud$ and $cs$ quarks,
 as expected from a right handed $W^{\prime}$ model, $W_R$.  This was the only exclusion of a $W_R$ model in dijet
 resonance searches, as all other searches were for a more standard model like left-handed $W^{\prime}$.
 Using similar techniques, UA2 also excluded at 90\% CL $Z^{\prime}$ bosons decaying to dijets 
in the mass region $130 < M <252$ GeV, and mass degenerate excited quarks ($q^*$) in the mass region $140 < M < 288$ GeV. 
Fig.~\ref{CDFModelSpecific} shows the UA2 $q^*$ limits in the coupling vs. mass plane.
UA2 also presented mass limits for a few other variations of the $W^{\prime}$, $Z^{\prime}$, and $q^*$ models~\cite{UA21993}. 
The cross section of the $W^{\prime}$ and $Z^{\prime}$ models came from an $O(\alpha_s^2)$ NNLO calculation, and the 
cross section for the $q^*$ model came from a LO Born level calculation.

\subsubsection{Limits from the Fermilab Tevatron Collider Experiments}

The search from CDF in 1990~\cite{CDF1990} did not publish upper limits on the cross section.
A $\chi^2$ was calculated for a fit of an axigluon plus LO QCD background to the data, and a jet energy
scale uncertainty ranging from 5\% to 9\% was included by allowing both edges of each dijet mass bin to
vary, and CDF used the smallest resulting $\chi^2$ to exclude axigluons with $N=5$ decay channels 
in the mass range $120<m_A<210$ GeV at 95\% CL.
In 1993 CDF~\cite{CDF1993} published a table of upper limits on the cross section, the first example of
publishing generic upper limits, but 
did not compare them with any models in a figure. The 1993 paper directly excluded axigluons with 
$N=10$ decay channels in the mass range $220<m_A<640$ GeV at 95\% CL, with an unspecified statistical 
test. The paper also set limits on axigluons with $N=20$ decay channels, and presented limits for two choices of PDFs.
Both the 1990 and 1993 papers from CDF used a coherent sum of the axigluon signal and the LO QCD background,
normalized to the data at low dijet mass.

\begin{figure}[htbp]
\centerline{
\psfig{file=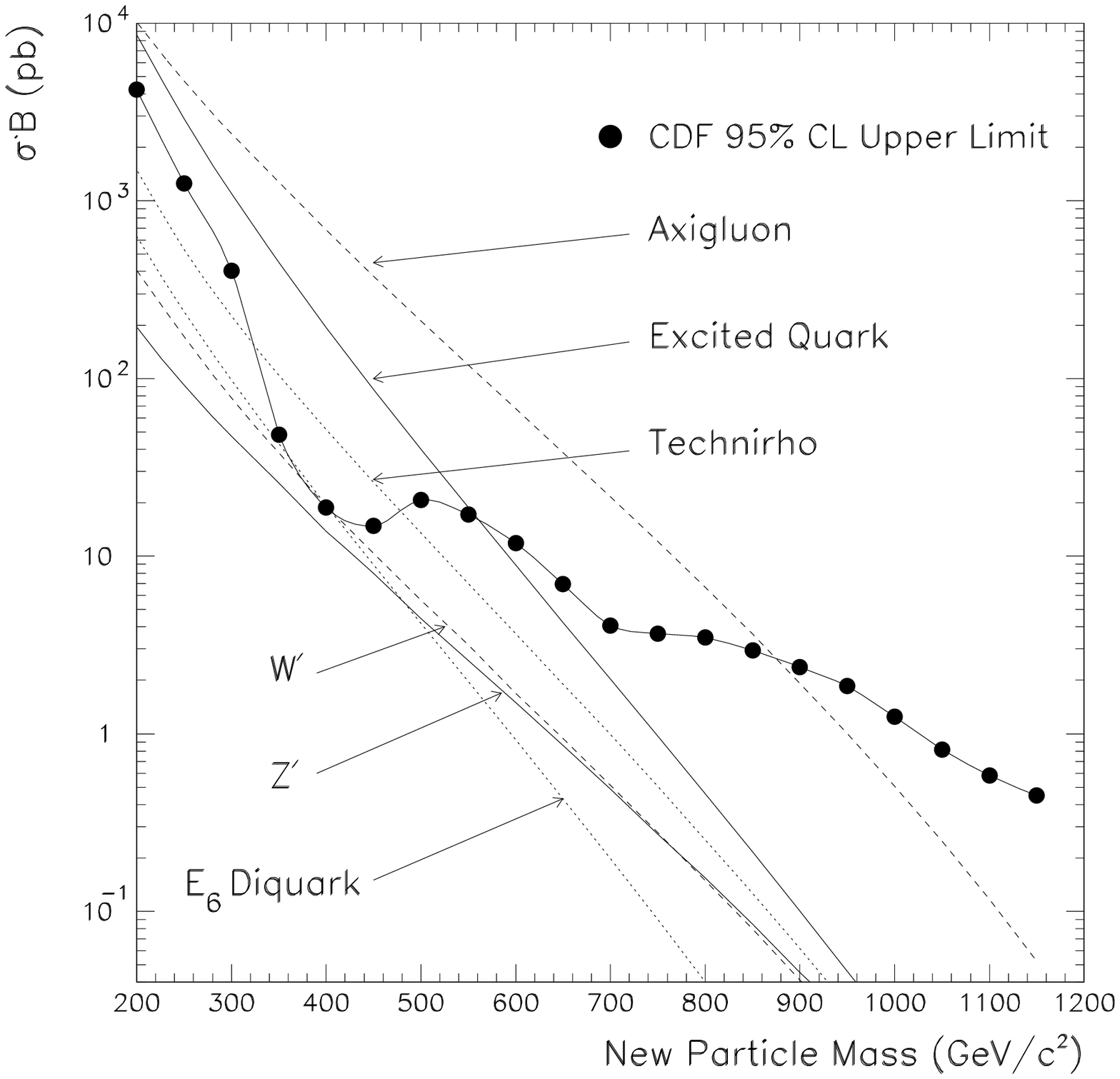,width=2.5in}
\psfig{file=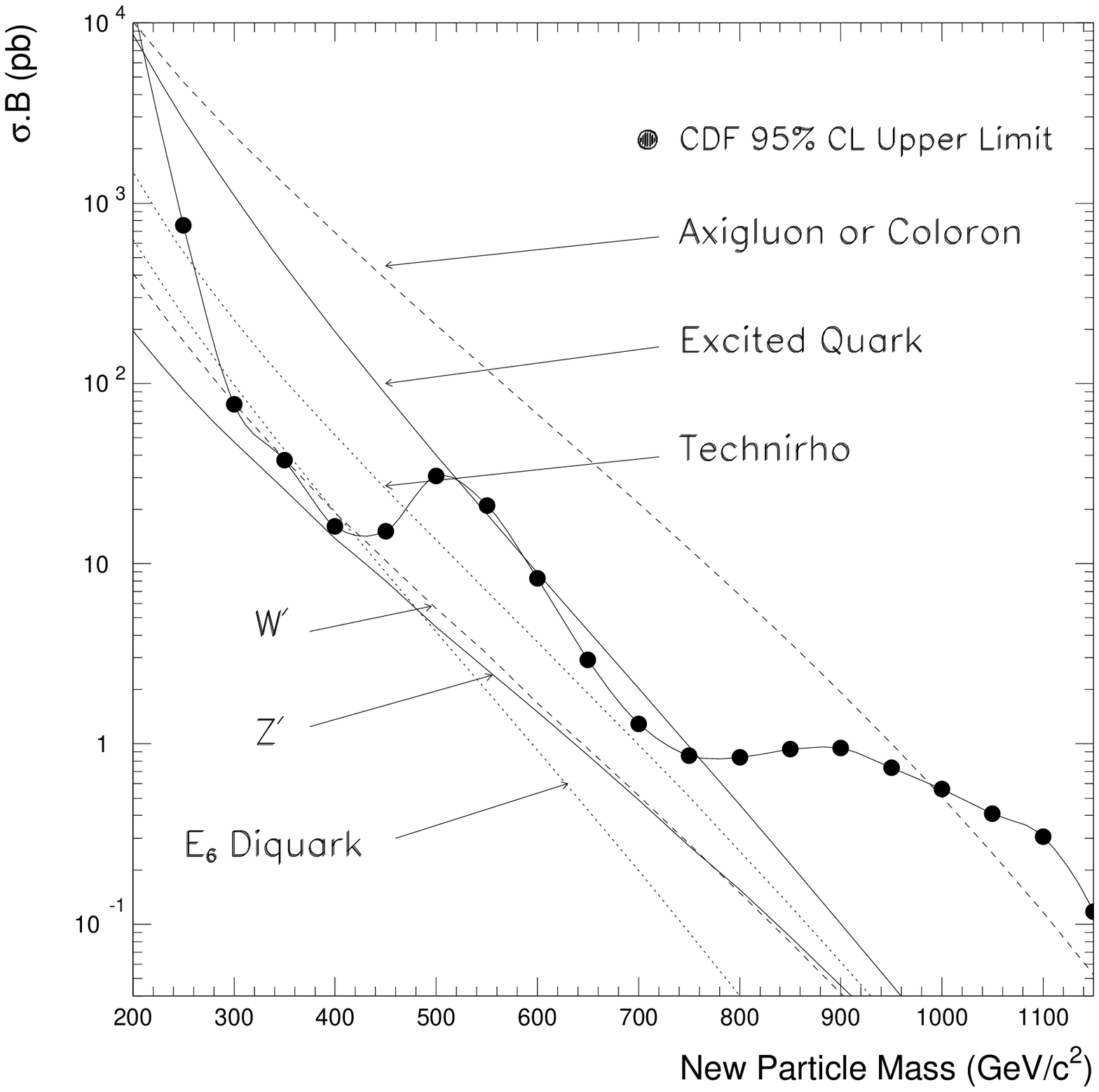,width=2.5in}
}
\vspace*{8pt}
\caption{Limits from CDF in 1995 and 1997. 
(left) Upper limits at 95\% CL on dijet resonance cross section times branching ratio times acceptance 
using 19 pb${-1}$, compared to calculations of the same observable
in six models, from Ref.~\protect\refcite{CDF1995}, Copyright 1995, and  (right) the same using 
106 $pb^{-1}$, from Ref.~\protect\refcite{CDF1997}, Copyright 1997 by the
American Physical Society.
\label{CDFMid90sLimits}}
\end{figure}

\begin{figure}[htbp]
\centerline{
\psfig{file=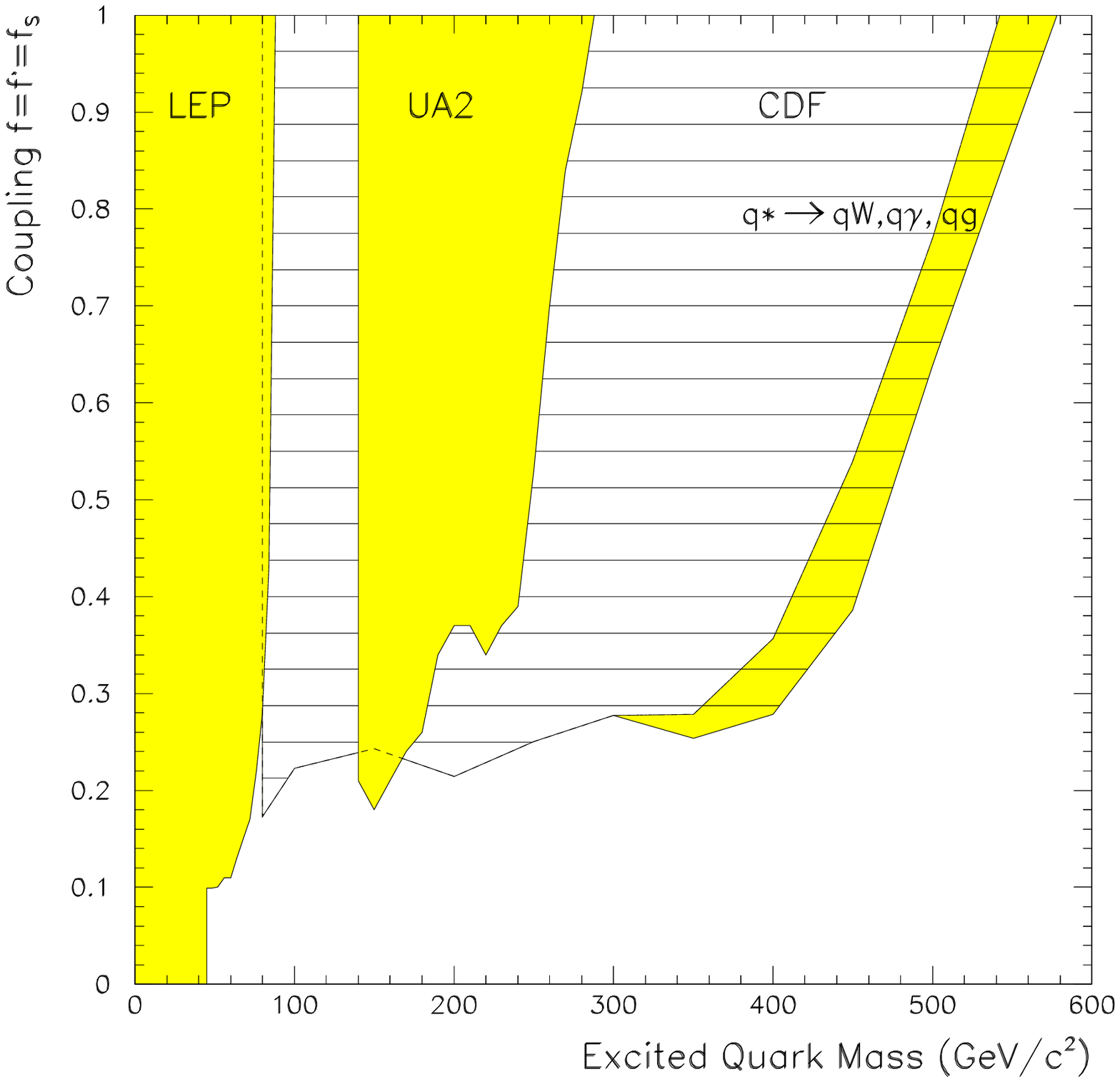,width=2.5in}
\psfig{file=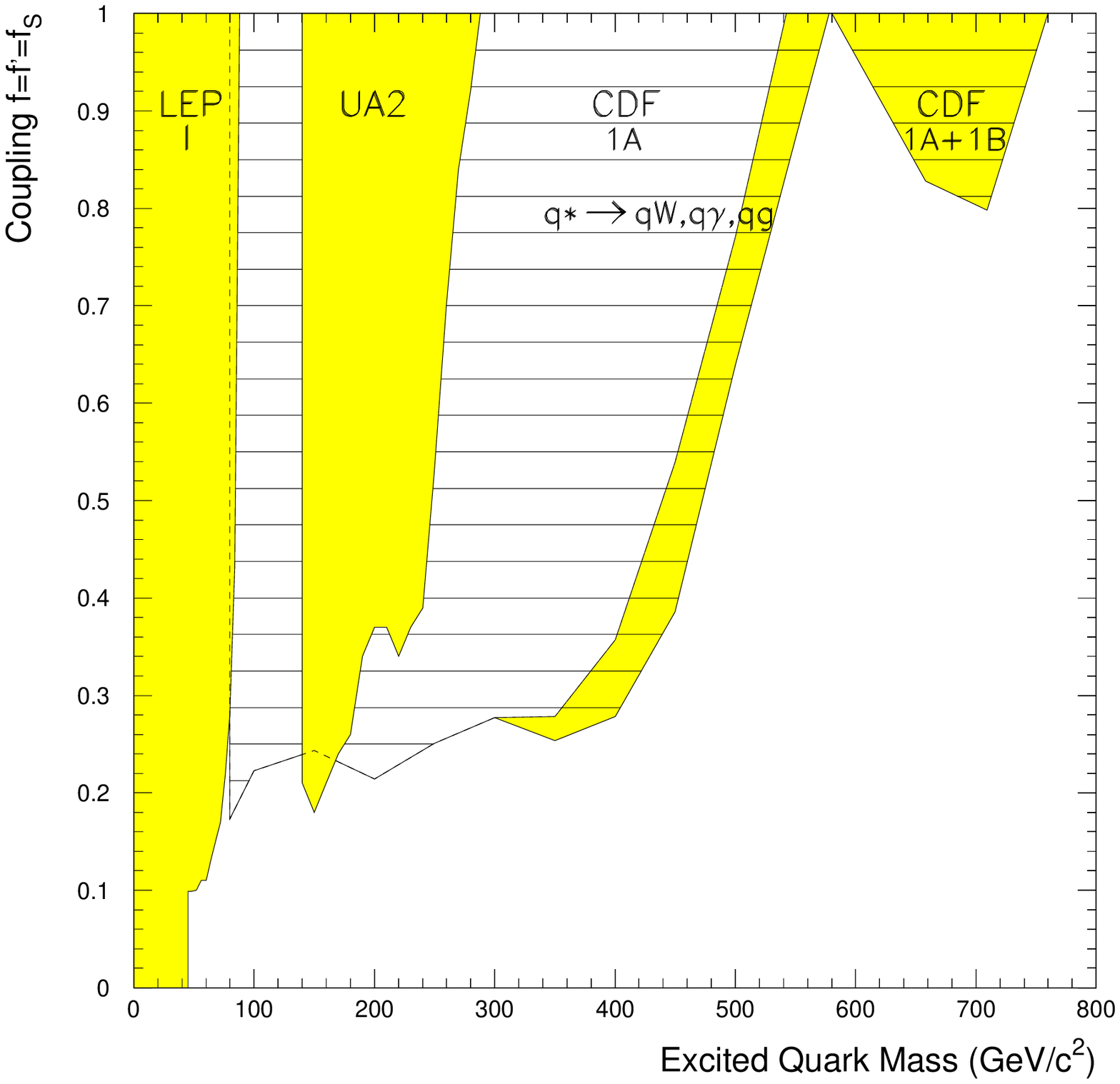,width=2.5in}
}
\vspace*{8pt}
\caption{Excited quark excluded regions from CDF in 1995 and 1997 including UA2 from 1993.
(left) Region of the excited quark coupling vs. mass plan excluded by CDF using 19 pb${-1}$ in the $qg$
channel, combined with exclusions in the $qW$ and $q\gamma$ channel, and compared with the
dijet exclusions from UA2 in 1993 and also from LEP, from Ref.~\protect\refcite{CDF1995}, 
Copyright 1995, and (right) same using 106 pb${-1}$, from Ref.~\protect\refcite{CDF1997}, Copyright 1997 
by the American Physical Society.
\label{CDFModelSpecific}}
\end{figure}

In 1995~\cite{CDF1995} and 1997~\cite{CDF1997} CDF published cross-section upper limits on 
dijet resonances. The statistical technique was described in the prior CDF search for 
excited quarks in the $\gamma$ + jet and $W$+jet channel~\cite{CDFqstarPaper}. Using a
binned likelihood for each value of resonance mass, the best fit to the data of the parametrized 
background plus a floating signal were found. For this background  
the binned likelihood, $L$, was written as a function of the signal normalization $\alpha$:
\\
\begin{equation}
L = \prod_{i} \frac{\mu_{i}^{n_{i}}e^{-\mu_{i}}}{n_{i}!} \ \ \ \ \ \mbox{where} \ \ \ \ \ \mu_{i} = \alpha n_{i}(S) + n_{i}(B).
\label{Likelihood}
\end{equation}
\\
Here $n_i$ is the number of events observed in the data in dijet mass bin $i$, $n_{i}(S)$ is the number of events predicted for the signal using the resonance
shape in the same bin, and $n_{i}(B)$ is the number of events predicted for the background in the same bin.
The background parameterization used in 1995 was Eq.~\ref{CDFparam1} and in 1997 it was Eq.~\ref{CDFparam2}.
The 95\% quantile of the likelihood distribution in Eq.~\ref{Likelihood} was found, defining an upper limit
on the signal cross section with statistical uncertainties only. The dominant sources of systematic uncertainty 
considered by CDF were the 5\% jet energy scale uncertainty, the QCD radiation effects on the the mass resonance 
line shape, and the background parameterization, while other sources were also included. For each $1\sigma$- 
shift in the source of systematic uncertainty, the background plus signal fit was repeated, giving a new limit.
The change in the limit defined a $1\sigma$ uncertainty in the signal cross section for each source of systematic, 
and a total systematic was derived by adding these in quadrature, as shown in the insets in Fig.~\ref{CDFmiddleFit}. The likelihood 
distribution of Eq.~\ref{Likelihood} was then convolved with the total Gaussian systematic uncertainty in 
the cross section, and the final limit including systematics was defined as the 95\% quantile of this smeared
likelihood. The cross-section upper limits on narrow dijet resonances are shown in Fig.~\ref{CDFMid90sLimits},
where the effect of the small upward fluctuation in the data near a dijet mass of 550 GeV in both publications 
produced a noticeable bump in the upper limit. The upper limits were published in a table, and can be compared to 
the cross section for any model of dijet resonances decaying to two partons with $|\eta|<2$ and $|\cos\theta^*|<2/3$.
CDF explicitly did the comparison with a few model cross sections, shown in Fig.~\ref{CDFMid90sLimits}, and the 
reader can note that the strong processes, like $q^*$, have much larger cross sections than the weak processes like
$W^{\prime}$. From this figure CDF excluded in 1995 and 1997 the mass intervals 
listed in table~\ref{tabLimits} for axigluons, excited quarks, $W^{\prime}$, $Z^{\prime}$ and $E_6$ diquarks. 
CDF also excluded color octet technirhos in the mass interval $320<M<480$ GeV in 1995 and 
$260<M<470$ GeV in 1997. CDF used its own LO calculations of the cross section for all 
models~\cite{resonanceCmsNote}, and 
included $O(\alpha_s)$ $K$-factors for the $W^{\prime}$ and $Z^{\prime}$ models.
The calculations for axigluons, $W^{\prime}$, $Z^{\prime}$, and $E_6$ diquarks employed the narrow 
width approximation discussed in section~\ref{secCalc}.
In 1997 CDF noted that the cross section for the coloron model was always greater than or 
equal to that for the axigluon model, and so all the 95\% CL upper limits for axigluons also applied
to colorons. For the excited quark model, CDF plotted the exclusion in the coupling vs. mass 
plane in Fig.~\ref{CDFModelSpecific}, comparing with the dijet resonance search from UA2 in 1993. 
Figure~\ref{CDFModelSpecific} also shows prior CDF exclusions of $q^*$ decays from the $\gamma q$ and $Wq$ channel,
which were combined with the limit
from the dijet channel in 1995 to extend the $q^*$ mass limit from 560 GeV with dijets alone to 570 GeV including
$\gamma q$, $Wq$ and $gq$ decays.

\begin{figure}[htbp]
\centerline{
\psfig{file=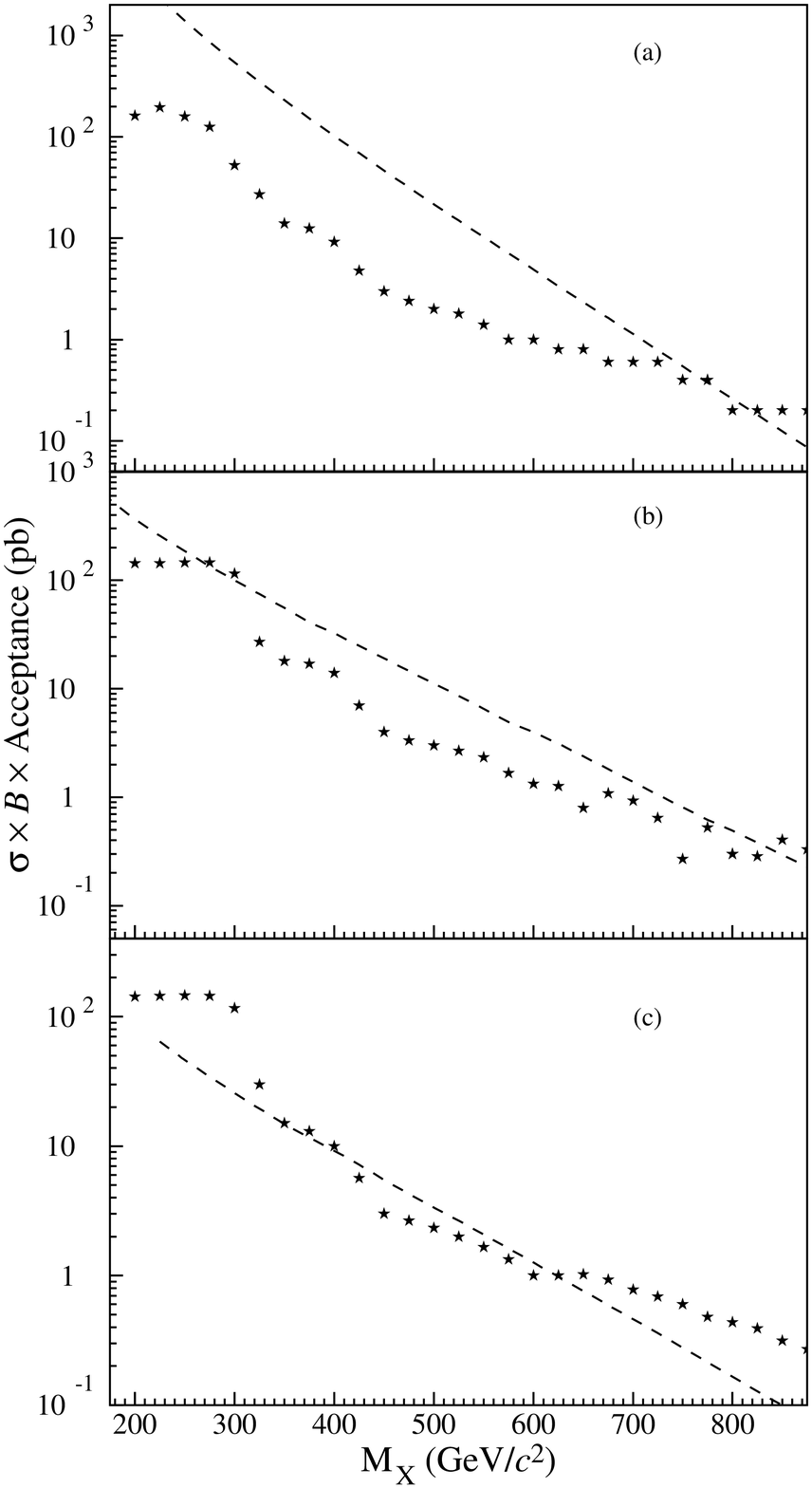,width=2.27in}
\psfig{file=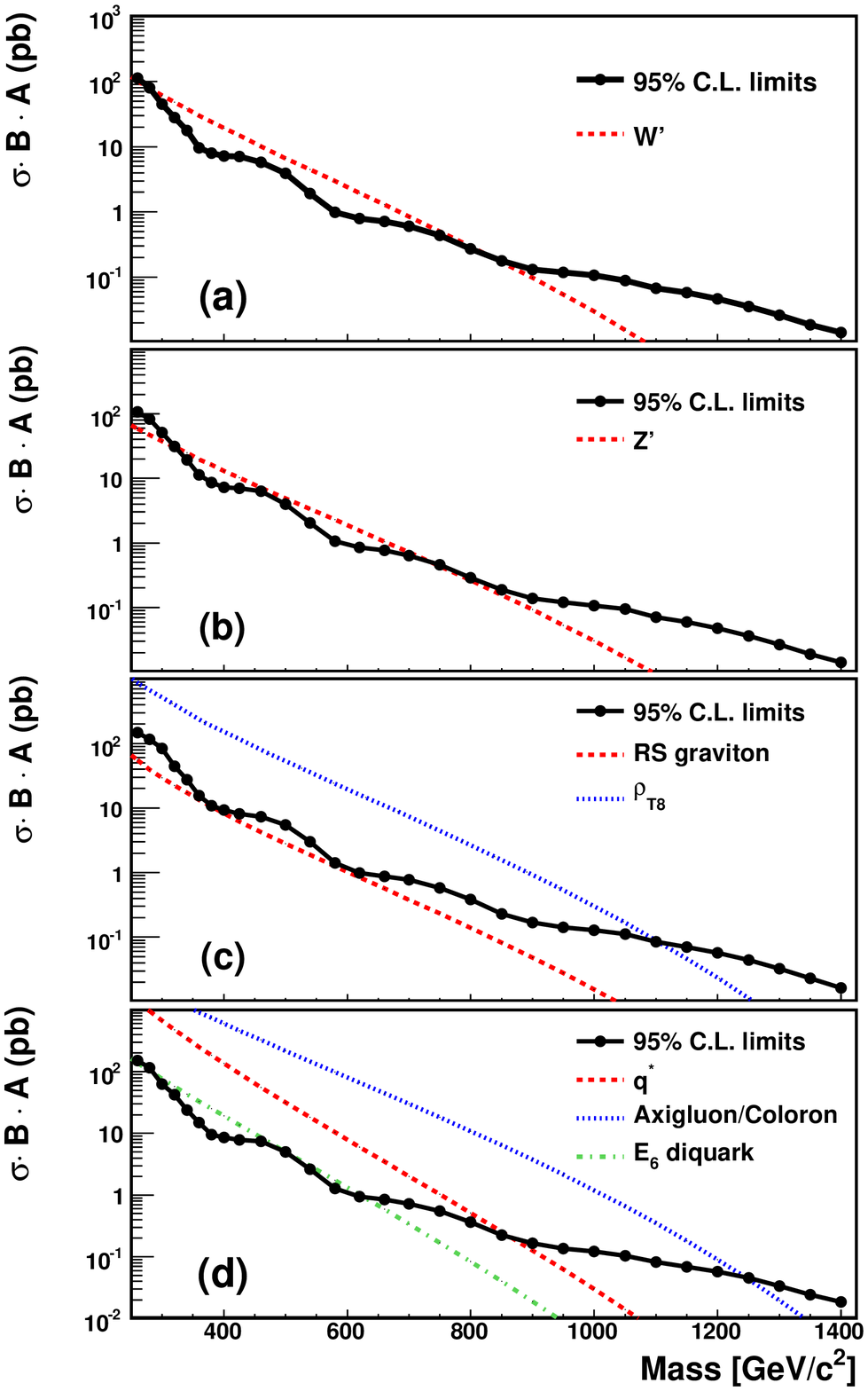,width=2.73in}
}
\vspace*{8pt}
\caption{Limits from D0 in 2004 and CDF in 2009.
(left) Upper limits at 95\% CL on dijet resonance cross section times branching ratio times acceptance 
from D0 using 109 pb$^{-1}$, compared with three model calculations and using resonance shapes from 
three models: a) excited quarks, b) $W^{\prime}$, c) $Z^{\prime}$, 
from Ref.~\protect\refcite{D02004}, Copyright 2004, and 
(right) same from CDF using 1.13 fb$^{-1}$, compared with seven model calculations and using
resonance shape from four models: a)  $W^{\prime}$, b) $Z^{\prime}$, c) RS graviton, and
d) excited quarks, from Ref.~\protect\refcite{CDF2009}, Copyright 2009 by the American Physical Society.
\label{D0CDFLimits}}
\end{figure}

The search from D0 in 2004~\cite{D02004} published limits on three models of dijet resonances: $W^{\prime}$, $Z^{\prime}$ and $q^*$.
The handling of statistical uncertainties in the limit was similar to the technique used by CDF in 1995 
and 1997, which D0 noted as a Bayesian technique with a flat prior for the signal~\cite{D0stat}. 
D0, unlike CDF, applied a truly Bayesian methodology to the treatment of the systematic uncertainties.
The systematics considered were the jet energy scale (2\%), resolution, efficiency, and luminosity, with
no uncertainty on the background from the NLO QCD calculation. All these
nuisance parameters had Gaussian prior probability distributions with widths given by their uncertainties.
Eq.~\ref{Likelihood} was multiplied by all priors, and then integrated over the
nuisance parameters to obtain the posterior probability density as a function of the amount of signal,
from which the limit was found as the 95\% quantile. This relatively modern procedure likely resulted 
in a smaller and more correct effect of the experimental systematic uncertainties 
on the limit than the conservative procedure used by CDF in 1995 and 1997. In Fig.~\ref{D0CDFLimits} the 
upper limits at 95\% CL on 
cross section time branching fraction times acceptance are shown separately for each of the three models,
and compared to the model cross sections. D0 obtained the $q^*$, $W^{\prime}$ and $Z^{\prime}$ LO 
cross sections from \textsc{Pythia}, and also applied NLO correction factors of about $1.3$ to the $W^{\prime}$ 
and $Z^{\prime}$. The mass intervals excluded for the the three models are listed in table~\ref{tabLimits}.
Comparing to the CDF limits from 1997, which used a similar sized dataset from the same running period, 
we note that D0 was able to obtain a $Z^{\prime}$ mass limit while CDF was not, D0 $W^{\prime}$ mass limits were 
significantly better than the corresponding CDF mass limits, and the D0 $q^*$ mass limit was a little better than
CDF and filled in the CDF gap in the $q^*$ mass limit around 550 GeV. Comparisons of mass limits are perilous,
as the model cross sections presented by the two experiments do not agree. For example, at a resonance mass
of 700 GeV, where the acceptance of the two experiments was the same, the D0 $W^{\prime}$ model cross section in 
Fig.~\ref{D0CDFLimits} is around 1.3pb, about a factor of 2.6 times the CDF  $W^{\prime}$ model cross section, and equal to the 
D0 $q^*$ model cross section at the same mass. The upper limits on the cross section are generally better to
compare than mass limits, particularly in regions where the acceptance is the same, and the D0 cross-section upper 
limits at 700 GeV in Fig.~\ref{D0CDFLimits} are around $0.8-1$ pb for all three models which is better 
than the CDF limit of 1.3 pb.

The search from CDF in 2009~\cite{CDF2009} published upper limits on the cross section for dijet resonances.  
Similar to the 1997 search, CDF also applied a Bayesian methodology starting from Eq.~\ref{Likelihood} 
for the binned likelihood including statistical uncertainties only. Unlike the CDF 1997 search but 
similar to the D0 search, the method 
for incorporating systematic uncertainties was truly Bayesian~\cite{statLuc}. 
The systematic uncertainties considered were the jet energy scale (2-3\%), resolution, and luminosity. 
CDF used the parameterization in Eq.~\ref{CDFparam3} to describe the background. To account for systematic 
uncertainties in the background, CDF utilized a profile likelihood method, in which for each value of the 
signal cross section considered the parameters of the background were found again by maximizing the likelihood.
This is in contrast to the previous CDF analysis in 1997, where the background parameters were held fixed as 
the signal cross section was varied. CDF used the shapes from four models of dijet resonances 
$W^{\prime}$, $Z^{\prime}$, RS graviton, and $q^*$, available in \textsc{Pythia}, to set upper limits at 95\% CL on the cross section 
and noted that the limits get progressively worse as more gluons are included in the final state. CDF then compared these
upper limits to the model cross sections for these four models, from \textsc{Pythia},
and to its own LO calculations~\cite{resonanceCmsNote} of the cross section for 
color octet technirhos, axigluons, colorons, and $E_6$ diquarks in Fig.~\ref{D0CDFLimits}. 
The resulting mass limits are shown in Table~\ref{tabLimits} for most models.
The CDF cross-section calculations for axigluons, colorons, and $E_6$ diquarks used the same technique as in its 1997 search, and CDF included the customary NLO k-factor
of 1.3 for the $W^{\prime}$ and $Z^{\prime}$ models. The color octet technirho cross section came from 
\textsc{Pythia}, where the model included the QCD background, and which had to be subtracted off in order to obtain the net resonance
cross section~\cite{resonanceCmsNote}. CDF excluded color octet technirhos in the mass interval
$260<M<1100$ GeV in 2009. 
No mass limits were set on Randall-Sundrum gravitons by this search, or
by any other dijet resonance search up to this date.
CDF presented the upper limits in a table for future use, which allowed the CMS experiment 
to compare the CDF upper limits on the cross section 
to that expected for a string resonance and determine a CDF mass limit of 1.4 TeV on string 
resonances~\cite{CMS2010}.  In 2009 CDF published the most stringent mass limits to date 
on color octet technirhos and the dijet decays of $Z^{\prime}$.

\subsubsection{Limits from the CERN Large Hadron Collider Experiments}

\begin{figure}[htbp]
\centerline{
\psfig{file=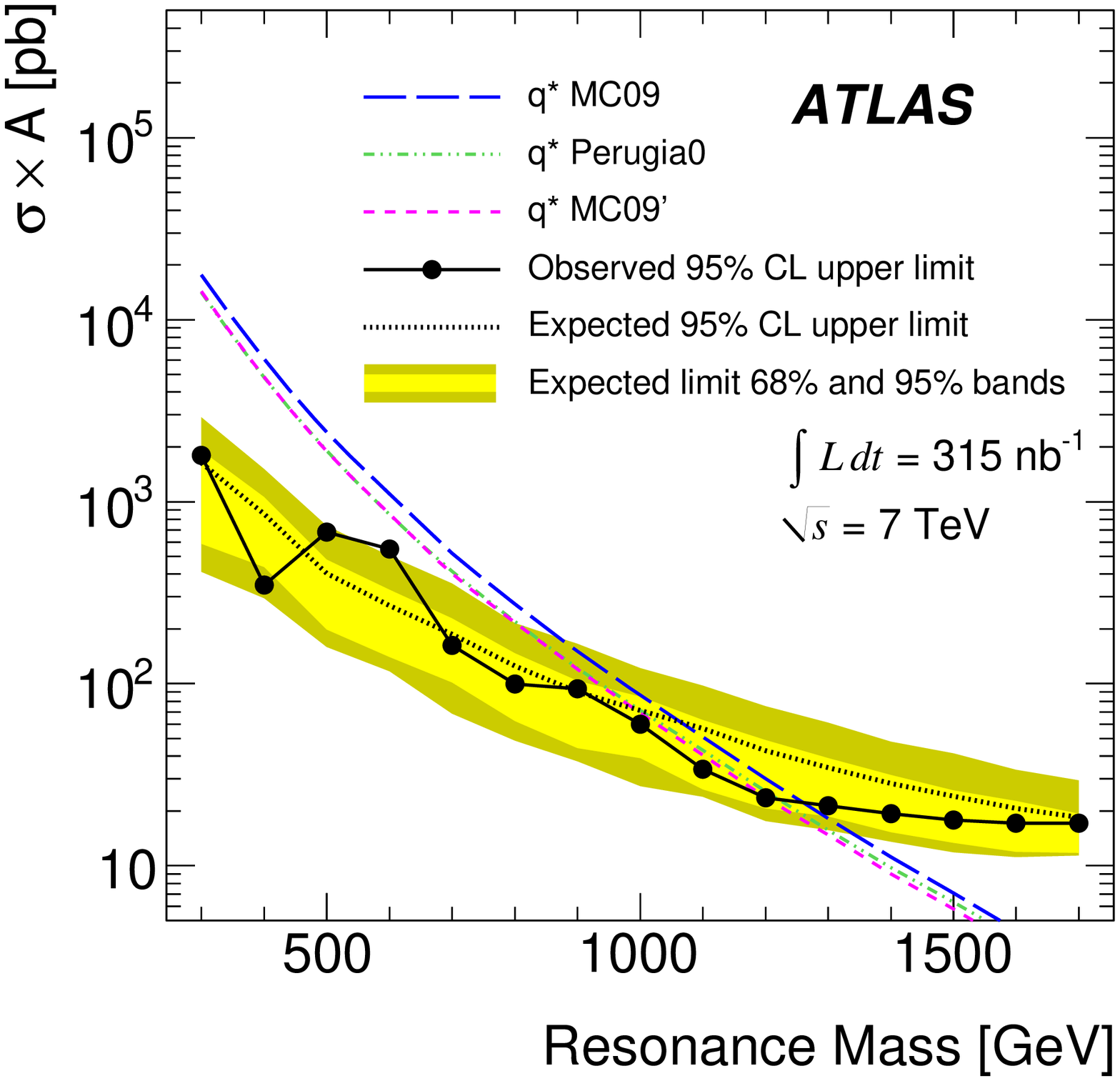,width=2.5in}
\psfig{file=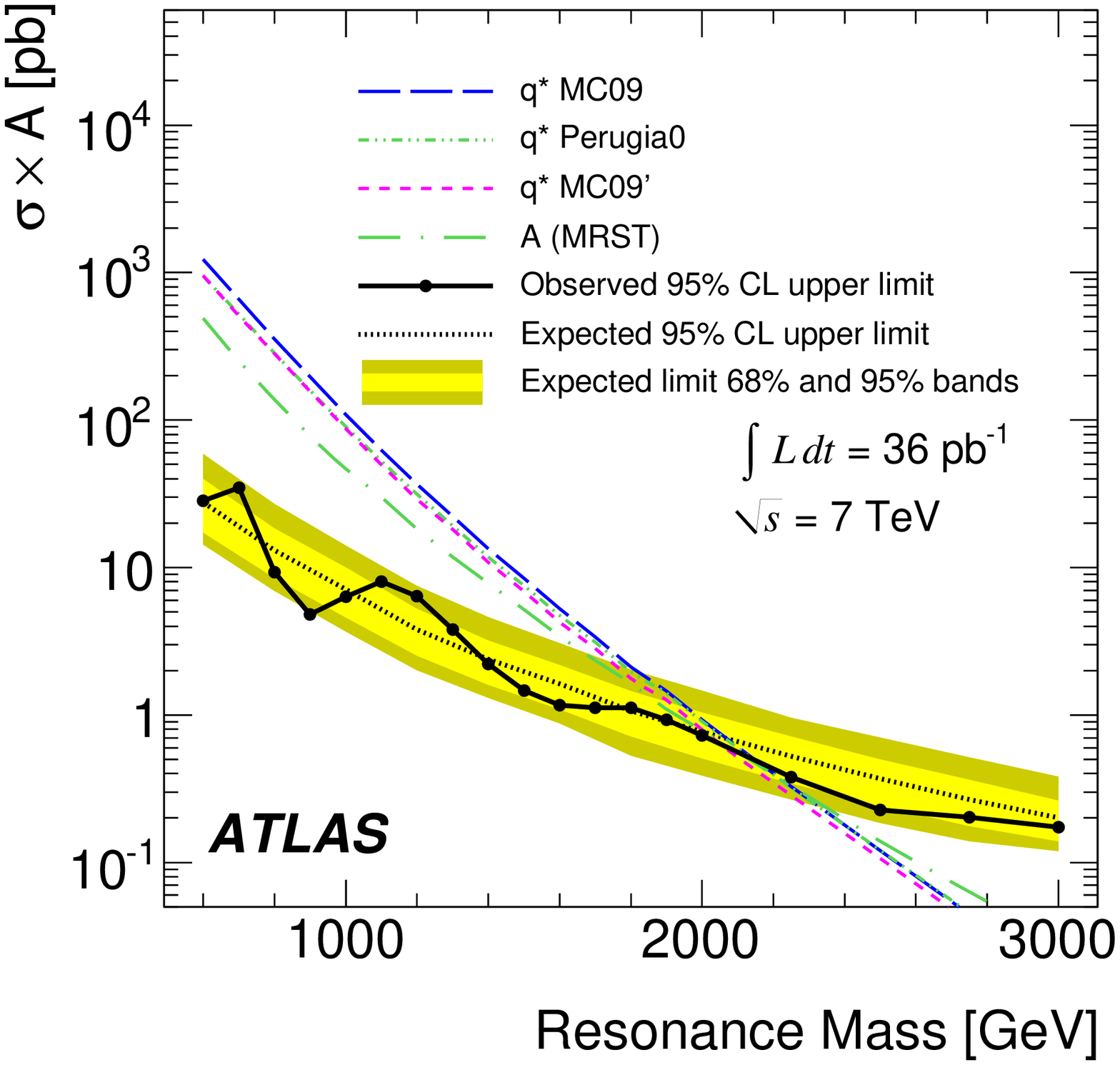,width=2.5in}
}
\centerline{
\psfig{file=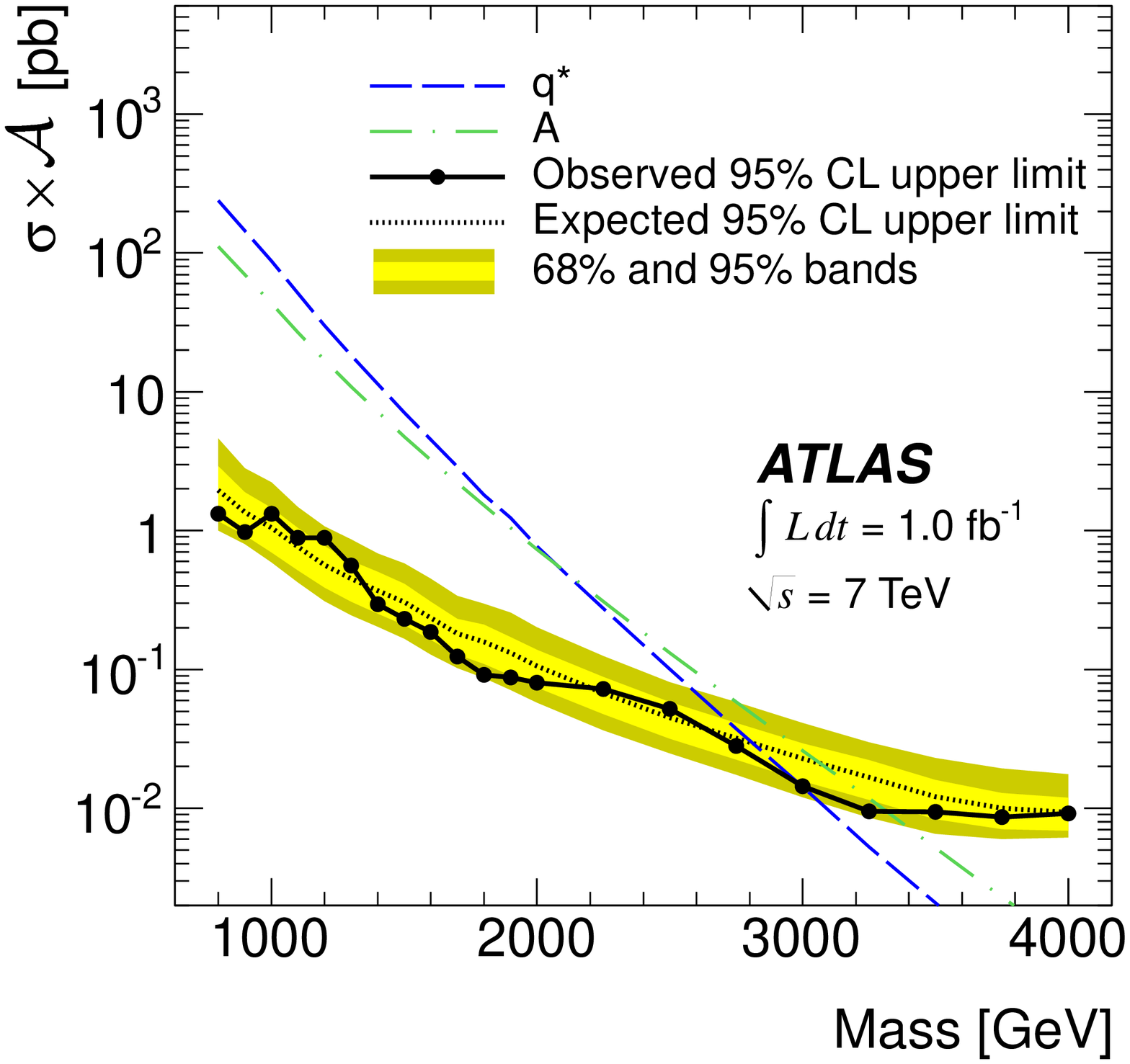,width=2.5in}
\psfig{file=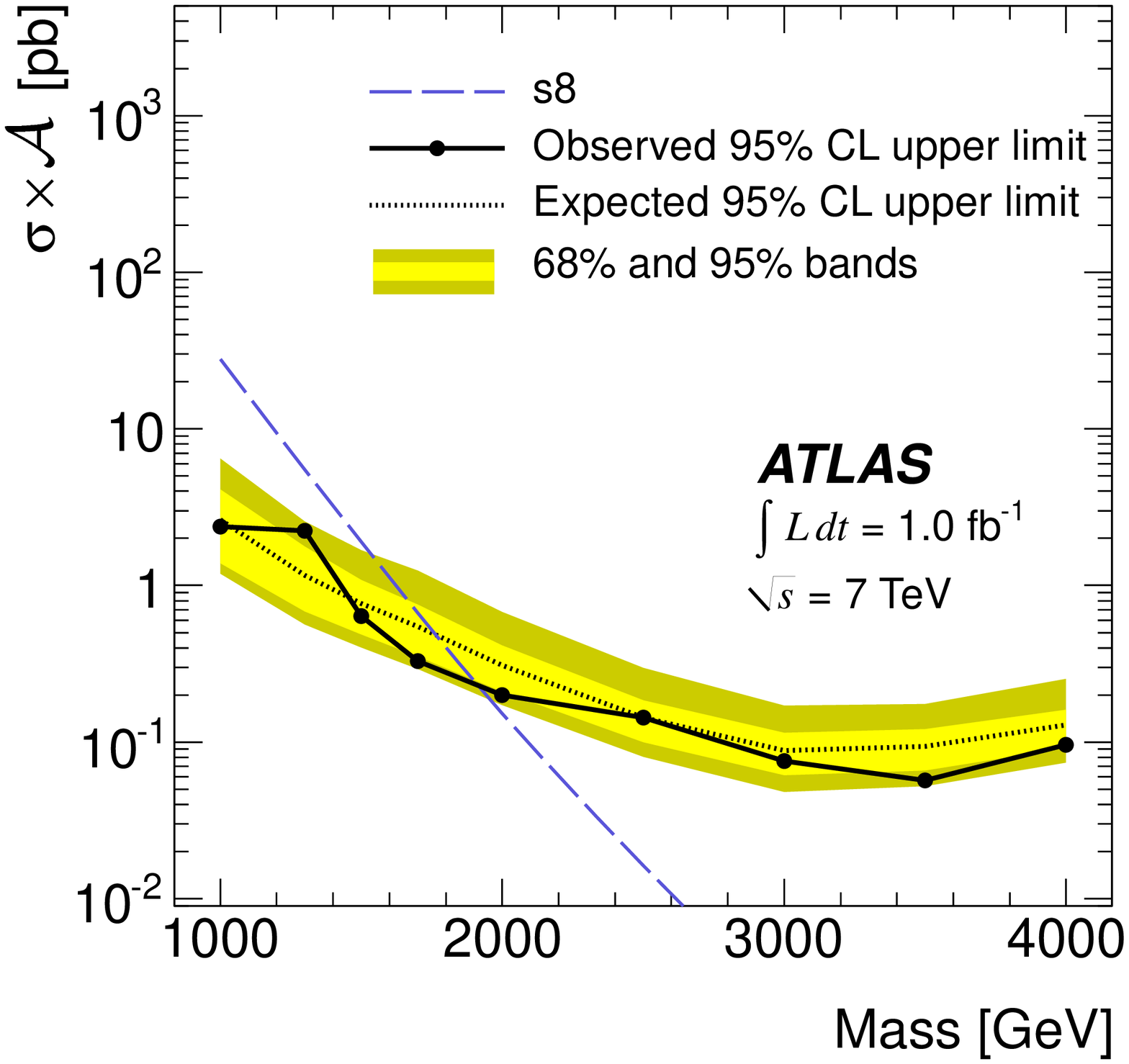,width=2.5in}
}
\vspace*{8pt}
\caption{Limits from ATLAS in 2010, winter 2011, and summer 2011.
(top left) Observed and expected upper limits at 95\% CL on excited quark cross section times acceptance 
using 315 nb${-1}$, compared to predictions of excited quarks with various tunes and PDFs. Shaded bands are 
$1\sigma$ and $2\sigma$ variations in the expected limit. From Ref.~\protect\refcite{ATLAS2010}, Copyright 2010 
by the American Physical Society. (top right) Same using 36 pb$^{-1}$ and in addition comparing to an axigluon 
prediction.  From Ref.~\protect\refcite{ATLAS2011}. (bottom left) Same using 1.0 fb$^{-1}$ and (bottom right) 
same, except limits and predictions are specifically for a color octet scalar resonance decaying to $gg$, 
from Ref.~\protect\refcite{ATLAS2011summer}.
\label{ATLASlimits}}
\end{figure}

 ATLAS published limits from three datasets: in 2010~\cite{ATLAS2010}, winter 2011~\cite{ATLAS2011} and summer 2011~\cite{ATLAS2011summer}, with 
0.3, 36, and 1000 pb$^{-1}$ of integrated luminosity. Limits were set using a Bayesian 
statistical technique with uniform prior for the signal cross section, and systematic uncertainties were incorporated
in a fully Bayesian treatment using Gaussian priors. The dominant source of 
systematic uncertainty was the jet energy scale in all three publications: 6\%-9\% in 2010,
3.2\%-5.7\% in winter 2011, and less than 4\% in summer 2011. The background was determined using 
the last CDF parameterization in Eq.~\ref{CDFparam3}. The systematic uncertainty on the 
background was taken from the statistical uncertainty in the fitted 
parameters, and for the winter 2011 search the uncertainty increased with resonance mass 
from 3\%-40\%. The luminosity
uncertainty was 11\%, 11\%, and 3.7\% for the three searches respectively. In all three searches
the uncertainty in the jet energy resolution had a negligible effect on the limits. 

ATLAS published
the first dijet resonance searches that included expected limits and their variation
at the $1\sigma$ and $2\sigma$ levels. These are limits determined from pseudo-experiments 
generated from the smooth background prediction. Expected limits vary smoothly as a function 
of resonance mass, while observed limits have wiggles that reflect statistical fluctuations in 
the data. In Fig.~\ref{ATLASlimits} the ATLAS observed and expected limits at 95\% CL are compared with the 
calculations of the model cross sections. In 2010 the fluctuation at $0.55$ TeV mentioned previously 
produced a $2\sigma$
fluctuation in the observed limit, their largest upward fluctuation. 

In all three publications 
ATLAS compared their cross-section upper limits from a $q^*$ shape to the cross section times 
acceptance for the $q^*$ model from \textsc{Pythia} including all decays: dominantly $qg$, but also 
including roughly an additional 20\% ``dijet'' cross section resulting from $qW$, $qZ$, 
and $q\gamma$. ATLAS explored variations in the $q^*$ cross section with different Monte Carlo tunes and associated 
PDFs, and chose to quote as the main result the MC09 tune~\cite{MC09} with MRST2007 PDF~\cite{MRST2007}. 
In 2011 ATLAS also
compared its cross-section upper limits to an axigluon calculation using \textsc{CalcHEP}~\cite{CalcHEP}, again with
MRST2007 PDF. We discuss the ATLAS calculations more in~\ref{sectionCompare}.
The excluded mass intervals for $q^*$ and axigluons are listed in table~\ref{tabLimits},
and the expected and observed mass limits are compared in table~\ref{tabMassLimits}. 
In the summer of 2011 ATLAS set cross-section upper limits on color octet scalars, which decay to $gg$ and
have a wider line shape than $q^*$ or axigluons. ATLAS compared with the model cross section from 
a MADGRAPH~\cite{MADGRAPH} plus \textsc{Pythia} calculation to exclude color octet scalars in the mass interval 
$1.0<M<1.92$ TeV when the mass limit $1.77$ TeV was expected.  All observed mass limits from ATLAS 
are larger than their expected mass limits, due to downward fluctuations in the data.  ATLAS in the summer of
2011 has published the most stringent limits to date on axigluons, excited quarks and color octet scalars.

\begin{figure}[htbp]
\centerline{
\psfig{file=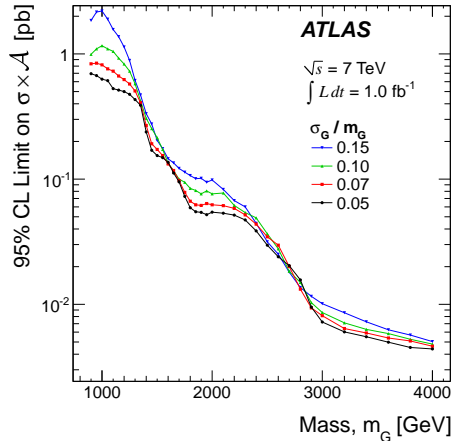,width=2.5in}
}
\vspace*{8pt}
\caption{Model independent limits from ATLAS in summer 2011.  Observed upper limits at 95\% CL 
for different Gaussian widths, from  Ref.~\protect\refcite{ATLAS2011summer}.
\label{ATLASgenericlimit}}
\end{figure}

In 2011 ATLAS introduced generic upper limits for wide dijet resonances based 
on a Gaussian line shape. In winter 2011 ATLAS published~\cite{ATLAS2011} a table of their 95\% CL upper limits on the cross section
for dijet resonances with measured Gaussian widths of 3\%, 5\%, 7\%, 10\% and 15\% of the peak mass.  
In summer 2011~\cite{ATLAS2011summer} ATLAS updated all the limits, except they dropped the 3\% Gaussian width, and published
them in a table and also in Fig.~\ref{ATLASgenericlimit}. Notice that the resonances in Fig.~\ref{ATLASgenericlimit} are 
not narrow, and the upper limits on the cross section increases as the Gaussian width of these resonance increases. 
In summer 2011 ATLAS published detailed 
instructions for how to obtain mass limits on new particles from these generic upper limits.
In short, they recommended the production of a Monte Carlo sample for the calculation of the acceptance at parton level,
and for the smearing of a dijet resonance signal with the ATLAS dijet resolution, which should then
be truncated within $\pm$ 20\% of the resonance peak before summing to find the total cross section.
This is the first publication of generic limits on wide resonances since the 1993 publication from CDF~\cite{CDF1993}.

While this review is only covering searches in the dijet mass spectrum, we note
that in 2011 ATLAS published the only search for dijet resonances at hadron colliders 
using the dijet angular distribution~\cite{ATLAS2011}, and used it to set limits on excited quarks.
The excited quark limits that we quote for ATLAS in table~\ref{tabLimits} and table~\ref{tabMassLimits} 
are from the search in the dijet mass spectrum only, not the search in the dijet angular distribution.

\begin{figure}[htbp]
\centerline{
\psfig{file=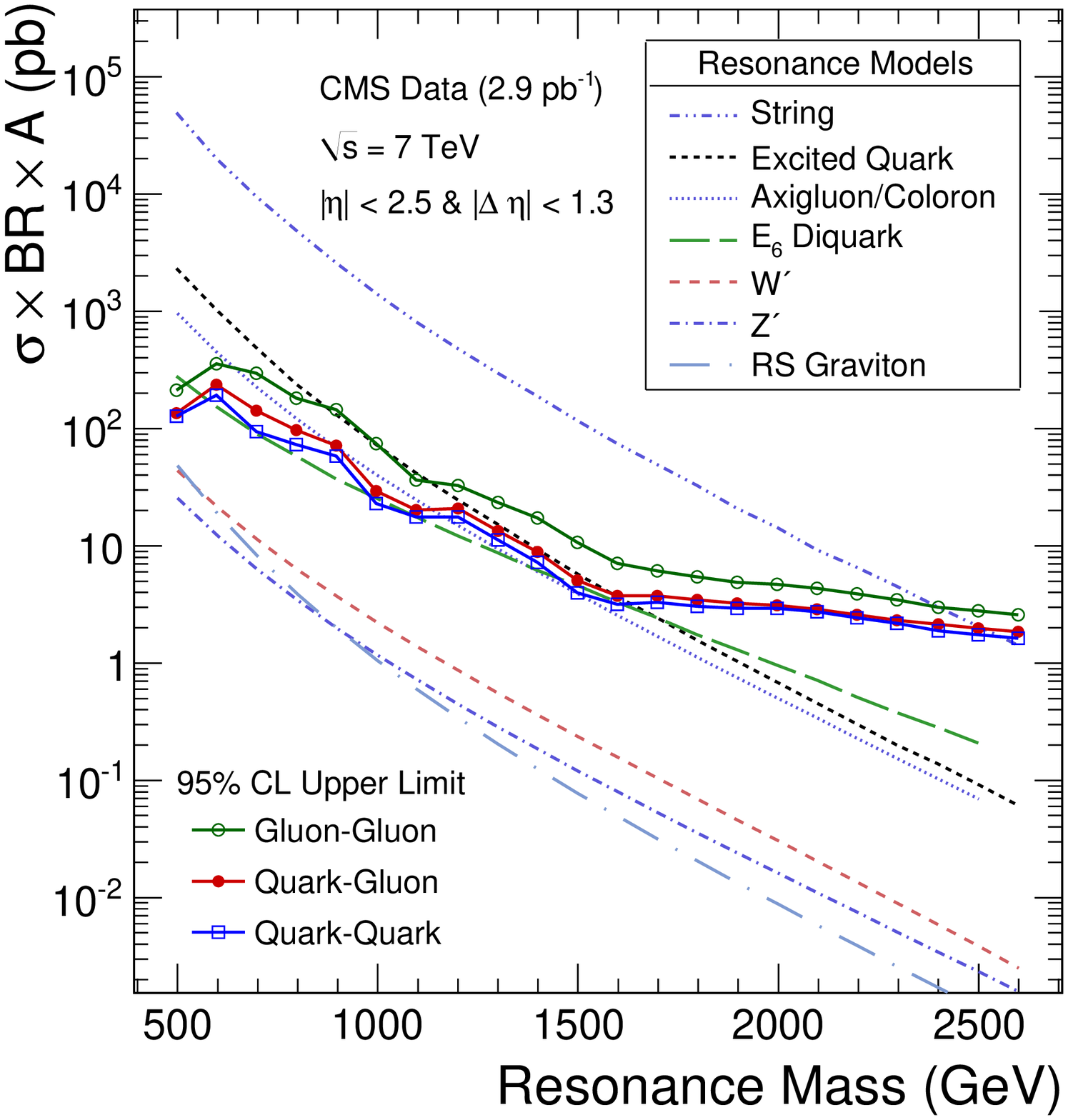,width=2.4in}
\psfig{file=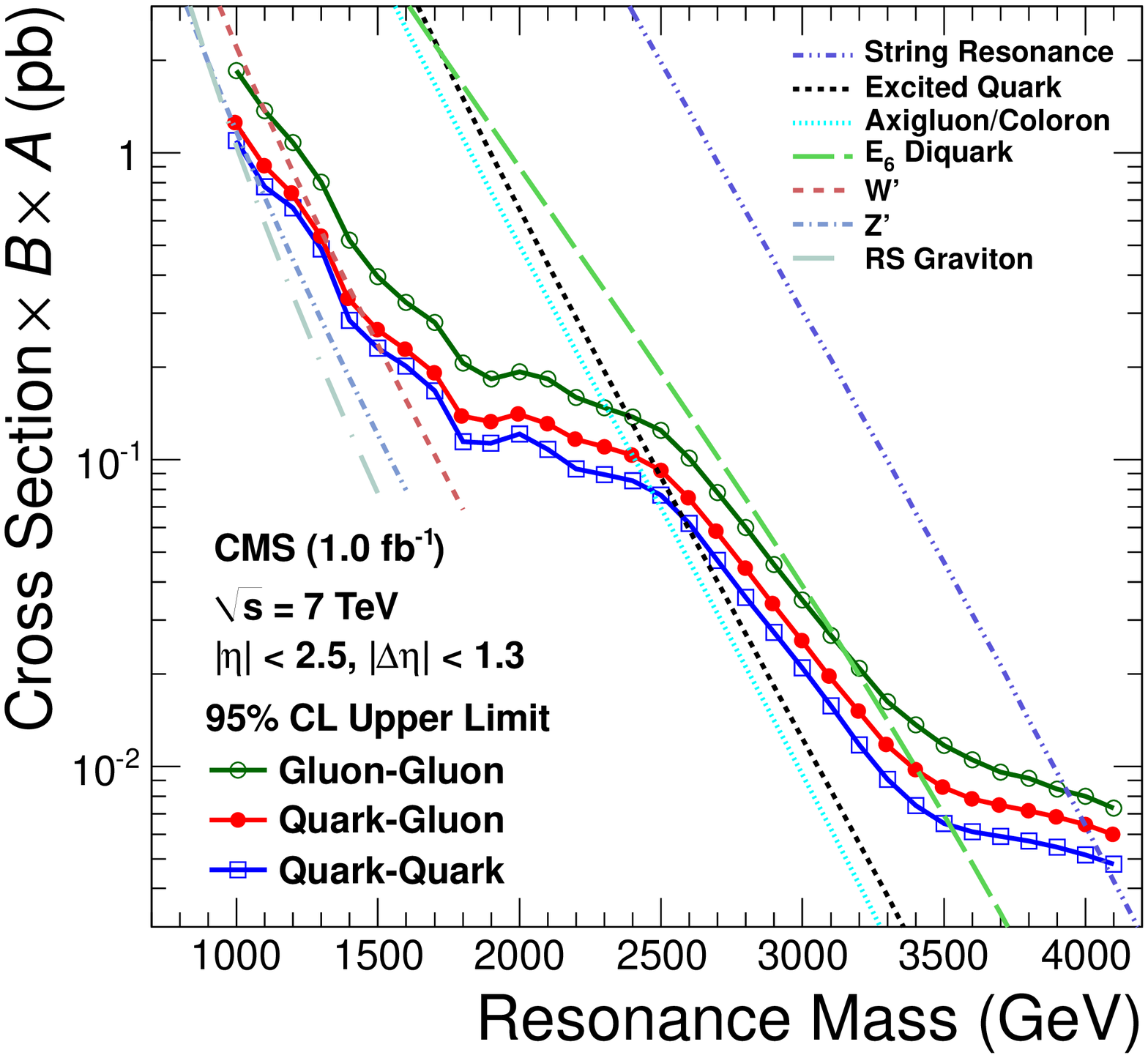,width=2.6in}
}
\centerline{
\psfig{file=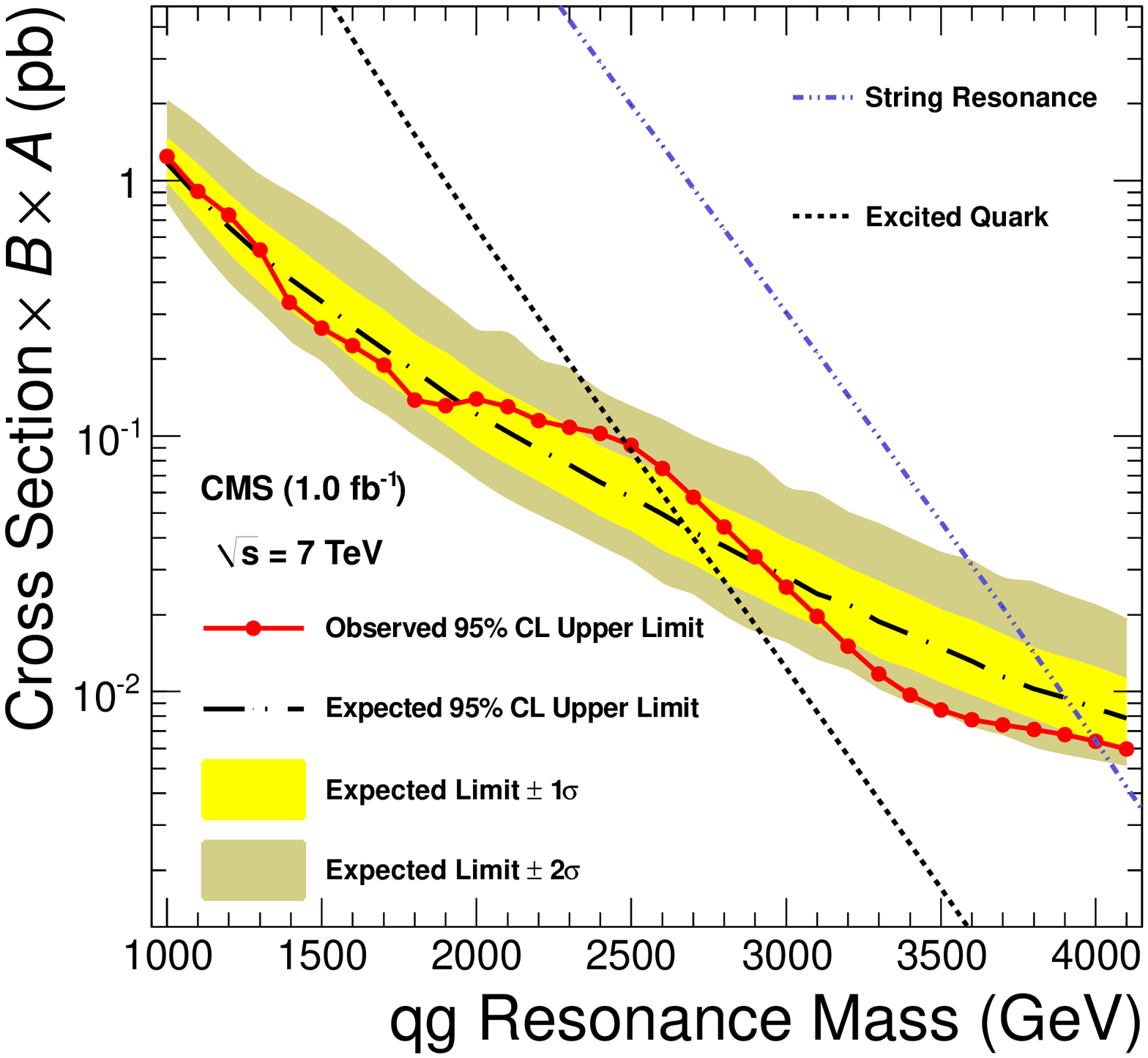,width=2.5in}
}
\vspace*{8pt}
\caption{Limits from CMS in 2010 and 2011.
(top left) ) Observed upper limits at 95\% CL on dijet resonance cross section times branching ratio times acceptance 
using 2.9 nb$^{-1}$, shown separately for gluon-gluon, quark-gluon, and quark-quark resonances, and compared with 
predictions from various models. From Ref.~\protect\refcite{CMS2010}, Copyright 2010 by the American Physical Society. 
(top right) Same, using 1.0 fb$^{-1}$ and (bottom) same for only quark-gluon resonances and including expected limits
and the $1\sigma$ and $2\sigma$ variations in the expected limits, from  Ref.~\protect\refcite{CMS2011}, Copyright 2011, 
with permission from Elsevier. 
\label{CMSlimits}}
\end{figure}

CMS published limits in 2010~\cite{CMS2010} using 2.9 pb$^{-1}$ and in 2011~\cite{CMS2011} using 1 fb$^{-1}$.
As for CDF, D0, and ATLAS, the handling of statistical uncertainties in the limits used a Bayesian procedure 
with uniform prior on the cross section. Systematic uncertainties considered were on the jet energy scale 
(10\% in 2010 and 2.2\% in 2011), jet energy resolution (10\% producing negligile effect), luminosity 
(11\% in 2010, 6\% in 2011), and the background. The background parameterization used was Eq.~\ref{CDFparam3}, the same
as used by CDF in 2009 and ATLAS. The uncertainty on the background was determined in 2010
by varying the choice of the parameterization similar to CDF in 1997, and in 2011 by varying the fit parameters within their 
statistical uncertainty, similar to ATLAS. The procedure for incorporating systematic uncertainties used by CMS in 2010 was the same
as the method employed by CDF in 1997, and was described as {\it ``an approximate technique, which in our application
is generally more conservative than a fully Bayesian treatment''}. In 2010 the systematic uncertainties 
increased the cross-section upper limits by 17\%-49\%. In 2011 CMS used a fully Bayesian procedure with Gaussian priors for incorporating systematics. The updated procedure, along with the reduced
systematics on the jet energy scale, had a much smaller effect on the observed cross-section upper limits than in 2010.

CMS used the shapes of narrow $qq$, $qg$ and $gg$ resonances to set generic upper limits at 95\% CL on 
the cross section times branching ratio into dijets times acceptance, and published the numbers in 
tables in both 2010 and 2011. In Fig.~\ref{CMSlimits} the generic upper limits are compared to the dijet 
decays of string resonances, excited quarks, axigluons, colorons, $E_6$ diquarks, $W^{\prime}$, $Z^{\prime}$, and
Randall-Sundrum gravitons. The model cross sections were LO calculations using CTEQ6L parton 
distributions, the same calculation as at CDF in 1997, and included the same ``K-factors'' for $W^{\prime}$ and
$Z^{\prime}$~\cite{resonanceCmsNote}. The calculations for axigluons, $W^{\prime}$, $Z^{\prime}$, and 
$E_6$ diquarks employed the narrow width approximation discussed in section~\ref{secCalc}. The CMS axigluon calculation was very different from the
calculation done by ATLAS, and the CMS $q^*$ calculation was slightly different than ATLAS, as discussed 
in \ref{sectionCompare}. The excluded regions for
all models are shown in table~\ref{tabLimits}, and the expected and observed limits are compared with ATLAS
in table~\ref{tabMassLimits}. The differences for a data sample of 1 fb$^{-1}$ are discussed in the next paragraph. 
In 2011, CMS also measured the $1\sigma$ and $2\sigma$ variations in the
expected limit, and showed these for $qg$ resonances in Fig.~\ref{CMSlimits}. The largest upward fluctuation in
the 2011 data was near $2.5$ TeV which caused the observed limit to be less than the expected limit for
both excited quarks and axigluons. In 2011 CMS published the most stringent limits to date on string resonances,
$E_6$ diquarks, and the dijet decays of $W^{\prime}$.

\begin{table}[hbt]
\tbl{Expected and observed mass limits at ATLAS and CMS from searches in the dijet mass spectrum.}
{\begin{tabular}{@{}lllcccc@{}} \toprule
Expt. & Year &  $\int L dt$  & \multicolumn{2}{c}{Axigluon} & \multicolumn{2}{c}{Excited Quark}\\
      &      &  ($pb^{-1}$)  & Expected & Observed  & Expected & Observed \\ 
      &      &               & (TeV)    & (TeV)    &  (TeV)   & (TeV)     \\ \colrule
ATLAS & 2010 &  $3.2\times 10^{-1}$& -- &  --      & 1.06     & 1.26    \\
CMS & 2010   &  $2.9\times 10^{0}$ & 1.32 & 1.58   & 1.23     & 1.17  \\
ATLAS & 2011w&  $3.6\times 10^{1}$ & 2.01 & 2.10   & 2.07     & 2.15 \\
CMS & 2011   &  $1.0\times 10^{3}$ & 2.66 & 2.47   & 2.68     & 2.49     \\
ATLAS & 2011s&  $1.0\times 10^{3}$ & 3.07 & 3.32   & 2.81     & 2.99   \\ 
 \botrule
\end{tabular} \label{tabMassLimits}}

\end{table}

The CMS publication in 2011~\cite{CMS2011} and the ATLAS publication in summer 2011~\cite{ATLAS2011}, 
both using an integrated luminosity of 1 fb$^{-1}$, reported different upper limits for the same model 
as shown in table~\ref{tabMassLimits}. The greatest difference lies
in the observed limits for axigluons (CMS 2.47 TeV, ATLAS 3.32 TeV) but there is also a 
significant difference for excited quarks (CMS 2.49 TeV, ATLAS 2.99 TeV).
As noted above, the observed limits by ATLAS were greater than 
their expected limits due to a downward fluctuation in the data, and the observed limits by CMS were
less than the expected limits due to an upward fluctuation in the data, so these fluctuations clearly
contributed to the difference in the observed limits between the two experiments. The fairest comparison
should be in the expected limits. However, there is still a significant
difference in the expected mass limits for axigluons (CMS 2.66 TeV, ATLAS 3.07 TeV) and a smaller 
but non-negligible difference
in the expected mass limits for excited quarks (CMS 2.68 TeV, ATLAS 2.81 TeV). For both models, the 
vast majority of the difference in the mass limits appears to result in differences in the calculated 
cross section of the model, which we discuss in detail in \ref{sectionCompare}. 

The most direct and natural comparison between the performance of ATLAS and CMS is in 
their expected limits on the cross section, as opposed to their expected mass limits on models.
At a resonance mass of 3 TeV, the ATLAS expected upper limit on cross section time acceptance 
for excited quarks is 0.023 pb, while the CMS expected upper limit on cross 
section times acceptance for $qg$ resonances is 0.028 pb. For excited quark signals at high masses 
the CMS acceptance is 14\% greater than the ATLAS acceptance. After correcting for acceptance 
the ATLAS expected limit on
the cross section is about 10\% more stringent than the CMS limit, which would give an expected mass limit
about 1\% more stringent. Compared this way the ATLAS and CMS performance is similar. 
This is expected given the comparable capabilities of the two experiments with the same integrated 
luminosity at the same center-of-mass energy.

%% file: Conclusions.tex
\section{Summary}
\label{secConclusions}

Dijet resonance searches at hadron colliders have constrained a
rich variety of models of new physics. 
Limits have been set on models motivated by grand unification, 
string theory, technicolor, compositeness, and ideas for 
new color interactions. Mass limits on the majority
of models constrained are summarized in table~\ref{tabLimits}. 
Limits on many colored resonances, like axigluons, colorons, and 
excited quarks, are now around 3 TeV and limits as large as 4 TeV
have been set. In addition to model specific limits, the searches
have provided model independent limits on the cross section for 
dijet resonances that can be used to constrain future models of
new particle production.

The most important factors influencing the sensitivity of dijet resonance
searches are the center-of-mass energy of the collider and the integrated 
luminosity of the dataset. The largest experimental uncertainty, the jet 
energy scale, now contributes little to upper limits on the cross section.
Nevertheless, the experiments must remain vigilant that uncertainties in 
the jet energy scale do not manufacture a dijet resonance signal or hide one.
Uncertainties in jet resolution at the experiments have usually had negligible 
effect. The searches are always dominated by statistical uncertainties, and 
the experiments have therefore for the most part done the most important thing, 
which is to rapidly search and publish when the energy or integrated 
luminosity have increased significantly. Keeping the search simple and model independent has made this easier. 

In the course of time, the techniques and ideas employed by the experiments
have varied significantly. However, an evolution can be observed towards
standard practices, and it is worth noting the most important ones 
which may be useful for future searches as well.

The existence of a permanent record of fluctuations in the data for future reference, 
is helpful when experiments publish detailed comparisons of their dijet mass data
to the background prediction. This was attempted by 
CDF and D0 with a ratio plot, but the plot of the bin-by-bin 
significance of the difference between data and background, introduced
by the ATLAS collaboration and adopted by CMS, is ideal for this purpose.
Estimations of the global significance of fluctuations introduced by ATLAS 
are instructive, and it would be even more helpful if they were accompanied 
by the undiluted estimates of the local significance of the fluctuation.

Both the ATLAS and CMS experiments have adopted a parametrization from CDF to 
model their QCD background. This is because a parameterization fit to the data 
generally gives a better model of the background than QCD calculations. 

The experimental practice of reporting cross-section upper limits on dijet resonances, in 
addition to mass limits on specific models, has been important to allow 
continued use of the data to constrain models (see e.g. Ref~\refcite{S8}). 
We have also shown that the cross section limits are important to understand and compare the 
results of the experiments,because differences in the calculation of the model cross sections 
often affect the mass limits significantly, as we saw in the comparison of the CDF and D0 
limits with 100 pb$^{-1}$ and the ATLAS and CMS limits with 1 fb$^{-1}$. Clearly it 
would also be helpful if the experiments used a common method of calculating the model cross sections.

Finally, to understand the results of the experiments and to compare them, it has been important 
to have expected upper limits accompanied by their statistical variations, as well as observed upper 
limits. This practice was introduced by ATLAS and adopted by CMS, and is a visible benefit of the 
modern statistical techniques that are now commonly employed in the searches.

We look forward to future searches for dijet resonances at hadron colliders 
and their discovery of new physics beyond the standard model.

%% file: Acknowledgements.tex
\section*{Acknowledgments}

We gratefully acknowledge discussions with our colleagues who conducted 
dijet resonance searches at hadron colliders.
In particular, to compare CMS and ATLAS, critical information was
provided by Georgios Choudalakis and Alfonso Zerwekh from ATLAS, and Maxime 
Gouzevitch and Chiyoung Jeong from CMS. For sharing information and useful
discussions over the years, we also thank Iain Bertram, John Paul Chou, 
Kenichi Hatakeyama, Emilio Meschi, Maurizio Pierini, Sertac Ozturk and 
Pierre Savard.
For assistance with calculations and concepts of the models over the
years we thank our colleagues in the theory community, including Uli Baur, 
Estia Eichten, Ken Lane, Can Kilic, Steve Mrenna, Tom Rizzo and Scott Thomas.
This research was supported by Fermilab, operated by Fermi Research Alliance, 
LLC under Contract No. De-AC02-07CH11359 with the United States Department 
of Energy.

%% file: AppCMSandATLAS.tex
\section{Axigluon and Excited Quark Calculations from ATLAS and CMS}
\label{sectionCompare}

\begin{figure}[hbtp]
\centerline{
\psfig{file=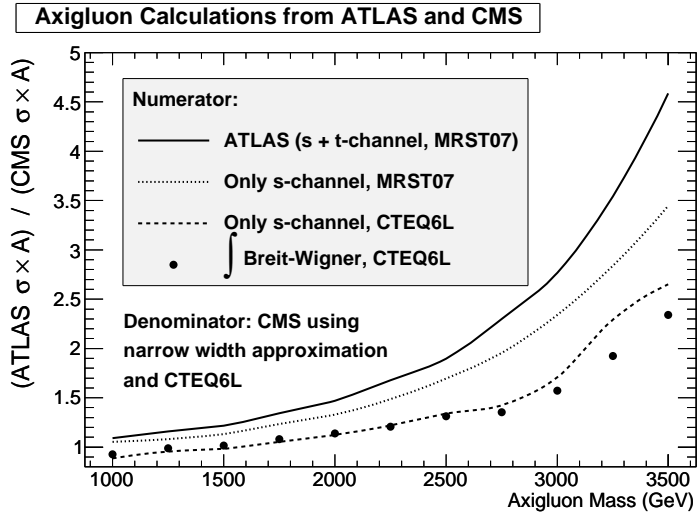,width=4.0in}
}
\vspace*{8pt}
\caption{Ratio of the ATLAS and CMS calculations of the axigluon cross section times branching ratio
times acceptance.
\label{axigluonRatio}}
\end{figure}

Fig.~\ref{axigluonRatio} shows the ratio ATLAS / CMS for the calculated cross section times acceptance 
for axigluons decaying to dijets in each experiment.  The solid curve in Fig.~\ref{axigluonRatio} shows
the ratio of the cross section times acceptance curves published by the two experiments 
in Fig.~\ref{ATLASlimits} and Fig.~\ref{CMSlimits} for 1 fb$^{-1}$. 
At an axigluon mass of 1 TeV the calculations agree, and the ratio is the relative acceptance
of the two experiments, but the ratio increases rapidly with mass.
We will use a 3 TeV axigluon resonance as an
example, where the ratio of the cross section times acceptance published by ATLAS and CMS is a factor of 2.8. 
The Axigluon models employed by the two experiments are the same but there are differences in the way the two 
experiments perform the cross section calculations. The major sources of difference are 
\begin{romanlist}[(iv)]
\item  Resonance tail. CMS uses the narrow width approximation which accounts for only the cross section at 
the resonant pole, while ATLAS uses \textsc{CalcHEP} and integrates the cross section over a dijet mass window 
within $\pm$30\% of the resonance pole ($0.7M<m<1.3M$) picking up a large tail at masses beneath the pole.  
The affect of ATLAS integration over the mass window is shown by both a
\textsc{CalcHEP} calculation~\cite{alfonso} 
(dashed curve in Fig.~\ref{axigluonRatio}) and by our own calculation of the affect of integrating a 
Breit-Wigner on the CMS calculation (points in Fig.~\ref{axigluonRatio}). After taking out differences due 
to acceptance, this increases the ATLAS cross section by a factor of 1.92 for a 3 TeV axigluon according to 
the \textsc{CalcHEP} calculation.

\item Parton distributions. CMS uses CTEQ6L and ATLAS uses MRST07. Comparing the dashed and dotted curves
in Fig.~\ref{axigluonRatio}, which come from a \textsc{CalcHEP} calculation~\cite{alfonso}, we find this increases the ATLAS cross section
by a factor of 1.37 for a 3 TeV axigluon.

\item Processes included.  CMS includes only the process where axigluons are produced resonantly in
the s-channel, while ATLAS also includes processes where axigluons are exchanged in the t-channel~\cite{alfonso},
which increases the ATLAS cross section by a factor of 1.18 for a 3 TeV axigluon.

\item Acceptance. The acceptance of the two experiments for Axigluons is dominated by the 
$|\cos\theta^*|$ cut which is $0.57$ for CMS and $0.54$ for ATLAS.   The combined affect of this cut and
the ATLAS removal of a narrow region of the detector decreases the ATLAS acceptance by a factor of $0.89$.
\end{romanlist}

The differences in the axigluon cross section calculations at ATLAS and CMS are the largest
source of the difference in the expected limits and requires further discussion.
The resonance tail at low mass is the largest single effect. How to handle its contribution 
to the new particle cross section,
is a common problem faced by every resonance search in hadron colliders.  It is caused by the 
decrease of the parton distributions of the proton with increasing fractional momentum 
$x\sim m/\sqrt{s}$, and is a particularly severe problem for $q\bar{q}$ resonances at $pp$ 
colliders.  The narrow width approximation used by CMS, and discussed in section~\ref{secCalc}, 
is commonly employed by theorists to calculate the cross section for new particles. It replaces 
the true shape of the resonance with a delta function at the resonance pole, and therefore 
matches well the experimental approximation of both the ATLAS and CMS experiments when 
employing narrow resonance shapes to search for new physics and set limits on the cross section. 
However, the narrow width approximation clearly underestimates the true axigluon cross section, 
a significant fraction of which is off the pole at lower masses.  The ATLAS choice to integrate 
over the actual line shape gives a total axigluon cross section that may be more realistic. 
However, this may not match as well the assumption of a narrow resonance shape peaked at the 
pole which was used to set cross section limits. The ATLAS choice to include t-channel
processes in the total cross section again gives a result that may be more realistic than the 
CMS choice to only include s-channel processes. However, it is unclear whether the ATLAS and 
CMS methodology of fitting the data for the 
background shape and normalization effectively absorbs some of the cross section of these 
additional processes into the background, setting limits on only the bump like 
component of the resonances. The same can be said for the remainder of the axigluon tail
at masses outside the search window which is ignored by both experiments.  Further, both
experiments ignore coherence between QCD and axigluons, which may significantly affect the 
tails of the distribution, and is likely an issue when the tail has a larger cross section 
than the peak.  Finally we note that the different choice of PDFs by the two
experiment is a source of variation in the mass limits that is easy to remove
in the future.

 A similar comparison between the model cross sections for excited quarks has been done at 
CMS and ATLAS.  Here the calculated values of $(\sigma \times A)$ at ATLAS at a mass of 3 TeV 
are only 16\% greater than the reported values of $(\sigma \times B \times A)$ at CMS.  
There are three factors making up this difference:
\begin{romanlist}[(iv)]

\item Parton distributions. CMS uses CTEQ6L and ATLAS uses MRST07, making the ATLAS cross section
greater by about 10\% at M=3 TeV.

\item Branching Fraction.  CMS includes only $q^*\rightarrow qg$ decays in the dijet cross section, 
which make up 84\% of the branching fraction. ATLAS also includes the $qW$, $qZ$ and $q\gamma$ 
decays.  This choice increases the ATLAS cross section by a factor of roughly 1/0.84 = 20\%

\item Acceptance. The acceptance of the two experiments for excited quarks, which at resonance mass
of 3 TeV is $0.57$ for CMS and $0.50$ for ATLAS.  The ATLAS acceptance is smaller by 12\%.
\end{romanlist}

%% file: ReviewPreprint.bbl
\begin{thebibliography}{0}    

\bibitem{LOxsec}      B. L. Combridge, J. Kripfganz and J. Ranft, {\it Phys. Lett. B} {\bf 70}, 234, (1977).
\bibitem{QCDColliderPhysics} R. K. Ellis, W. J. Stirling and B. R. Webber, {\it Cambridge University Press}, QCD and Collider Physics (1996).
\bibitem{Qstar1}      U. Baur, I. Hinchliffe and D. Zeppenfeld, {\it Int. J. Mod. Phys. A} {\bf 2}, 1285, (1987).
\bibitem{Qstar2}      U. Baur, M. Spira and P. M. Zerwas, {\it Phys. Rev. D} {\bf 42}, 815 (1990).
\bibitem{RS1}         L. Randall and R. Sundrum, {\it Phys. Rev. Lett.} {\bf 83}, 3370 (1999).
\bibitem{RS2}         L. Randall and R. Sundrum, {\it Phys. Rev. Lett.} {\bf 83}, 4690 (1999).
\bibitem{RS3}         H. Davoudiasl, J. L. Hewett and T. G. Rizzo, {\it Phys. Rev. Lett.} {\bf 84}, 2080 (2000).
\bibitem{RS4}         J. Bijnens, P. Eerola, M. Maul, A. Mansson, T. Sjostrand, {\it Phys. Lett. B} {\bf 503},341 (2001).
\bibitem{Axigluon}    P. H. Frampton and S. L. Glashow, {\it Phys. Lett. B} {\bf 190}, 157 (1987).
\bibitem{Coloron1}    R. S. Chivukula, A. G. Cohen and E. H. Simmons, {\it Phys. Lett. B} {\bf 380}, 92, (1986).  
\bibitem{Coloron2}    E. H. Simmons, {\it Phys. Rev. D} {\bf 55}, 1678 (1997).
\bibitem{UA1ptLimit} J. Bagger, C. Schmidt and S. King, {\it Phys. Rev. D} {\bf 37}, 1188 (1988).
\bibitem{S8}          T. Han, I. Lewis and Z. Liu, {\it JHEP} {\bf 12}, 085 (2010).
\bibitem{WZprime}     E. Eichten, I. Hinchliffe, K. D. Lane and C. Quigg, {\it Rev. Mod. Phys.} {\bf 56}, 579 (1984).
\bibitem{String1}     L. A. Anchordoqui et al., {\it Phys. Rev. Lett.} {\bf 101}, 241803, (2008) 
\bibitem{String2}     S. Cullen, M. Perelstein and M. E. Peskin, {\it Phys. Rev. D} {\bf 62}, 055012, (2000).
\bibitem{E6}          P. Candelas, G. T. Horowitz, A. Strominger and E. Witten, {\it Nucl. Phys. B} {\bf 258}, 46, (1985).
\bibitem{E6Diquarks1} J. L. Hewett and T. G. Rizzo, {\it Phys. Rept.} {\bf 183}, 193 (1989).
\bibitem{E6Diquarks2} G. Katsilieris, O. Korakiantitis and S. Vlassopulos, {\it Phys. Lett. B} {\bf 288}, 221, (1992). 
\bibitem{E6Diquarks3} V. D. Angelopoulos, John R. Ellis, H. Kowalski, D. V. Nanopoulos, N. D. Tracas , F. Zwirner, {\it Nucl.Phys.B} {\bf 292}, 59, (1987).
\bibitem{WalkingTC}   B. Holdom, {\it Phys. Rev. D} {\bf 24}, 1441 (1981); {\it Phys. Lett.B} {\bf 150}, 301 (1985); T. Appelquist, D. Karabali and L. C. R. Wijewardhana, {\it Phys. Rev. Lett.} {\bf 57}, 957 (1986); T. Appelquist and L. C. R. Wijewardhana, {\it Phys. Rev. D} {\bf 36}, 568 (1987); K. Yamawaki, M. Bando and K. Matumoto, {\it Phys. Rev. Lett.} {\bf 56}, 1335 (1986);T. Akiba and T. Yanagida, {\it Phys. Lett. B} {\bf 169}, 432 (1986).
\bibitem{Technirho1}  K. Lane and M. Ramana, {\it Phys. Rev. D} {\bf 44}, 2678 (1991).
\bibitem{Technirho2}  K. Lane and S. Mrenna, {\it Phys. Rev. D} {\bf 67}, 115011 (2003).

\bibitem{UA1ptData}  UA1 Collab. (G. Arnison {\it et al}.), {\it Phys. Lett. B\/} {\bf 172}, 461 (1986).
\bibitem{UA1mass} UA1 Collab. (C. Albajar {\it et al}.), {\it Phys. Lett. B\/} {\bf 209}, 127 (1988).
\bibitem{CDF1990} CDF Collab. (F. Abe {\it et al}.), {\it Phys. Rev. D\/} {\bf 41}, 1722 (1990).
\bibitem{UA21990} UA2 Collab. (J. Alitti {\it et al}.), {\it Z. Phys. C\/} {\bf 49},17 (1991).
\bibitem{CDF1993} CDF Collab. (F. Abe {\it et al}.), {\it Phys. Rev. Lett.} {\bf 71}, 2542 (1993).
\bibitem{UA21993} UA2 Collab.  (J. Alitti {\it et al}.), {\it Nucl. Phys. B\/}{\bf 400}, 3 (1993).
\bibitem{CDF1995} CDF Collab. (F. Abe {\it et al}.), {\it Phys. Rev. Lett.} {\bf 74}, 3538 (1995).
\bibitem{CDF1997} CDF Collab. (F. Abe {\it et al}.), {\it Phys. Rev. D.} {\bf 55}, R5263 (1997).
\bibitem{D02004} D0 Collab. (V.M. Abazov {\it et al}.), {\it Phys. Rev. D.} {\bf 69}, R111101 (2004).
\bibitem{CDF2009} CDF Collab. (T. Aaltonen {\it et al}.), {\it Phys. Rev. D.} {\bf 79}, 112002 (2009).
\bibitem{ATLAS2010} ATLAS Collab. (G. Aad {\it et al}.), {\it Phys. Rev. Lett.} {\bf 105}, 161801 (2010).
\bibitem{CMS2010} CMS Collab. (V. Khachatryan {\it et al}.), {\it Phys. Rev. Lett.} {\bf 105}, 211801 (2010).
\bibitem{ATLAS2011} ATLAS Collab. (G. Aad {\it et al}.), {\it New J.  Phys.} {\bf 13}, 053044 (2011).
\bibitem{CMS2011} CMS Collab. (S. Chatrchyan {\it et al}.), {\it Phys. Lett. B} {\bf 704}, 123 (2011).
\bibitem{ATLAS2011summer} ATLAS Collab. (G. Aad {\it et al}.), {\it Submitted to Phys. Lett. B}, arXiv:1108.6311.
\bibitem{ISAJET}  F. Paige {\it et al}, arXiv:hep-ph/0312045.
\bibitem{HERWIG} G. Corcella {\it et al}, {\it JHEP} {\bf 0101}, 010 (2001) [arXiv:hep-ph/0011363].
\bibitem{PYTHIA} T. Sjostrand {\it et al}, Comput. Phys. Commun. {\bf 135}, 238 (2001).
\bibitem{refMaxime} M. Gouzevitch from CMS Collaboration, private communication.
\bibitem{anti-kt} M. Cacciari, G.P. Salam, G. Soyez, JHEP {\bf 0804} 063 (2008); M. Cacciari and G.P. Salam, {\it Phys. Lett. B} {\bf 641} 57 (2006) [hep-ph/0512210]; M. Cacciari, G.P. Salam and G. Soyez, URL http://www.fastjet.fr
\bibitem{OccamsRazor} H. G. Gauch, {\it Scientific Method in Practice}, (Cambridge University Press, New York, 2003).
\bibitem{bertram} I. Bertram, arXiv:hep-ph/9811445v1.
\bibitem{CDFpt} CDF Collab. (F. Abe {\it et al}.),{\it Phys. Rev. Lett.}{\bf 77}, 438 (1996).
\bibitem{LEE} E. Gross and O. Vitells, {\it Eur. Phys. J.}{\bf C70}, 525 (2010).
\bibitem{JETRAD} W. T. Giele, E.W.N. Glover, and D.A. Kosower, {\it Phys. Rev. Lett.} {\bf 73}, 2019 (1994).
\bibitem{CTEQ6} J. Pumplin {\it et al}., {\it JHEP} {\bf 07}, 12 (2002).
\bibitem{fastNLO} T. Kluge, K. Rabbertz, and  M. Wobisch, arXiv:hep-ph/0609285v2.
\bibitem{CTEQ6.1} D. Stump {\it et al}., {\it JHEP} {\bf 10}, 46 (2003).
\bibitem{bumphunter} CDF Collab. (T. Aaltonen {\it et al}.) {\it Phys. Rev. D}{\bf 79}, 011101 (2009); G. Choudalakis, arXiv:physics.data-an/1101.0390.
\bibitem{CDFqstarPaper} CDF Collab. (F. Abe {\it et al}.),{\it Phys. Rev. Lett.}{\bf 72}, 3004 (1994).
\bibitem{resonanceCmsNote} K. Gumus {\it et al}., {\it CMS sensitivity to dijet resonances}, CMS Note 2006/070
(2006), URL http://cdsweb.cern.ch/record/962025.
\bibitem{D0stat} I. Bertram {\it et al},``A Recipe for the Construction of Confidence Limits'', FERMILAB-TM-2104 (2000).
\bibitem{statLuc} L. Dermortier, in {\it Proceedings of the Conference on Advanced Statistical Techniques
in Particle Physics} (Institute for Particle Physics Phenomenology, University of Durham, UK, 2002),p. 18.
\bibitem{MC09} ATLAS Collab., {\it ATLAS Monte Carlo turnes for MC09}, ATLAS Note ATL-PHY-PUB-2010-002.
\bibitem{MRST2007} A. Sherstnev and R. S. Thorne, {\it Eur. Phys. J. C.} {\bf 55}, 553 (2008).
\bibitem{CalcHEP} A. Pukhov, arXiv:hep-ph/0412191.
\bibitem{MADGRAPH} J. Alwall {\it et al.}, arXiv:hep-ph/1106.0522.
\bibitem{alfonso} A. Zerwekh from ATLAS Collaboration, private communication.
\end{thebibliography}
